\documentclass[11pt,a4paper]{article}
\usepackage{jheppub,xcolor}
\pdfoutput=1
\usepackage[utf8]{inputenc}
\graphicspath{{./figs/}}

\usepackage{amssymb}
\usepackage{dcolumn}	% Align table columns on decimal point
\usepackage{bm}			% bold math
\usepackage{bbm} 		% blackboard math for \mathbbm{1}
\usepackage{multirow}
\usepackage{slashed}
\usepackage{blkarray}
\usepackage{booktabs}
\usepackage{makecell}
\usepackage{longtable}
\usepackage{placeins}
\usepackage{graphicx}
\usepackage{float}
\usepackage[font=small,labelfont=bf]{caption}

\definecolor{Red}{rgb}{1.,0.,0.}

\newcommand{\lag}{\mathcal{L}}
\newcommand{\mcM}{\mathcal{M}}
\newcommand{\mcO}{\mathcal{O}}

\newcommand{\phSpa}{\ensuremath{\phantom{\frac{C^Q}{1}}}\hspace{-1.75em}}

\newcommand{\Ohq}{\mcO_{\varphi q}^{(1)}}
\newcommand{\Ohqt}{\mcO_{\varphi q}^{(3)}}
\newcommand{\Ohu}{\mcO_{\varphi u}}
\newcommand{\Ohd}{\mcO_{\varphi d}}

\newcommand{\chq}{c_{\varphi q}^{(1)}}
\newcommand{\chqt}{c_{\varphi q}^{(3)}}
\newcommand{\chu}{c_{\varphi u}}
\newcommand{\chd}{c_{\varphi d}}

\newcommand{\cfqtrip}{c_{\varphi q}^{(3)}}

\newcommand{\bc}{\begin{center}}
\newcommand{\ec}{\end{center}}
\newcommand{\ba}{\begin{array}}
\newcommand{\ea}{\end{array}}

\newcommand{\TeV}{\,\mathrm{TeV}}
\newcommand{\GeV}{\,\mathrm{GeV}}

\def\Re{{\rm Re\,}}

\title{Revisiting $\mathbf{Vh(\rightarrow b\bar b)}$ at the LHC and FCC-hh}

\author[a]{Fady~Bishara,}
\author[a,b]{Philipp~Englert,}
\author[a,b]{Christophe~Grojean,}
\author[c]{Giuliano~Panico}
\author[d]{and Alejo~N.~Rossia}
\affiliation[a]{Deutsches Elektronen-Synchrotron DESY, Notkestr. 85, 22607 Hamburg, Germany}
\affiliation[c]{Universit\`{a} di Firenze and INFN Firenze, Via Sansone 1, 50019 Sesto Fiorentino, Florence, Italy}
\affiliation[b]{Institut f{\"u}r Physik, Humboldt-Universit{\"a}t zu Berlin, 12489 Berlin, Germany}
\affiliation[d]{Dept. of Physics and Astronomy, University of Manchester, Manchester M13 9PL, UK}

\emailAdd{fady.bishara@desy.de}
\emailAdd{philipp.englert@desy.de}
\emailAdd{christophe.grojean@desy.de}
\emailAdd{giuliano.panico@unifi.it}
\emailAdd{alejo.rossia@manchester.ac.uk}

\date{\today}
\abstract{
Diboson production processes provide good targets for precision measurements at present and future hadron colliders.
We consider $Vh$ production, focusing on the $h \to b\bar b$ decay channel, whose sizeable cross section makes it accessible at the LHC.
We perform an improved analysis by combining the 0-, 1- and 2-lepton channels with a scale-invariant $b$-tagging algorithm that allows us to exploit events with either a boosted Higgs via mass-drop tagging or resolved $b$-jets.
This strategy gives sensitivity to 4 dimension-6 SMEFT operators that modify the $W$ and $Z$ couplings to quarks and is competitive with the bounds obtained from global fits.
The benefit of the $h\to b\bar b$ decay channel is the fact that it is the only $Vh$ channel accessible at the LHC Run 3 and HL-LHC, while at FCC-hh it is competitive with the effectively background-free $h\to \gamma\gamma$ channel assuming $\lesssim 5\%$ systematic uncertainty.
Combining the boosted and resolved categories yields a 17\% improvement on the most strongly bounded Wilson coefficient at the LHC Run 3 with respect to the boosted category alone (and a 7\% improvement at FCC-hh).
We also show that, at FCC-hh, a binning in the rapidity of the $Vh$ system can significantly reduce correlations between some EFT operators.
The bounds we obtain translate to a lower bound on the new physics scale of $5$, $8$, and $20$\;TeV at the LHC Run 3, HL-LHC, and FCC-hh respectively, assuming
new-physics couplings of order unity.
Finally, we assess the impact of the $Vh$ production channel on anomalous triple gauge coupling measurements, comparing with their determination at lepton colliders.
}


\begin{document}
\begin{flushright}
DESY 22-136\\
HU-EP-22/27\\
\end{flushright}

\maketitle
\flushbottom

%%%%%%%%%%%%%%%%%%%%%%%%%%%%%%%%%%%%%%%%%%%%%%%%%%%%%%%%%%%%%%%%%%%%%%%%
\section{Introduction}
\label{sec.intro}
%%%%%%%%%%%%%%%%%%%%%%%%%%%%%%%%%%%%%%%%%%%%%%%%%%%%%%%%%%%%%%%%%%%%%%%%

Precision measurements of electroweak (EW) and Higgs processes provide a fruitful approach for testing the Standard Model (SM) and exploring the landscape of Beyond-the-SM (BSM) theories. Thanks to significant theoretical and experimental improvements, their relevance and impact at the LHC is steadily growing and is expected to become even more prominent with the high-luminosity LHC program (HL-LHC). Electroweak processes are also one of the primary targets of future hadron and lepton colliders, thus their study is essential to assess the potential physics reach of these machines.

Some of the most powerful indirect probes of BSM dynamics rely on new-physics effects that grow with energy and are more easily accessible in the tails of the kinematic distributions.
Hadron colliders have a potential advantage in this case thanks to their extended energy range~\cite{Farina:2016rws,deBlas:2013qqa}. 
However a careful identification of suitable processes (and analysis strategies) is essential to ensure that experimental and theoretical systematic uncertainties can be kept under control.

Provided that the threshold for direct production of new particles is high enough, new-physics effects can be captured by a finite set of effective field theory (EFT) operators. This approach allows one to probe new-physics in a largely model independent way.
An interesting target of EW precision measurements at hadron colliders is given by diboson production channels. Such processes can be exploited to study Higgs dynamics at high energies which are modified in a large class of BSM scenarios.
Thus, several EFT operators involving the Higgs can be tested in diboson production and a subset of them generate new-physics effects that grow with energy. In particular the four so-called primary dimension-6 operators~\cite{Franceschini:2017xkh}, $\Ohqt$, $\Ohq$, $\Ohu$ and $\Ohd$ in the Warsaw basis, give rise to amplitudes whose interference with the SM grows quadratically with the center-of-mass energy of the event.

In this paper we focus on two diboson channels involving the Higgs boson.
Namely, associated production with a $W$ or a $Z$ boson. The peculiarity of these channels is the fact that their leading SM amplitude is the one involving a longitudinally polarized vector boson. This helicity configuration is present at leading order in the EFT expansion of the squared amplitude.

Since we are interested in performing precision measurements in the high-energy tails of the kinematic distributions, we are forced to consider Higgs decay channels with large branching ratios. This is especially true at the LHC (and HL-LHC), where the number of events is relatively small. For this reason we will focus on the $h \to b \bar b$ decay channel.\footnote{Cleaner decay channels, such as $h\to\gamma\gamma$, can be measured at LHC but only in the low-energy regime and with very low statistics, making them of limited interest for BSM searches~\cite{CMS:2021kom}.}
Consequently, to suppress the backgrounds as much as possible, we consider only vector boson decays into charged leptons and neutrinos.

At future high-energy hadron colliders, thanks to larger cross sections and increased integrated luminosities, additional decay channels could be accessible for precision measurements in the tail of the $Vh$ distributions. In refs.~\cite{Bishara:2020pfx,Bishara:2020vix}, the leptonic $Vh(\to \gamma \gamma)$ processes were considered at FCC-hh, and it was found that they provide good sensitivity to energy-growing new-physics effects. In the present work we complement those studies by considering the leptonic $Vh(\to b\bar b)$ process as well. As we will see, depending on the achievable level of systematic uncertainty, the $h \to b \bar b$ final state can provide bounds competitive with the $h \to \gamma \gamma$ one.

The large, QCD induced, $Vbb$ background makes enhancing the sensitivity to the signal, $Vh(\to b\bar b)$, challenging. 
However, since we are interested in accessing very energetic events where the BSM contribution is sizeable, the $b$-quarks generated by Higgs decay tend to be boosted and collimated.
This, along with the large and peaked invariant mass of the $b$-quark pair from Higgs decays, makes jet substructure techniques, and in particular mass-drop tagging~\cite{Butterworth:2008iy}, a crucial tool in extracting the signal and suppressing the background.

Precision measurements in the $Vh$ production process with a boosted Higgs decaying to $b$-quarks have already been considered in refs.~\cite{Banerjee:2018bio,Liu:2018pkg,Banerjee:2019pks,Banerjee:2019twi,Banerjee:2021efl}, where the final state channels including either $1$ or $2$ charged leptons were studied.
A more complete analysis has been presented by the ATLAS Collaboration in refs.~\cite{ATLAS:2020fcp,ATLAS:2020jwz} for LHC Run 2 data. In these works, the final states with $0$ charged leptons were also included and the Higgs boson candidates were reconstructed from either resolved~\cite{ATLAS:2020fcp} or boosted jets~\cite{ATLAS:2020jwz}.

In the present work, we revisit the $Vh(\to b\bar b)$ production processes combining the study of the three leptonic decay channels ($0$, $1$ and $2$ charged leptons) with the characterization of events with either two resolved $b$-jets or a boosted Higgs candidate. For the event classification, we use a scale-invariant $b$-tagging strategy adapted from refs.~\cite{Gouzevitch:2013qca,Bishara:2016kjn}, which allows us to split the events in mutually exclusive categories.
We perform a detailed analysis for the current LHC run (LHC Run 3) and for the end of the HL-LHC program, comparing our results with the sensitivity expected from other diboson channels (in particular $WZ$)
and from global EFT fits. In addition, we assess the relevance of the $Vh(\to b \bar b)$ channels at FCC-hh, highlighting their interplay with the $Vh(\to \gamma\gamma)$ channel and their complementarity with global fits performed at future lepton colliders, specifically FCC-ee.

The rest of this paper is organised as follows. In section~\ref{sec:TheoBack}, we review the parametrization of BSM effects entering the $Vh$ processes in the framework of the Standard Model Effective Field Theory (SMEFT). We also briefly summarize the main features of the corresponding helicity amplitudes. In section~\ref{sec:signal_background}, we explain in detail our event simulation, $b$-tagging and analysis strategy. We also include a comparison of the signal yield and of the size of the main backgrounds. In section~\ref{sec:results} we collect and analyze our projected bounds at LHC Run 3, HL-LHC and FCC-hh, comparing them with the ones from other studies. Finally, we summarize and discuss our work in a broader context in section~\ref{sec:conclusions}.
Several appendices, in which we provide the full technical details of our analysis, as well as a more complete list of results, can be found at the end of the paper.

\section{Theoretical background}
\label{sec:TheoBack}
Effective Field Theories provide a powerful framework to parameterize deviations from the SM predictions in a model-independent way. In our analysis, we employ the Standard Model Effective Field Theory (SMEFT) in the Warsaw basis~\cite{Grzadkowski:2010es} restricted to operators of dimension $6$. For definiteness, we only consider CP-preserving operators and simplify the flavour structure of the effective operators assuming flavor universality (see discussion on the impact of these choices in ref.~\cite{Bishara:2020pfx}).

Under these assumptions, there exist only $4$ independent operators which modify the $Vh$ production process at leading order and give rise to interference terms with the SM that grow with energy. These are,
\begin{eqnarray}
{\cal O}_{\varphi q}^{(1)} &=&\left(\overline{Q}_{L}  \gamma^{\mu} Q_{L}\right)\left(i H^{\dagger}  \overset{\leftrightarrow}{D}_{\mu} H\right)\,,\label{eq:Ofq1} \\
{\cal O}_{\varphi q}^{(3)} &=&\left(\overline{Q}_{L} \sigma^{a} \gamma^{\mu} Q_{L}\right)\left(i H^{\dagger} \sigma^{a} \overset{\leftrightarrow}{D}_{\mu} H\right)\,, \\
\Ohu &=&\left(\overline{u}_{R}\gamma^{\mu} u_R\right)\left(i H^{\dagger} \overset{\leftrightarrow}{D}_{\mu} H\right)\,, \\
\Ohd &=&\left(\overline{d}_{R}\gamma^{\mu} d_R\right)\left(i H^{\dagger} \overset{\leftrightarrow}{D}_{\mu} H\right)\,,\label{eq:Ofd}
\end{eqnarray}
where $\overset{\leftrightarrow}{D}_{\mu}\; =\overset{\rightarrow}{D}_{\mu}- (\overset{\leftarrow}{D}_{\mu})^\dagger$ and $\sigma^a$ are the Pauli matrices. We write the corresponding Wilson coefficients as dimensionful quantities.

Although the effective operators we consider could modify the decays of the Higgs, $W$, and $Z$ bosons, we can safely neglect these effects since the branching ratios of these particles are already experimentally known to agree with the SM prediction with high accuracy. For sizeable Wilson coefficients, if the operators we consider induce large corrections to the SM particles decay fractions, these effects should be cancelled by correlated contributions from other effective operators such as  $\mcO_{\varphi\ell}^{(1)}$, $\mcO_{\varphi\ell}^{(3)}$, $\mcO_{\varphi e }$, $\mcO_{\varphi WB}$, $\mcO_{\ell\ell,1221}$, $\mcO_{\varphi D}$ and $\mcO_{b\varphi }$ (see their definition in ref.~\cite{Falkowski:2001958}). The latter, since they do not induce energy-growing effects in the processes we are considering, will not affect significantly our analysis.

It must be mentioned that several other operators can modify the $Vh$ production process with sub-leading effects, for instance the CP-violating purely bosonic operators.\footnote{A detailed discussion on those operators and why their contributions are suppressed can be found in ref.~\cite{Rossia:2021fsi}.} Some of these can be probed with a dedicated analysis of the $Wh$ channel as shown in ref.~\cite{Bishara:2020vix} for the $h \to \gamma\gamma$ decay channel at FCC-hh. We leave the extension of such studies to the $h\to b\bar{b}$ channel for future work.

To understand the possible impact of the proposed analysis on the search for new physics, it is important to characterize the BSM scenarios that can give rise to the operators we consider.
It is straightforward to check that ${\cal O}_{\varphi q}^{(1)}$, ${\cal O}_{\varphi q}^{(3)}$, $\Ohu$ and $\Ohd$ can be generated at tree-level via the exchange of EW-charged vector or fermionic resonances. According to the SILH power counting~\cite{Giudice:2007fh}, the expected size of their Wilson coefficients is
\begin{equation}
\chqt \sim \chq \sim \chu \sim \chd \sim \frac{g_{*}^2}{m^2}\,,    
\end{equation}
where $m$ is the mass of the exchanged resonance and $g_{*}$ is the coupling of the new particles to the SM ones. In weakly-coupled theories, one expects $g_{*}\sim g$ with $g$ the typical EW coupling (for definiteness we take the SM ${\rm SU}(2)_L$ coupling, $g = 0.65$), while strongly-coupled theories have $g_{*} \gtrsim 1$, reaching $g_{*}\sim 4\pi$ in the fully strongly-coupled case. A more detailed discussion about the BSM interpretation can be found in refs.~\cite{Bishara:2020vix,Bishara:2020pfx}.

\subsection{Interference patterns}\label{sec:Interference}

The scattering amplitude of the $Vh$ production process in the SM and its interference patterns with dimension-6 SMEFT operators have been studied in detail in refs.~\cite{Franceschini:2017xkh,Bishara:2020vix,Bishara:2020pfx}. Here, we will just summarize the main features and point out how they influence our analysis strategy. 

\begin{figure}[t]\centering
	\includegraphics[trim={0 0 3.5cm 0},clip,scale=1]{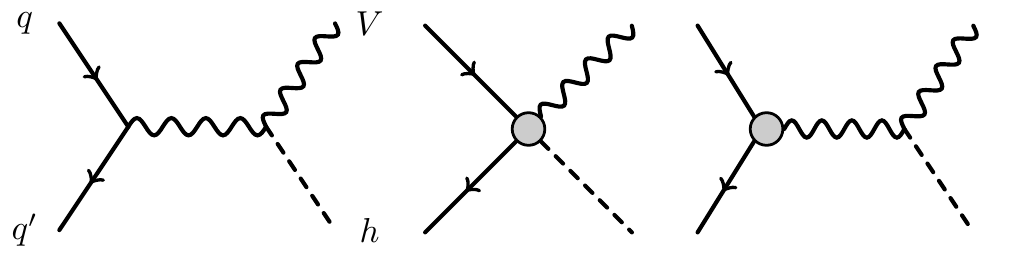}
	\caption{Feynman diagrams contributing to the $qq'\to Vh$ amplitude with up to one insertion of a dimension-6 operator from eqs.~\eqref{eq:Ofq1}--\eqref{eq:Ofd} as depicted by a gray circle.}
	\label{fig:mod_vh}
\end{figure}

The four dimension-6 operators we consider give rise to a modification of the $\bar{q} q V$ gauge couplings and also generate a contact term $\bar{q} q V h$ (see figure~\ref{fig:mod_vh}). They all interfere with the leading SM amplitude producing contributions that grow with the square of the centre-of-mass energy, $\hat{s}$, with respect to the SM squared amplitude. At high energy, the squared SM amplitude and the interference term for the different operators have the following behavior
\begin{equation}
	\begin{split}
		\left| \mcM_{SM} \right|^2 &\sim \sin^2 \theta \,,\\
		\Re\mcM_{SM} \, \mcM_{\textsc{bsm}}^* &\sim \frac{\hat s}{\Lambda^2} \sin^2 \theta\,,
	\end{split}\label{eq:main_amp_sq}
\end{equation}
where $\mcM_\textsc{bsm}\in\{\mcM_{\varphi q}^{(3)},\mcM_{\varphi q}^{(1)}, \mcM_{\varphi u}, \mcM_{\varphi d}\}$ and $\theta$ is the scattering angle of the vector boson with respect to the beam axis.
This is the same both in the $Zh$ and $Wh$ channels, although only $\Ohqt$ contributes to the latter.

Due to the presence of the Higgs boson in the final state, the leading amplitude at high energy is the one with a longitudinally polarized vector boson, both in the SM and in the amplitudes generated by the operators of interest.\footnote{The full expression of the helicity amplitudes can be found in refs.~\cite{Bishara:2020pfx,Bishara:2020vix,Rossia:2021fsi}.} 
Energy-growing interference effects therefore modify the leading SM helicity channel and are easily accessible through an analysis inclusive in the kinematics of the decay products.
Nevertheless, a differential analysis could be sensitive to the sub-leading contributions generated by other operators~\cite{Bishara:2020vix} but we will not consider such case in this work.

The helicity interference pattern is quite different from what happens in the other diboson channels, $WZ$ and $WW$, where the dominant SM amplitude is the one in which the gauge bosons have (opposite) transverse polarization. Since the main BSM effects are always confined to the longitudinally-polarized channels, the interference terms in the $WZ$ and $WW$ channels are suppressed with respect to the leading SM contributions. This makes their determination harder, requiring dedicated selection cuts to enhance the BSM effects.

Despite the fact that the $\Ohqt$, $\Ohq$, $\Ohu$, and $\Ohd$ operators generate amplitudes with a similar structure, there are striking differences in the size of the interference terms in the $Zh$ production channel (recall that only $\Ohqt$ contributes to the $Wh$ channel).
The interference generated by the $\Ohq$ term is affected by a cancellation between the contributions of the up- and down-type quarks. Additionally, a suppression affects the interference generated by $\Ohu$ and $\Ohd$ due to the SM coupling between the $Z$ boson and the right-handed quarks~\cite{Bishara:2020vix}. Hence, we expect to obtain a much better sensitivity to $\Ohqt$ than to the rest of Wilson coefficients.

The cancellation that affects $\Ohq$ can be partially lifted with a binning in the rapidity of the $Zh$ system~\cite{Bishara:2020pfx}. However, we will use this additional binning only in our analysis for FCC-hh, since its potential at (HL-)LHC is limited by the relatively low number of signal events.

\section{Event generation and analysis}\label{sec:signal_background}

In this section we concisely report the set-up of our analysis, including the
details regarding the Monte Carlo event generation. We discuss here only the most relevant points of our analysis strategy, and we refer the reader to the appendices for the complete technical details.

All the events were generated with \texttt{MadGraph5\_aMC@NLO} v.2.7.3~\cite{Alwall:2014hca} using the \texttt{NNPDF23} parton distribution functions~\cite{Ball:2013hta}. The simulation of the parton shower and the Higgs decay into $b\bar b$ pairs was performed with \texttt{Pythia8.24}~\cite{Sjostrand:2014zea}. The effective operators considered in our analysis were implemented at simulation level through the \texttt{SMEFTatNLO} \texttt{UFO} model~\cite{Degrande:2020evl,Degrande:2011ua}. All the signal processes were generated at NLO in the QCD coupling and we considered QED NLO effects via $k$-factors taken from ref.~\cite{Haisch:2022nwz}. Most of the background processes were simulated at NLO in the QCD coupling with a few exceptions specified in appendix~\ref{sec:AppMCdetails}. We made the conservative choice to not consider EW NLO/LO $k$-factors for the background processes.
More technical details about the simulations are reported in appendix \ref{sec:AppMCdetails}.

As we discussed in the previous section, thanks to the energy growth of BSM effects, most of the sensitivity to new physics comes from events in the high-energy tail of the kinematic distributions. In such region the Higgs boson decay products are significantly boosted, giving rise to very specific kinematic features and providing an additional handle to distinguish signal from background events.

To fully exploit the high-energy signal events, we find it advantageous to classify the events according to the presence of a boosted Higgs candidate or a pair of resolved $b$-jets.
To this end, we follow the scale-invariant tagging procedure~\cite{Gouzevitch:2013qca} as implemented in ref.~\cite{Bishara:2016kjn} to combine the two categories by looking for a boosted Higgs candidate first and then, if the event does not contain one, look for two resolved $b$-jets.
This strategy leads to an improvement on the bound for $\cfqtrip$ of 17\%, 11\%, and 7\% at the LHC Run 3, HL-LHC, and FCC-hh, respectively, with respect to bounds obtained from the boosted category alone.\footnote{Since the upper and lower bounds are not symmetric around zero, we choose to pick the weaker one in absolute value for this comparison to be consistent with the bounds shown later in Figs.~\ref{fig:fit_cphiq_HL_LHC} and~\ref{fig:fit_cphiq_FCC}.}
To reconstruct a boosted Higgs we use the mass-drop-tagging procedure~\cite{Butterworth:2008iy}  and require it to have exactly two $b$ tags -- see appendix~\ref{sec:btagger_details} for further details.

In a nutshell, the scale-invariant tagging strategy classifies an event into one of two mutually-exclusive categories: boosted and resolved. 
If an event contains a jet with a mass-drop-tag (MDT), the two constituents that triggered the mass-drop condition both carry a b-tag, and its mass falls into the Higgs window ($90<m<120$ GeV), then the event is classified into the boosted category.
Else, if the event contains two b-tagged jets with an invariant mass in the Higgs window, then it is classified as resolved.
Otherwise, if the events fails to qualify for either category, it is rejected.
This procedure can be illustrated schematically as follows:
\begin{equation*}
\includegraphics[scale=1]{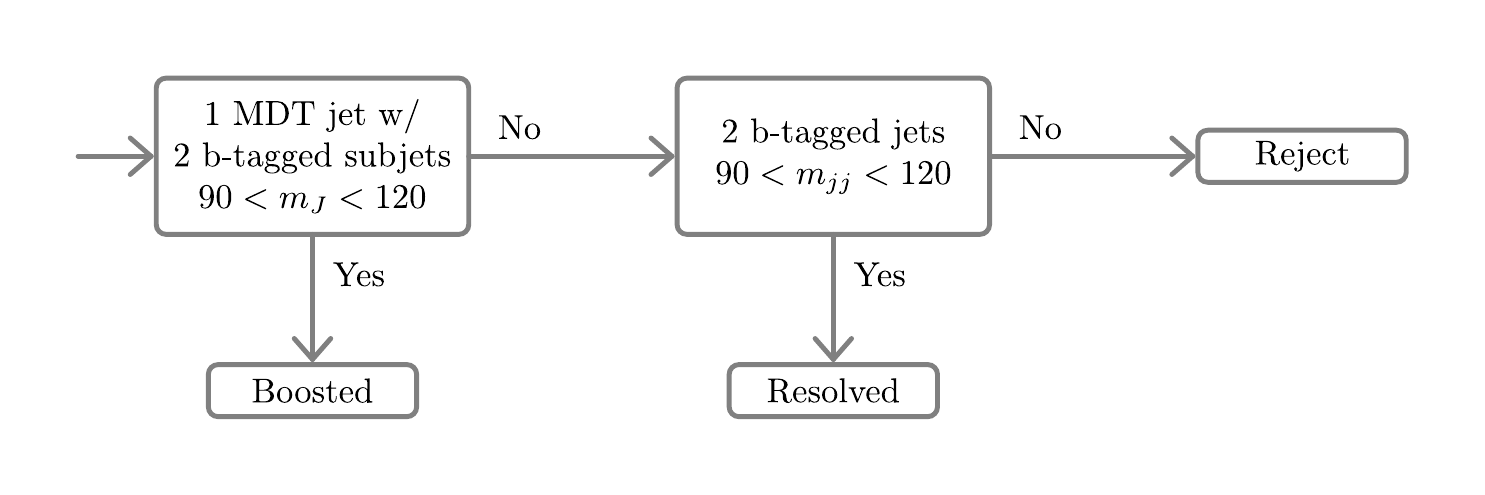}
\end{equation*}

The events considered in our analysis can be furthermore split in three categories, according to the number of final state leptons:
\begin{itemize}
\item \emph{Zero-lepton} category. The main signal process is $pp\rightarrow Z(\rightarrow \nu \bar \nu)h(\rightarrow b \bar b)$.\footnote{We did not include the $gg\rightarrow Zh$ channel in our simulations, for a discussion, see ref.~\cite{Bishara:2020pfx}. This is valid also for the two-lepton category. See ref.~\cite{Yan:2021veo} for a $\kappa$ framework analysis involving only third generation couplings.} An important additional contribution comes from $pp\rightarrow W(\rightarrow \ell \nu)h(\rightarrow b \bar b)$, where $\ell=e,\, \mu,\, \tau$ and $\nu=\nu_e,\, \nu_\mu,\, \nu_\tau$, with a missing charged lepton. Most of the $Wh$ contribution to the zero-lepton category comes from the $W\to\tau \nu_\tau$ decay channel (see discussion in ref.~\cite{Bishara:2020pfx}). The main backgrounds in this category come from $Wb\bar b$, $Zb\bar b$ and $t \bar t$ production.

\item \emph{One-lepton} category. The signal in this category comes from the $pp\rightarrow W(\rightarrow \ell \bar \nu)h(\rightarrow b \bar b)$ process, with  $\ell=e,\, \mu$ and $\nu=\nu_e,\, \nu_\mu$, where the charged lepton is detected.
The main backgrounds in this category are $Wb\bar b$ and $t \bar t$ production.

\item \emph{Two-lepton} category. The signal comes from the $pp\rightarrow Z(\rightarrow \ell^+ \ell^-)h(\rightarrow b \bar b)$ process, with $\ell=e,\, \mu$, while the main background is $Zb\bar b$ production.

\end{itemize}

\subsection{Selection cuts}
\label{sec:cuts}

The selection cuts we applied are mainly derived from the analogous studies performed by the ATLAS collaborations on LHC Run 2 data~\cite{ATLAS:2020fcp, ATLAS:2020jwz}. Some differences come from the fact that we partially optimized the cuts on the basis of the distributions we obtained from our simulations, which, in some cases, slightly differ from the ones reported in the ATLAS studies.
For the LHC analyses the differences with respect to the ATLAS cuts are very mild, while, as expected, significant modifications are needed for the FCC-hh analysis.

For the boosted and resolved categories described in the previous subsection, we apply different cuts and we treat them as uncorrelated observables when computing the $\chi^2$.
In the following, we summarize the cuts which are most effective in improving the signal-to-background ratio. Although the threshold values used for these cuts differ at (HL-)LHC and FCC-hh, their overall impact is similar.

As can be expected, since the $b$-quarks in the background processes do not come from a Higgs, one of the most efficient cuts for rejecting background events is the one on the invariant mass of the Higgs candidate ($m_{bb}$).
This cut leaves the signal almost unaffected regardless of which category the events belong to.

The $t\bar t$ background, whose cross section is particularly high in the 0- and 1-lepton categories, can be strongly reduced through the requirement of reconstructing a boosted or a resolved Higgs. Further reduction is achieved through a veto on additional untagged jets.
The jet veto also helps to control the $Wb\bar b$ and $Zb \bar b$ backgrounds in the 0- and 1-lepton categories, although it significantly reduces the $Wh$ signal in the 0-lepton category.
Finally, a cut on the maximum $p_{T}^{\ell}$ imbalance, defined as $|p_{T}^{\ell_1}-p_{T}^{\ell_2}|/p_{T}^{Z}$, is very helpful for improving the signal-to-background ratio in the analysis of the boosted events in the 2-lepton category. We summarize the most important selection cuts in Table~\ref{tab:sel_cuts_summary}.
More details on the selection cuts and acceptance regions, as well as the cut efficiencies can be found in appendix~\ref{sec:AppCuts}.

\begin{table}[htb!]
	\centering{
		\renewcommand{\arraystretch}{1.37}
    \begin{tabular}{ >{\centering\arraybackslash} m {0.18\textwidth} |  >{\centering\arraybackslash}m{0.3\textwidth} | >{\centering\arraybackslash}m{0.3\textwidth}  }
			\toprule
			\multirow{2}*{Channel} & \multicolumn{2}{ c }{Selection cuts at HL-LHC (FCC-hh)}\\
			\cline{2-3} & Boosted & Resolved \\ \midrule
			\multirow{3}*{0-lepton} & \multicolumn{2}{c}{$m_{h_{cand}} \in [90,120] \GeV$}\\
                                        & \multicolumn{2}{c}{$0$ untagged jets in accept. region}\\
			                          &  &  $\Delta R_{bb}\leqslant 1.5$\\
                             \hline
                \multirow{4}*{1-lepton} & \multicolumn{2}{c}{$m_{h_{cand}} \in [90,120] \GeV$}\\
                                        & \multicolumn{2}{c}{$0$ untagged jets in accept. region}\\
			                          & $|\Delta y(W,h_{cand})|\leqslant 1.4\, (1.2)$ &  $\Delta R_{bb}\leqslant 2.0$\\
                                        & $|\eta^{h_{cand}}|\leqslant 2.0\, (4.5)$ & \\
                                        \hline
                \multirow{3}*{2-lepton} & \multicolumn{2}{c}{$m_{h_{cand}} \in [90,120] \GeV$}\\
                                        & $\frac{p_{T}^{\ell_1}-p_{T}^{\ell_2}}{p_{T}^{Z}}\leqslant 0.8\, (0.5)$ & $0$ untagged jets\\
			                          & $|\Delta y (Z,h_{cand})|\leqslant 1.0$ &  in accept. region \\
	\bottomrule
		\end{tabular}}
		\caption{Summary of the most important selection cuts in all the channels and categories the LHC and FCC-hh analyses. Values between parenthesis were used for the FCC-hh analyses. Numbers outside parenthesis were used for the LHC analysis or both. The jet acceptance region is defined in appendix~\ref{sec:AppCuts}.}
		\label{tab:sel_cuts_summary}
	\end{table}

\subsection{Binning}
\label{sec:binning}
As can be seen from eq.~(\ref{eq:main_amp_sq}), new physics effects are maximal in the central scattering region. This suggests that binning in the transverse momentum of the Higgs and vector boson could provide better sensitivity than binning in the center of mass energy.
Analogously to the analysis in ref.~\cite{Bishara:2020pfx}, for both the 0- and the 2-lepton channels, we use as binning variable the minimum $p_T$ of the Higgs and the vector boson
\begin{align}
    p_{T,\mathrm{min}} \equiv \min\{p_T^h, p_T^Z\}\,.
\end{align}
In the 1-lepton channel, instead, we bin in the transverse momentum of the Higgs boson $p_T^h$. The definition of the bins is reported in Table \ref{tab:bin_boundaries}.
\begin{table}[t]
	\centering{
		\renewcommand{\arraystretch}{1.25}
		\begin{tabular}{  c | c | c | c | c  }
		\toprule
		\multicolumn{2}{c|}{Categories} & Variable & (HL-)LHC & FCC-hh \\
		\midrule
		 & boosted & \multirow{2}*{$p_{T,\mathrm{min}}\, \mathrm{[GeV]}$ } & $\{0, 300, 350,\infty\}$ & $\{0, 200, 400, 600, 800, \infty\}$ \\
		\multirow{-1}{*}[1.9ex]{0-lepton} & resolved & &$\{0, 160, 200, 250,\infty\}$ & $\{0, 200, 400,600, 800, \infty\}$ \\
		\midrule
		& boosted & \multirow{2}*{$p_{T}^h \, \mathrm{[GeV]}$} & $\{0, 175, 250, 300, \infty\}$ & $\{0, 200, 400, 600, 800, \infty\}$ \\
		\multirow{-1}{*}[1.9ex]{1-lepton} & resolved & &$\{0, 175, 250,\infty\}$ & $\{0, 200, 400, 600,\infty\}$ \\
		\midrule
		& boosted & \multirow{2}*{$p_{T,\mathrm{min}} \, \mathrm{[GeV]}$ } &$\{250,\infty\}$ & $\{0, 200, 400, 600,\infty\}$ \\
		\multirow{-1}{*}[1.9ex]{2-lepton} & resolved & & $\{175, 200,\infty\}$ & $\{0, 200, 400, 600,\infty\}$ \\
		\bottomrule
		\end{tabular}}
		\caption{Bins used in the analysis. In the 0- and 2-lepton cases, where both the vector and Higgs boson momenta can be reconstructed, the binning variable is the minimum of the two transverse momenta denoted $p_{T,\mathrm{min}}$ -- see main text for more detail.}
		\label{tab:bin_boundaries}
\end{table}

As explained in section~\ref{sec:Interference}, in the FCC-hh analysis, we introduce an additional binning in the rapidity of the Higgs in the 0-lepton category and in the rapidity of the $Zh$ system in the 2-lepton category:
\begin{align}
    |y_{\{h,Zh\}}|\in  [0,2],[2,6]\, .
\end{align}
The rapidity binning significantly enhances the sensitivity to the $\chq$ Wilson coefficient.

\begin{figure}[t]
	\centering
	\includegraphics[width=0.47\linewidth]{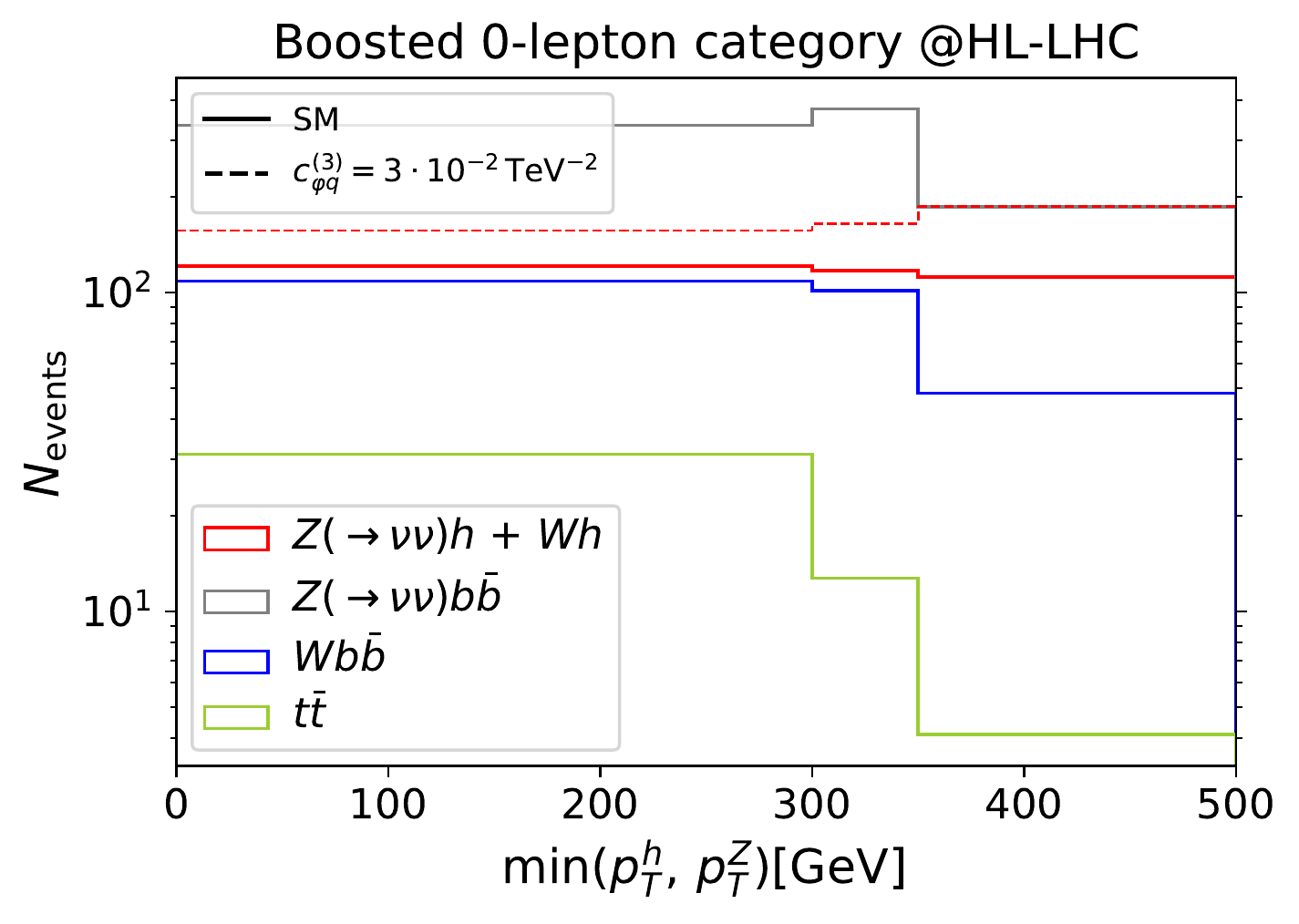}
	\hfill
	\includegraphics[width=0.47\linewidth]{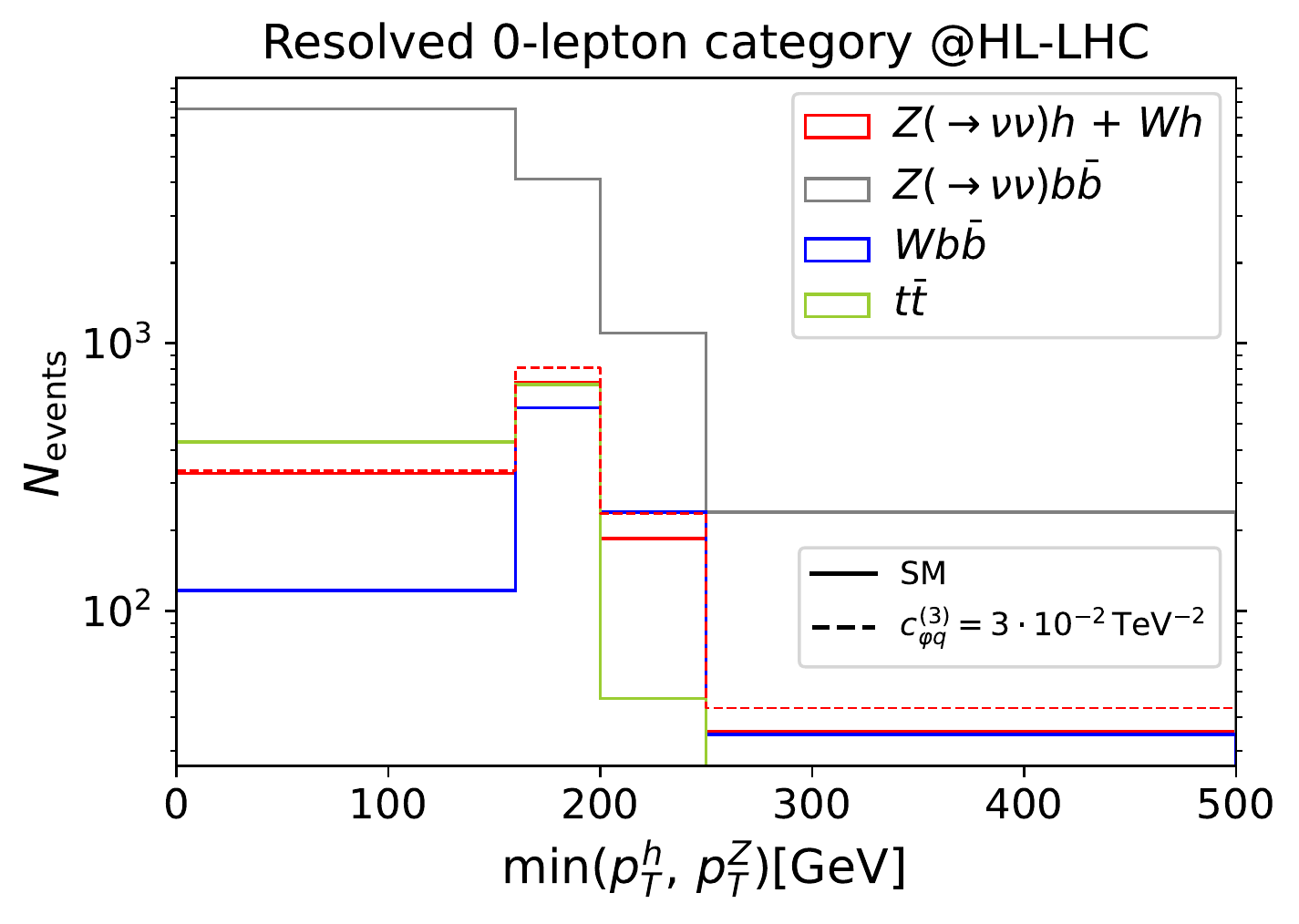}
	\caption{Number of SM signal and background events (solid lines) in the 0-lepton category after the selection cuts for HL-LHC. To visualize the BSM effects on the distributions, we are also plotting the signal setting ${\chqt=3\cdot 10^{-2}\,\mathrm{TeV}^{-2}}$ (dashed lines). The chosen value for the Wilson coefficient is slightly larger than the bounds we derived to increase the visibility of the BSM effects on the distributions. \,\,(\textbf{Left:} boosted category. \textbf{Right:} resolved category.)
 }
	\label{fig:histograms_vv}
\end{figure}

\begin{figure}[t]
	\centering
	\includegraphics[width=0.47\linewidth]{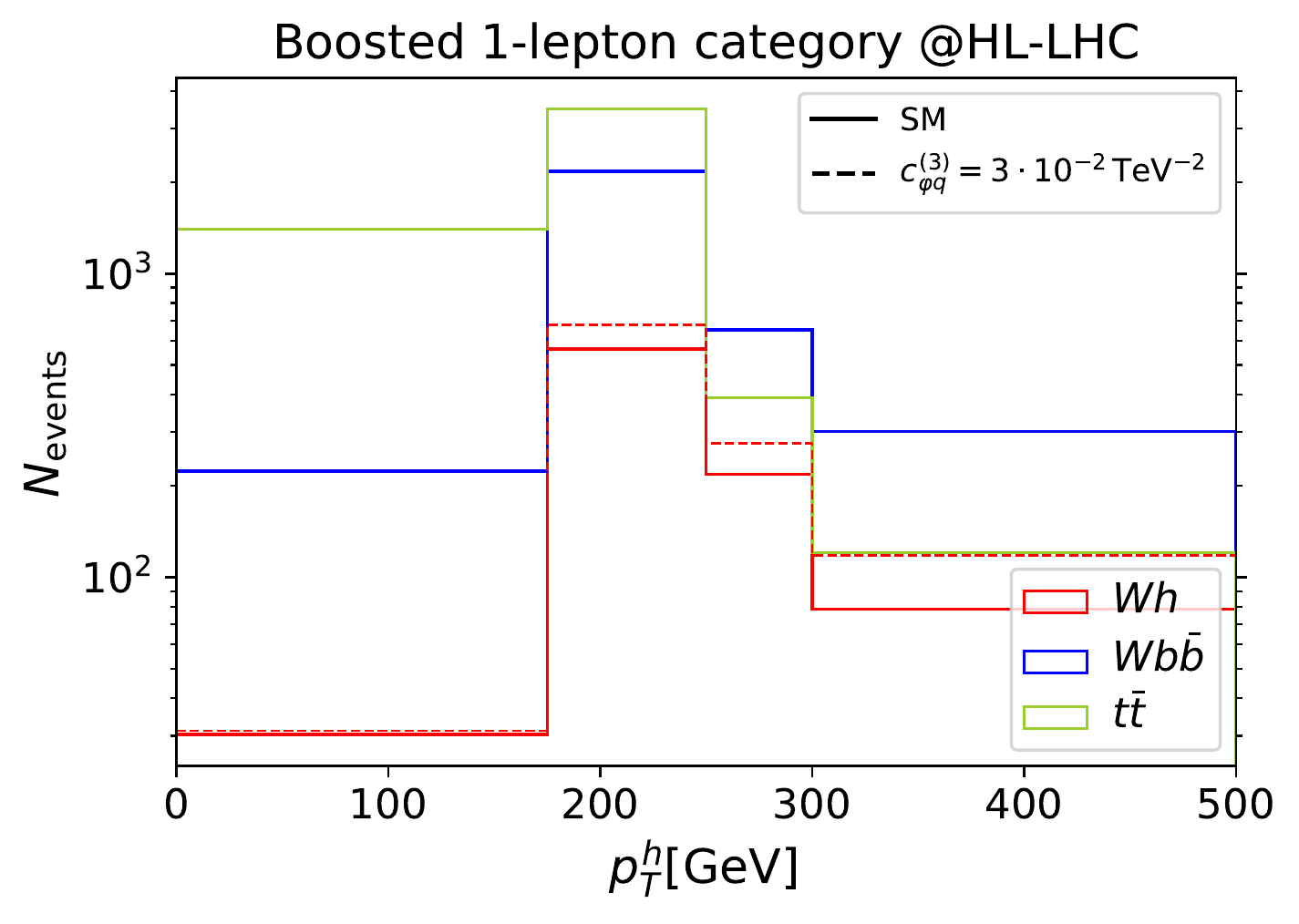}
	\hfill
	\includegraphics[width=0.47\linewidth]{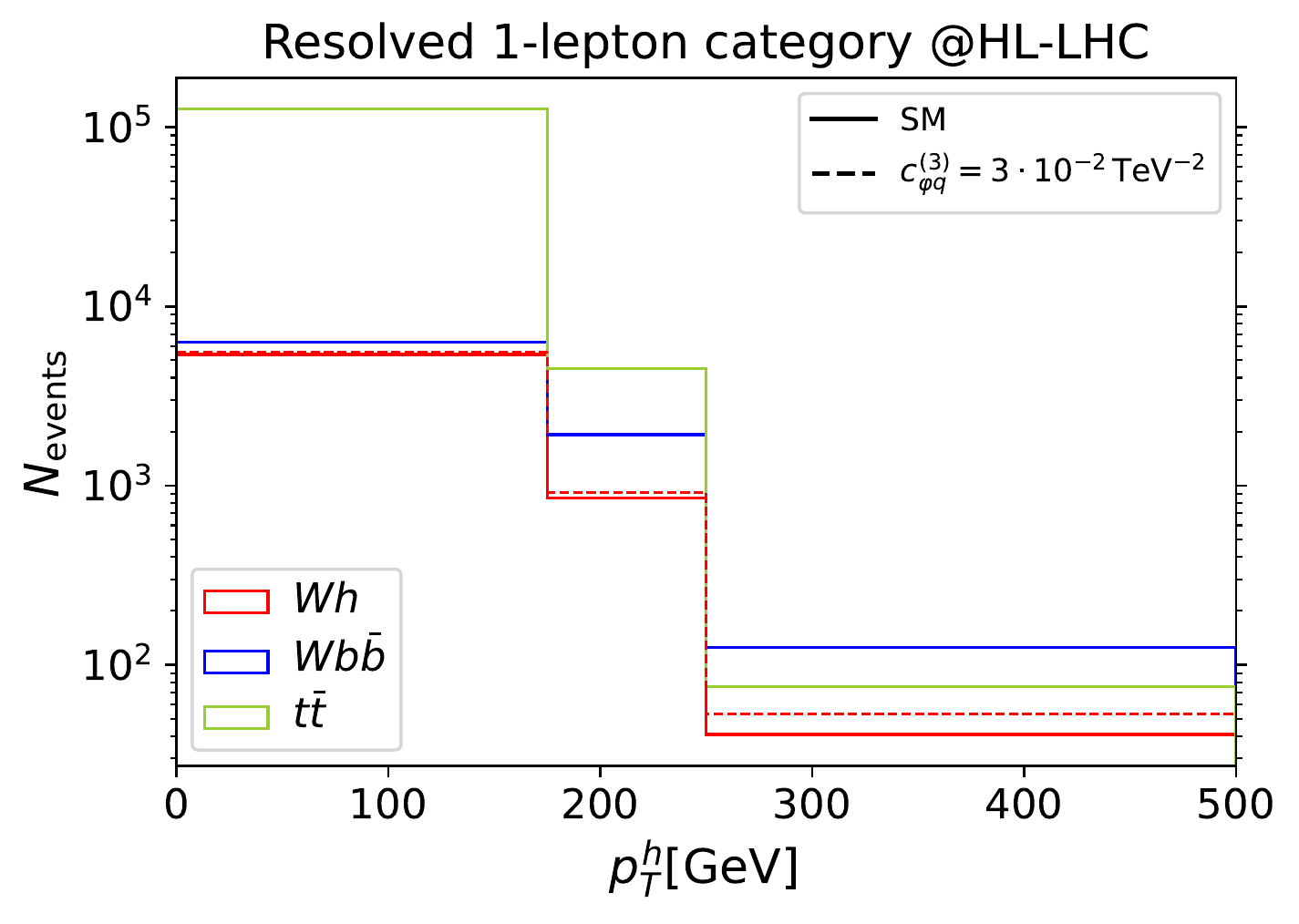}
	\caption{Number of SM signal and background events (solid lines) in the q-lepton category after the selection cuts for HL-LHC. To visualize the BSM effects on the distributions, we are also plotting the signal setting ${\chqt=3\cdot 10^{-2}\,\mathrm{TeV}^{-2}}$ (dashed lines). The chosen value for the Wilson coefficient is slightly larger than the bounds we derived to increase the visibility of the BSM effects on the distributions. \,\,(\textbf{Left:} boosted category. \textbf{Right:} resolved category.)
 }
	\label{fig:histograms_vl}
\end{figure}

\begin{figure}[!ht]
	\centering
	\includegraphics[width=0.47\linewidth]{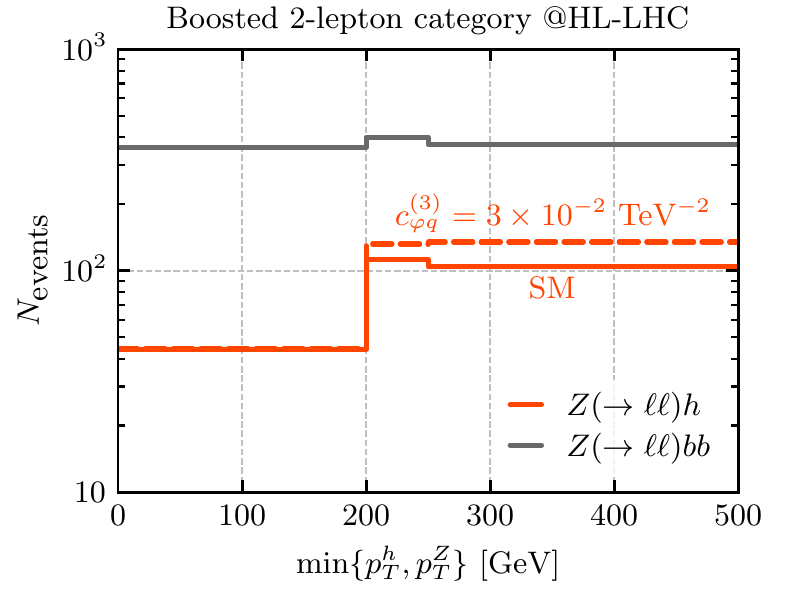}
	\hfill
	\includegraphics[width=0.47\linewidth]{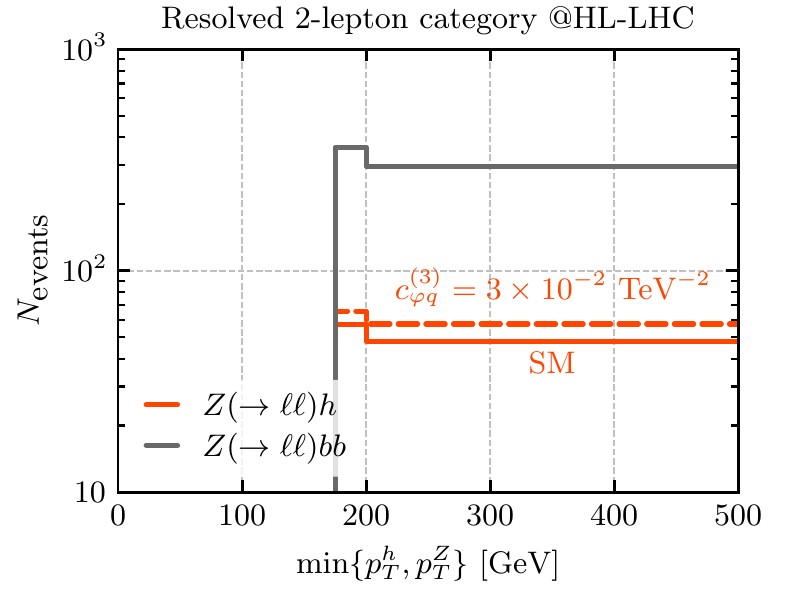}
	\caption{Number of SM signal and background events (solid lines) in the 2-lepton category after the selection cuts for HL-LHC. To visualize the BSM effects on the distributions, we are also plotting the signal setting ${\chqt=3\cdot 10^{-2}\,\mathrm{TeV}^{-2}}$ (dashed lines). The chosen value for the Wilson coefficient is slightly larger than the bounds we derived to increase the visibility of the BSM effects on the distributions. \,\,(\textbf{Left:} boosted category. \textbf{Right:} resolved category.) We do not show the BSM signal data for all the bins in these plots, since our fits are not reliable in the neglected bins due to limited statistics. However, we checked that the influence of these bins on the determination of the bounds is negligible.
 }
	\label{fig:histograms_ll}
\end{figure}

\begin{figure}[t]
	\centering
	\includegraphics[width=0.47\linewidth]{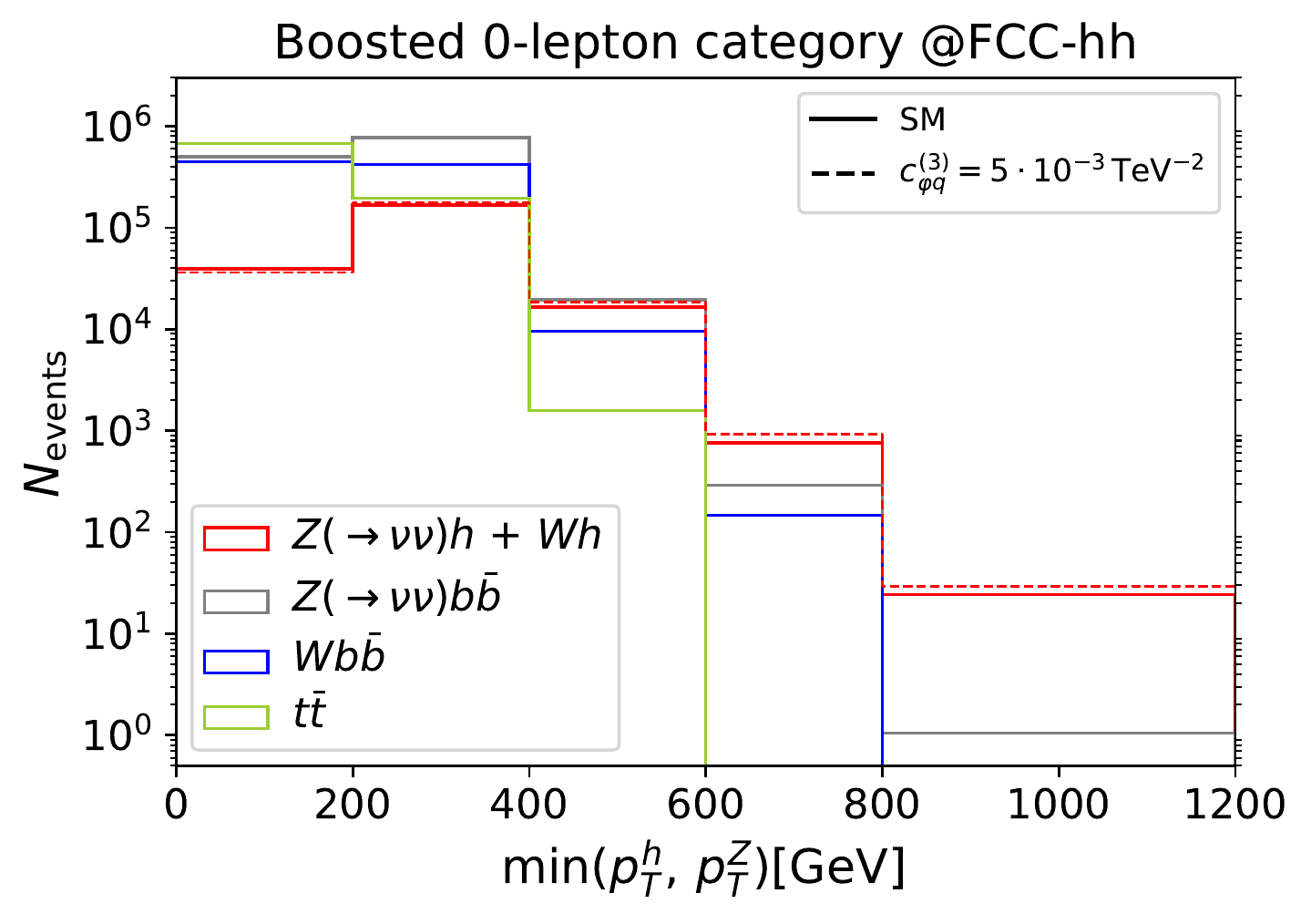}
	\hfill
	\includegraphics[width=0.47\linewidth]{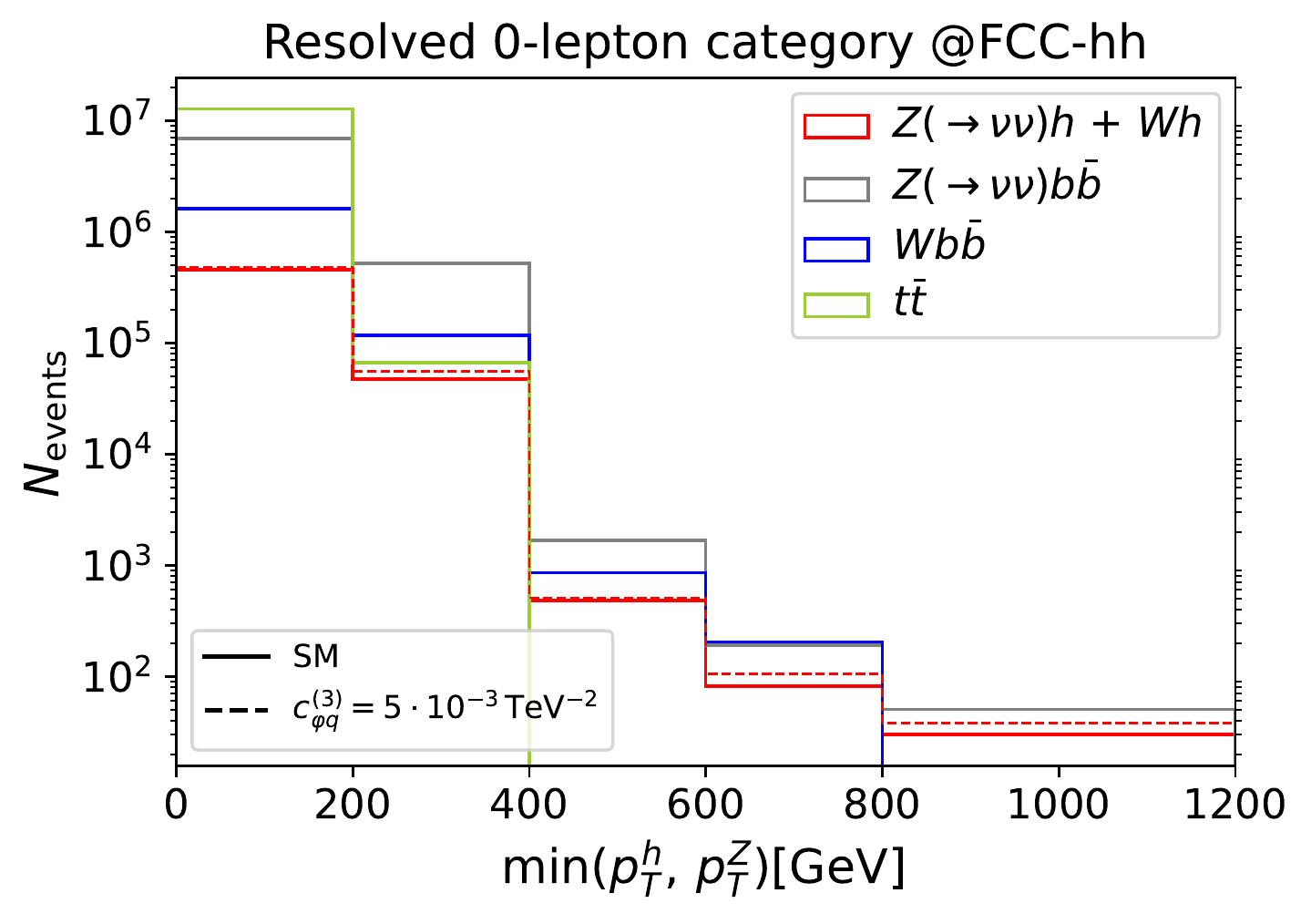}
	\caption{Number of SM signal and background events (solid lines) in the 0-lepton category after the selection cuts for FCC-hh. To visualize the BSM effects on the distributions, we are also plotting the signal setting ${\chqt=5\cdot 10^{-3}\,\mathrm{TeV}^{-2}}$ (dashed lines). The chosen value for the Wilson coefficient is slightly larger than the bounds we derived to increase the visibility of the BSM effects on the distributions. \,\,(\textbf{Left:} boosted category. \textbf{Right:} resolved category.)
 }
	\label{fig:histograms_vv_FCC}
\end{figure}

\begin{figure}[t]
	\centering
	\includegraphics[width=0.47\linewidth]{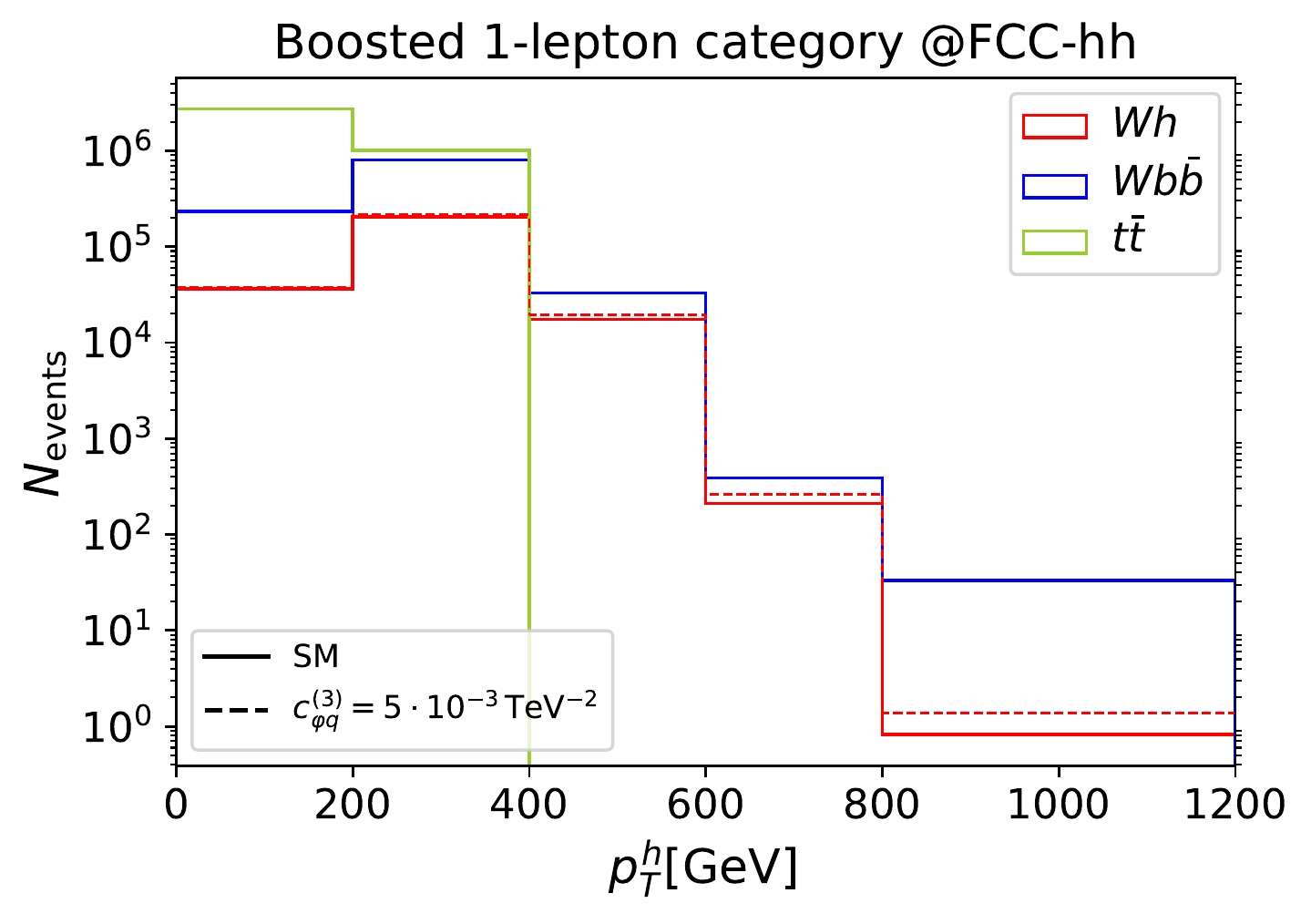}
	\hfill
	\includegraphics[width=0.47\linewidth]{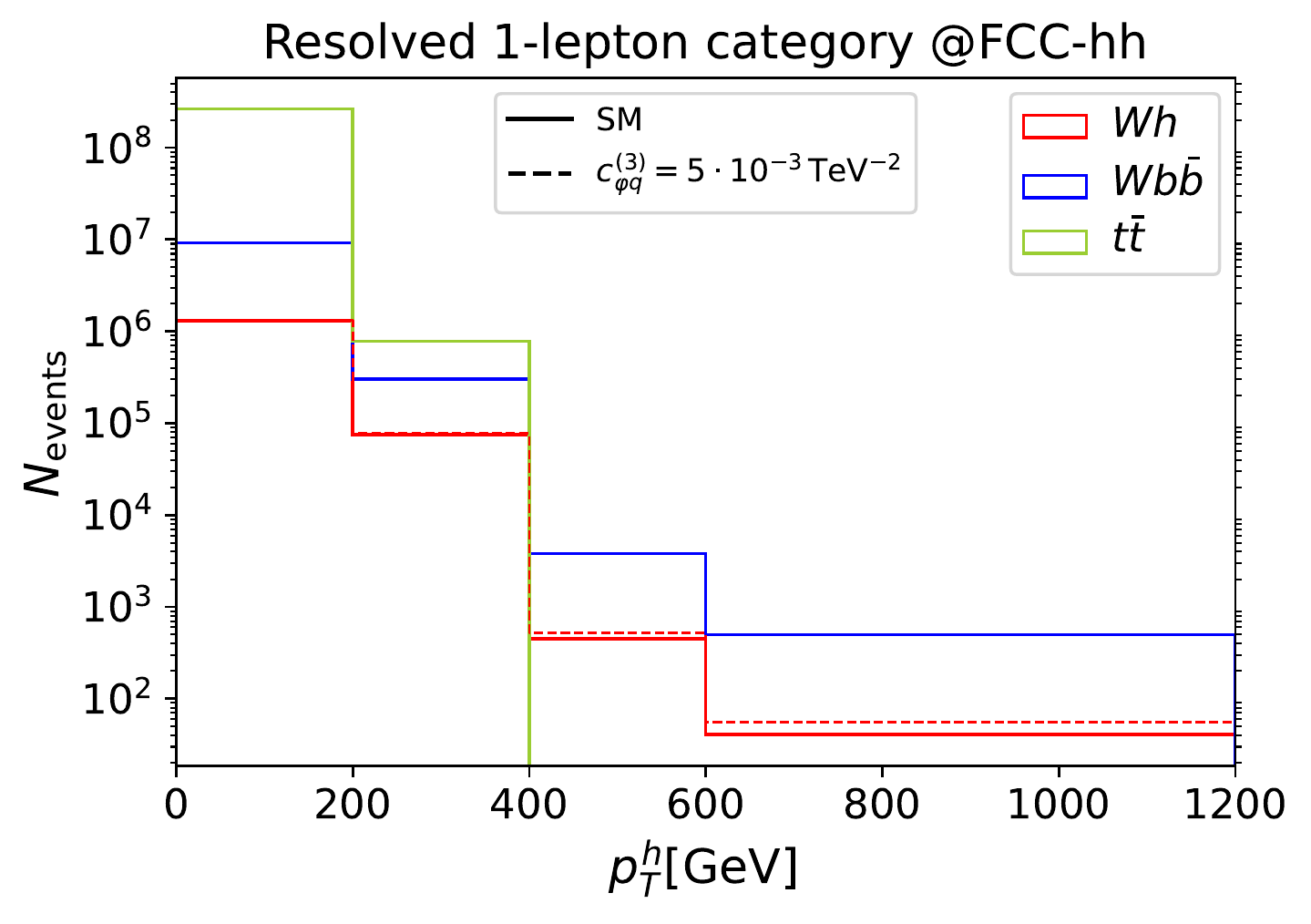}
	\caption{Number of SM signal and background events (solid lines) in the 1-lepton category after the selection cuts for FCC-hh. To visualize the BSM effects on the distributions, we are also plotting the signal setting ${\chqt=5\cdot 10^{-3}\,\mathrm{TeV}^{-2}}$ (dashed lines). The chosen value for the Wilson coefficient is slightly larger than the bounds we derived to increase the visibility of the BSM effects on the distributions. \,\,(\textbf{Left:} boosted category. \textbf{Right:} resolved category.)
	}
	\label{fig:histograms_vl_FCC}
\end{figure}

\begin{figure}[t]
	\centering
	\includegraphics[width=0.47\linewidth]{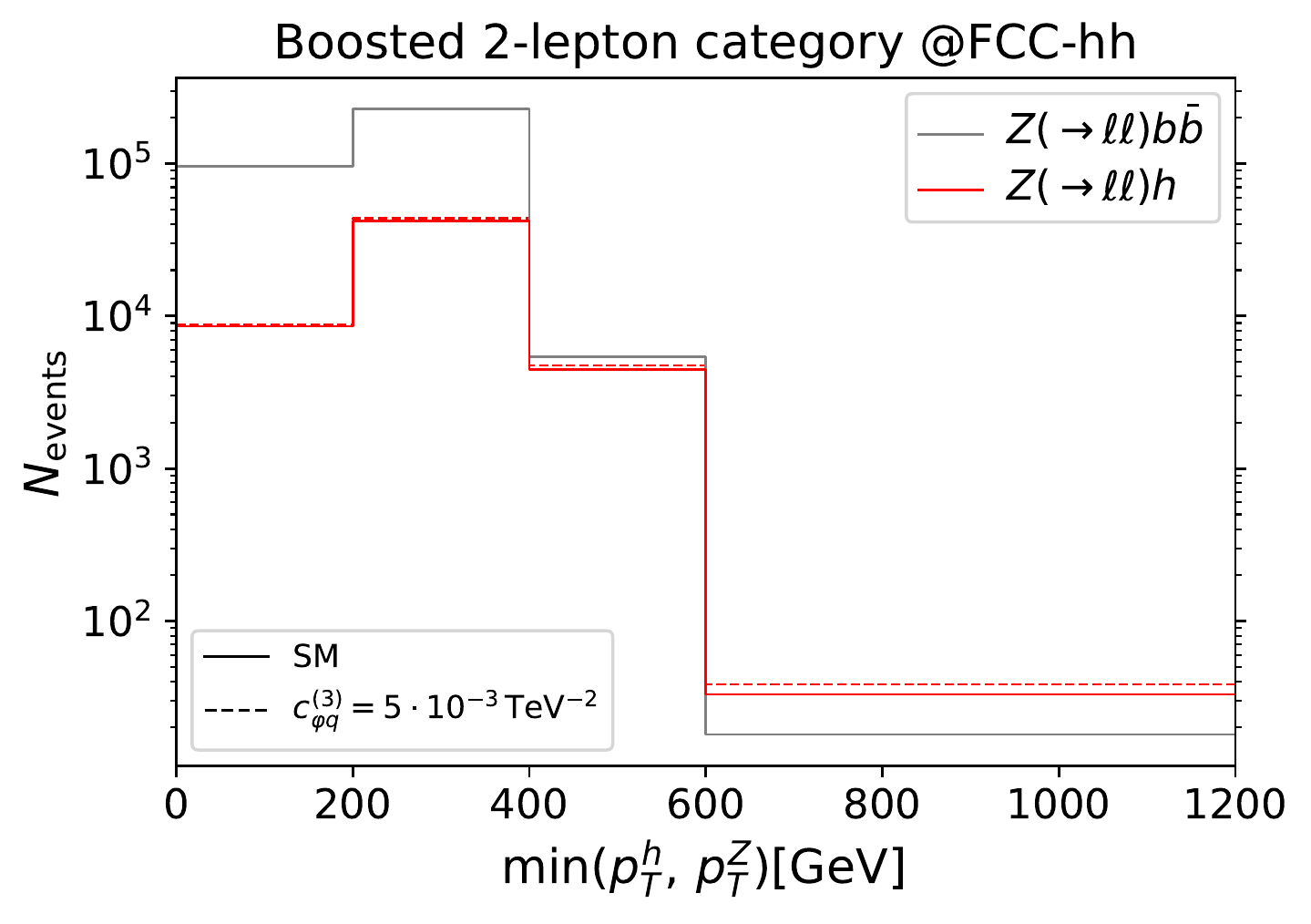}
	\hfill
	\includegraphics[width=0.47\linewidth]{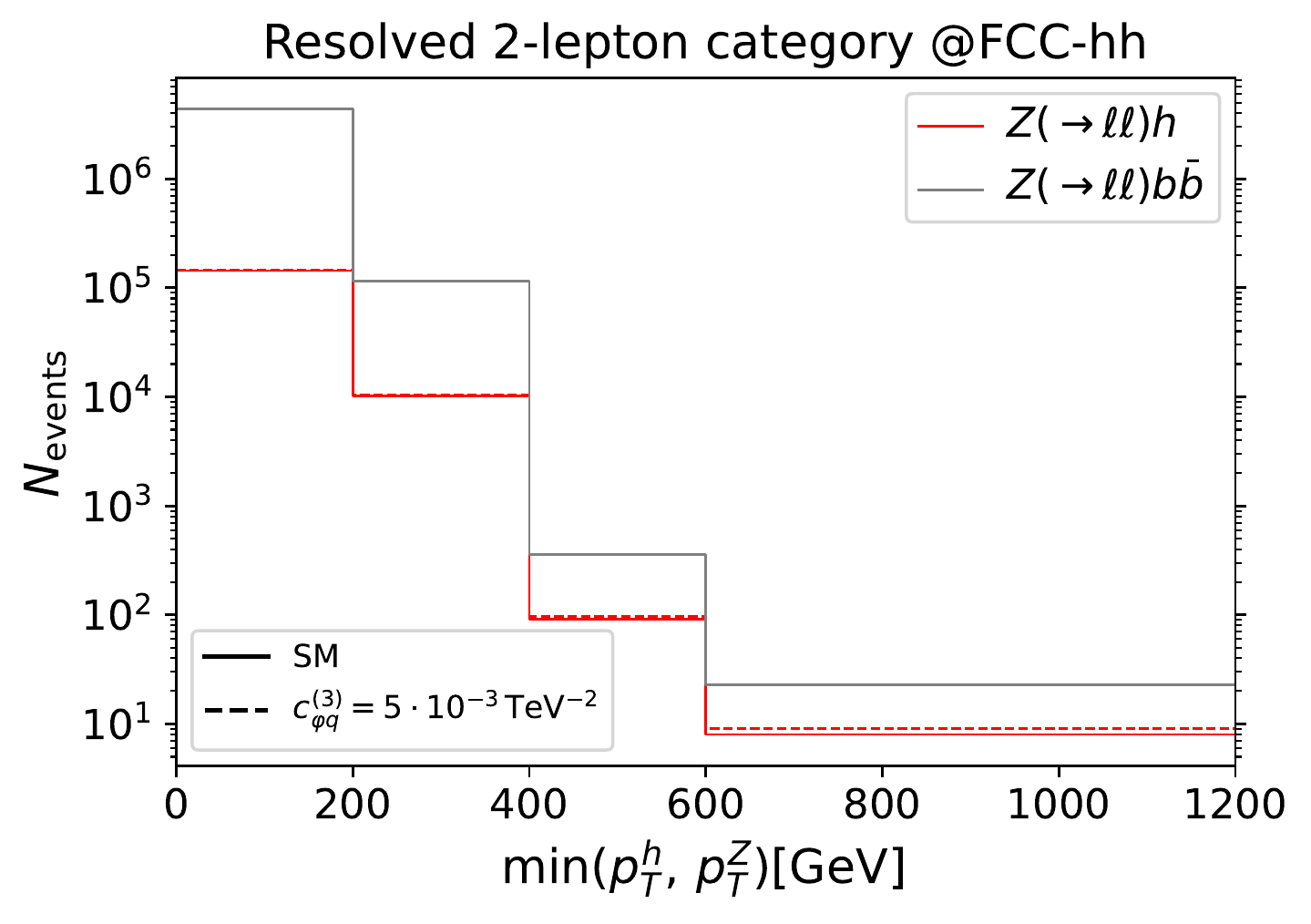}
	\caption{Number of SM signal and background events (solid lines) in the 2-lepton category after the selection cuts for FCC-hh. To visualize the BSM effects on the distributions, we are also plotting the signal setting ${\chqt=5\cdot 10^{-3}\,\mathrm{TeV}^{-2}}$ (dashed lines). The chosen value for the Wilson coefficient is slightly larger than the bounds we derived to increase the visibility of the BSM effects on the distributions. \,\,(\textbf{Left:} boosted category. \textbf{Right:} resolved category.)
	}
	\label{fig:histograms_ll_FCC}
\end{figure}

In figures~\ref{fig:histograms_vv}-\ref{fig:histograms_ll}, we show the number of SM signal and background events expected in each transverse momentum bin at HL-LHC.
Figures~\ref{fig:histograms_vv_FCC}-\ref{fig:histograms_ll_FCC} show analogous plots for FCC-hh (neglecting the binning in rapidity).


For HL-LHC (and LHC Run 3), in the 0-lepton category, we find that once all cuts are imposed,
the $Zb\bar b$ process is the dominant background, both in the boosted and in the resolved category. This background is larger than the signal in all bins. In the boosted category, $Zb\bar b$ events surpass the SM signal by a factor $2-3$, whereas in the
resolved category the difference is significantly larger, being roughly one order of magnitude. The second largest background is $Wb\bar b$, which is approximately of the same order as the SM signal, apart from the last bin in the boosted category and the first one in the resolved category, where it is suppressed by a factor $\sim 2$.
The $t\bar t$ background is relevant only in the lowest two bins in the resolved category, where it is of the same order as the signal, while it is fairly small in all other cases.

In the 1-lepton category, the background is always larger than the signal by roughly one order of magnitude. The main background in the high-$p_T$ bins comes from $W b \bar b$, whereas $t\bar t$ dominates for lower transverse momentum.

Finally, in the 2-lepton channel, the $Zb\bar b$ background surpasses the signal by a significant margin in all bins for both the boosted and the resolved categories.
Since the background is significantly larger than the SM signal in all bins, we expect the HL-LHC analysis (and even more the LHC Run 3 one) to be sensitive only to sizeable variations of the SM distributions. As can be checked from the results in section~\ref{sec:LHC} and appendix~\ref{app:EvtNumbersVH}, BSM contributions need to be larger than $20-30\%$ of the SM cross section in most bins to be detectable.

The impact of backgrounds is significantly different at FCC-hh (see figures~\ref{fig:histograms_vv_FCC}-\ref{fig:histograms_ll_FCC}), mainly due to the larger energy range that can be accessed. Bins at large transverse momentum tend to have a more favorable signal to background ratio, especially in the boosted category.
For instance, in the highest bins in the 0- and 2-lepton boosted categories the SM signal is found to be larger than the total background.

In the 0-lepton channel, the $Zb\bar b$ background is the leading background in most bins. The $Wb \bar b$ background is found to be roughly a factor of $2$ smaller than the $Zb\bar b$ background with the exception of the lowest bins in the resolved category where it is further suppressed. The $t \bar t$ background mainly contributes to the low-$p_T$ bins and is negligible in the high transverse momentum tail (for a detailed quantitative discussion of this feature see appendix~\ref{sec:fit_tt}). Differently from the HL-LHC case, in which the SM signal in the resolved category was always overwhelmed by the background, for FCC-hh a few high-$p_T$ bins with signal-to-background ratio close to $0.5$ are available. As already mentioned, the situation is even better for the boosted category, where the high-$p_T$ bins contain a sizeable number of SM events and are almost background free.

In the 1-lepton channel, the $Wb\bar b$ background is always larger than the SM signal. In most bins $Wb\bar b$ dominates by more than one order of magnitude, with the exception of the central bins in the boosted category where the SM signal is only a factor $2$ smaller than $Wb\bar b$. The $t \bar t$ background overwhelms the signal for $p^h_T < 200\,{\rm GeV}$, but it is negligible for higher transverse momentum.

Finally, in the 2-lepton channel, the main background, $Zb\bar b$, dominates over the SM signal in all bins in the resolved category and in the two lowest bins of the boosted one. The high-$p_T$ bins in the boosted category show instead a favourable signal-to-background ratio close or greater than $1$.

\section{Results}\label{sec:results}

In this section we present the expected exclusion bounds obtained
assuming that the measurements agree with the SM predictions. In subsection \ref{sec:LHC}, we report the results for the LHC at the end of Run 3 (with an integrated luminosity $\lag=300\,\mathrm{fb}^{-1}$) and for the HL-LHC ($\lag=3\,\mathrm{ab}^{-1}$). In subsection \ref{sec:FCChh}, we report the projected bounds at the end of FCC-hh ($\lag=30\,\mathrm{ab}^{-1}$). In both subsections we assume a flat and uncorrelated $5\%$ systematic uncertainty in each bin, which we introduce to mimic the expected theory and experimental uncertainties. The results for other systematic uncertainty values, namely $1\%$ and $10\%$, are presented in appendix~\ref{sec:AppFullResults}, along with the projections for the LHC Run 2 ($\lag=139\,\mathrm{fb}^{-1}$).

As we discussed in section~\ref{sec:cuts}, our predictions for the signal and background differential distributions show a small discrepancy with respect to the ATLAS results given in ref.~\cite{ATLAS:2020fcp,ATLAS:2020jwz}. In the following, we present the projections obtained using our determination of signal and background events. We report in appendix~\ref{sec:AppFullResults} the corresponding bounds obtained by rescaling the differential distributions to match the ATLAS results. Additionally, we compared our results for LHC Run 2 with the ones obtained by ATLAS in ref.~\cite{ATLAS:2021wqh} and found that our bounds are looser by $10-30\%$. This discrepancy is due in part to the differences in b-tagging techniques, the binning variable, and other simulation details. A more precise comparison is not possible due to methodological differences.

\subsection{LHC Run 3 and HL-LHC}\label{sec:LHC}

The projected bounds we obtained for the LHC Run 3 ($300$\,fb$^{-1}$) and HL-LHC ($3$\,ab$^{-1}$) are reported in table~\ref{tab:bounds_summary_LHC}. There, we present the bounds derived from a global fit profiling over the other coefficients (middle column) and from one-operator fits (last column).

\begin{table}[t]
\begin{centering}
\begin{tabular}{c|@{\;}c@{\;}|@{\;}c@{}}
\toprule
Coefficient & Profiled Fit  & One-Operator Fit \tabularnewline
\midrule
$c_{\varphi q}^{(3)}\,$[TeV$^{-2}$] &
\begin{tabular}{ll}
\rule{0pt}{1.25em}$[-7.9,\,3.5]\times10^{-2}$ & LHC Run 3 \\
\rule[-.65em]{0pt}{1.9em}$[-3.9,\,1.9]\times10^{-2}$ & HL-LHC
\end{tabular}
&
\begin{tabular}{ll}
\rule{0pt}{1.25em}$[- 4.3,\,3.2]\times10^{-2}$ & LHC Run 3 \\
\rule[-.65em]{0pt}{1.9em}$[-1.7,\,1.5]\times10^{-2}$ & HL-LHC \\
\end{tabular}
\tabularnewline

\hline
$c_{\varphi q}^{(1)}\,$[TeV$^{-2}$] &
\begin{tabular}{ll}
\rule{0pt}{1.25em}$[-1.2,\,1.3]\times10^{-1}$ & LHC Run 3 \\
\rule[-.65em]{0pt}{1.9em}$[-0.8,\,0.9]\times10^{-1}$ & HL-LHC
\end{tabular}
&
\begin{tabular}{ll}
\rule{0pt}{1.25em}$[-0.87,\,1.06]\times10^{-1}$ & LHC Run 3 \\
\rule[-.65em]{0pt}{1.9em}$[-0.54,\,0.72]\times10^{-1}$ & HL-LHC \\
\end{tabular}
\tabularnewline

\hline
$c_{\varphi u}\,$[TeV$^{-2}$] &
\begin{tabular}{ll}
\rule{0pt}{1.25em}$[-1.9,\,1.2]\times10^{-1}$ & LHC Run 3 \\
\rule[-.65em]{0pt}{1.9em}$[-1.35,\,0.82]\times10^{-1}$ & HL-LHC
\end{tabular}
&
\begin{tabular}{ll}
\rule{0pt}{1.25em}$[-1.68,\,0.9]\times10^{-1}$ & LHC Run 3 \\
\rule[-.65em]{0pt}{1.9em}$[-1.24,\,0.49]\times10^{-1}$ & HL-LHC \\
\end{tabular}
\tabularnewline

\hline
$c_{\varphi d}\,$[TeV$^{-2}$] &
\begin{tabular}{ll}
\rule{0pt}{1.25em}$[-1.8,\,2.1]\times10^{-1}$ & LHC Run 3 \\
\rule[-.65em]{0pt}{1.9em}$[-1.3,\,1.5]\times10^{-1}$ & HL-LHC
\end{tabular}
&
\begin{tabular}{ll}
\rule{0pt}{1.25em}$[-1.3,\,1.7]\times10^{-1}$ & LHC Run 3 \\
\rule[-.65em]{0pt}{1.9em}$[-0.8,\,1.2]\times10^{-1}$ & HL-LHC\\
\end{tabular} \\
\bottomrule

\end{tabular}
\par\end{centering}
\caption[caption]{Projected $95\%$ C.L. bounds at the LHC run 3 and HL-LHC on the coefficients of the $\Ohqt$, $\Ohq$, $\Ohu$ and $\Ohd$ operators. %(with the normalization $\Lambda = 1 \, \text{TeV}$).
We assume $5\%$ systematic uncertainty.
{\bf Left column:} Bounds from a global fit, profiled over the other coefficients. {\bf Right column:} Bounds from one-operator fits (i.e.~setting the other coefficients to zero).}
\label{tab:bounds_summary_LHC}
\end{table}

We point out that all the bounds at the LHC Run 3 are statistically limited, so that HL-LHC can give a significant improvement, tightening the bound on $\chqt$ by $\sim 50\%$ and the other ones by $\sim 30\%$. The larger improvement for $\chqt$ can be understood by taking into account that the bound at the HL-LHC is mainly driven by the SM-BSM interference terms in the squared amplitude (and thus scales roughly with the square root of the luminosity).
The bounds for the other operators, however, receive sizeable contributions from the squared BSM terms (and thus scale roughly with the fourth root of the luminosity).

At the HL-LHC, systematic uncertainties at the $5\%$ level start to dominate over the statistical ones, thus leading to a saturation of the bounds. A further increase in the integrated luminosity by a factor $2$ (as will be possible combining ATLAS and CMS) provides an improvement of order $5 - 15\%$ in the bounds.

\begin{figure}[t]
	\centering
	\includegraphics[width=0.6\linewidth]{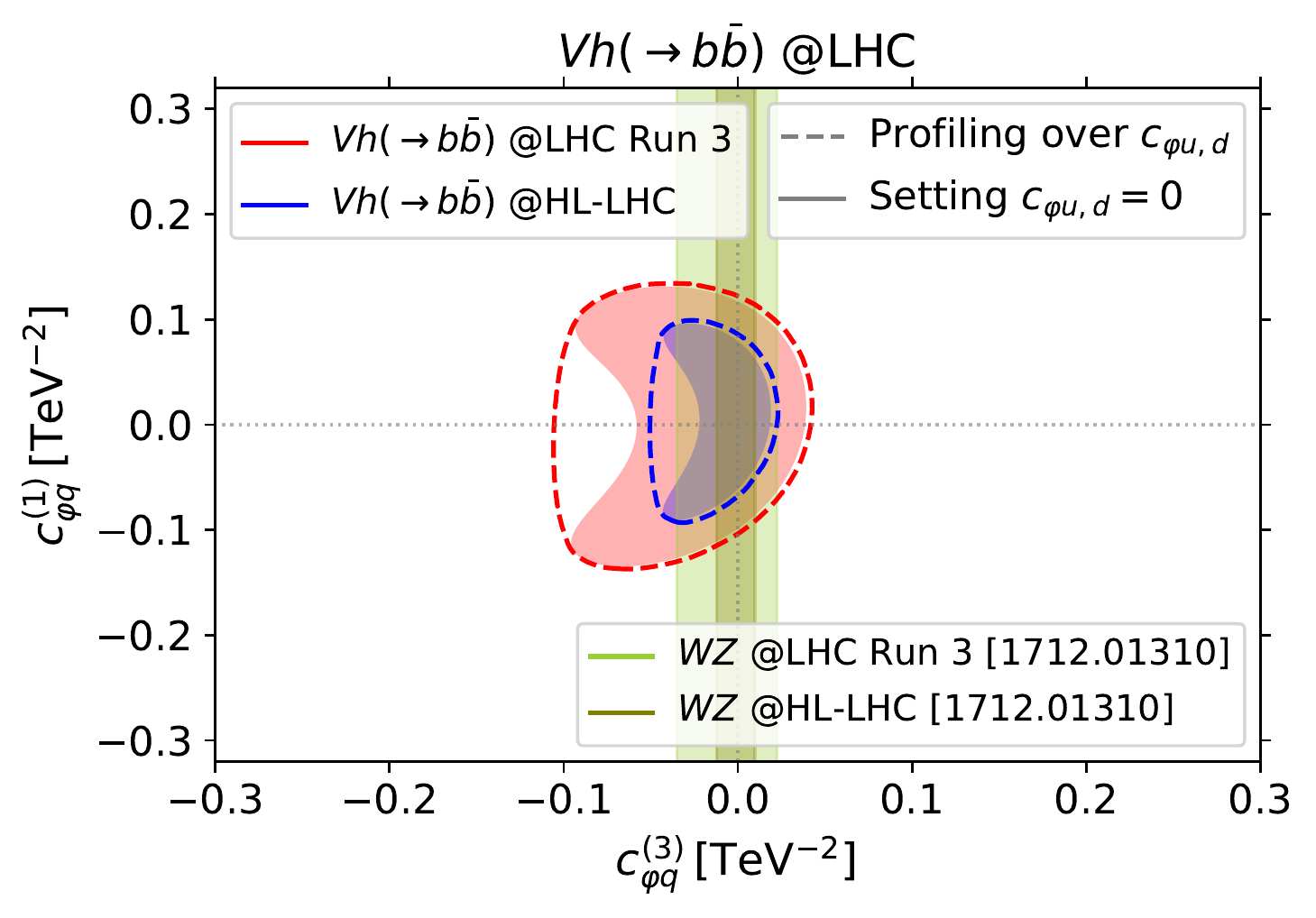}
	\caption{$95\%$ C.L. bounds on $\chq$ and $\chqt$, assuming 5\% systematics, at LHC Run 3 and HL-LHC. %, combining the 0-, 1- and 2-lepton category analyses.
	Solid (dashed) lines correspond to the bounds when profiling over (setting to zero) the Wilson coefficients not appearing in the plot. We compare our bounds to the bounds reported in ref.~\cite{Franceschini:2017xkh}. %\red{We could add on the plot the WZ bound at HL-LHC.}
	}
	\label{fig:2D_bounds_WhZh_Run3}
\end{figure}

It is interesting to compare our results with the bounds that can be obtained from other diboson channels. In particular, one clean channel with high sensitivity is $WZ$ production which is sensitive to $\chqt$. In figure~\ref{fig:2D_bounds_WhZh_Run3} we compare our projected bounds in the $\chq$-$\chqt$ plane with the ones from $WZ$ production with fully leptonic final state obtained in ref.~\cite{Franceschini:2017xkh}.
One can see that $WZ$ has slightly more constraining power on $\chqt$ than $Vh$ both at LHC Run 3 and HL-LHC, assuming the same level of systematic uncertainty (namely $5\%$). A combination of the $WZ$ and $Vh$ channels can thus provide a mild improvement in the 1-operator-only determination of $\chqt$ (i.e, the other operators are set to zero):
\begin{equation}\label{eq:bounds_1} 
\def\arraystretch{1.3}
\begin{array}{l@{\hspace{.8em}}l@{\hspace{2em}}l}
\textrm{LHC Run 3} &  \chqt\in[-3.1, 2.0] \times 10^{-2}\TeV^{-2}\;, \\
\textrm{HL-LHC} &  \chqt\in[-11.0, 8.9] \times 10^{-3}\TeV^{-2}\; ,
\end{array}
\end{equation}
The main advantage of considering the $Vh$ channels is the possibility to constrain the flat direction along $\chq$.
We also compared our one-operator-fit bounds on $\chqt$ with the ones derived in ref.~\cite{Banerjee:2018bio} finding reasonable overall agreement. However, a fully quantitative comparison can not be carried out due to methodological differences in the fits.

\begin{figure}[t]
	\centering
	\includegraphics[width=0.485\linewidth]{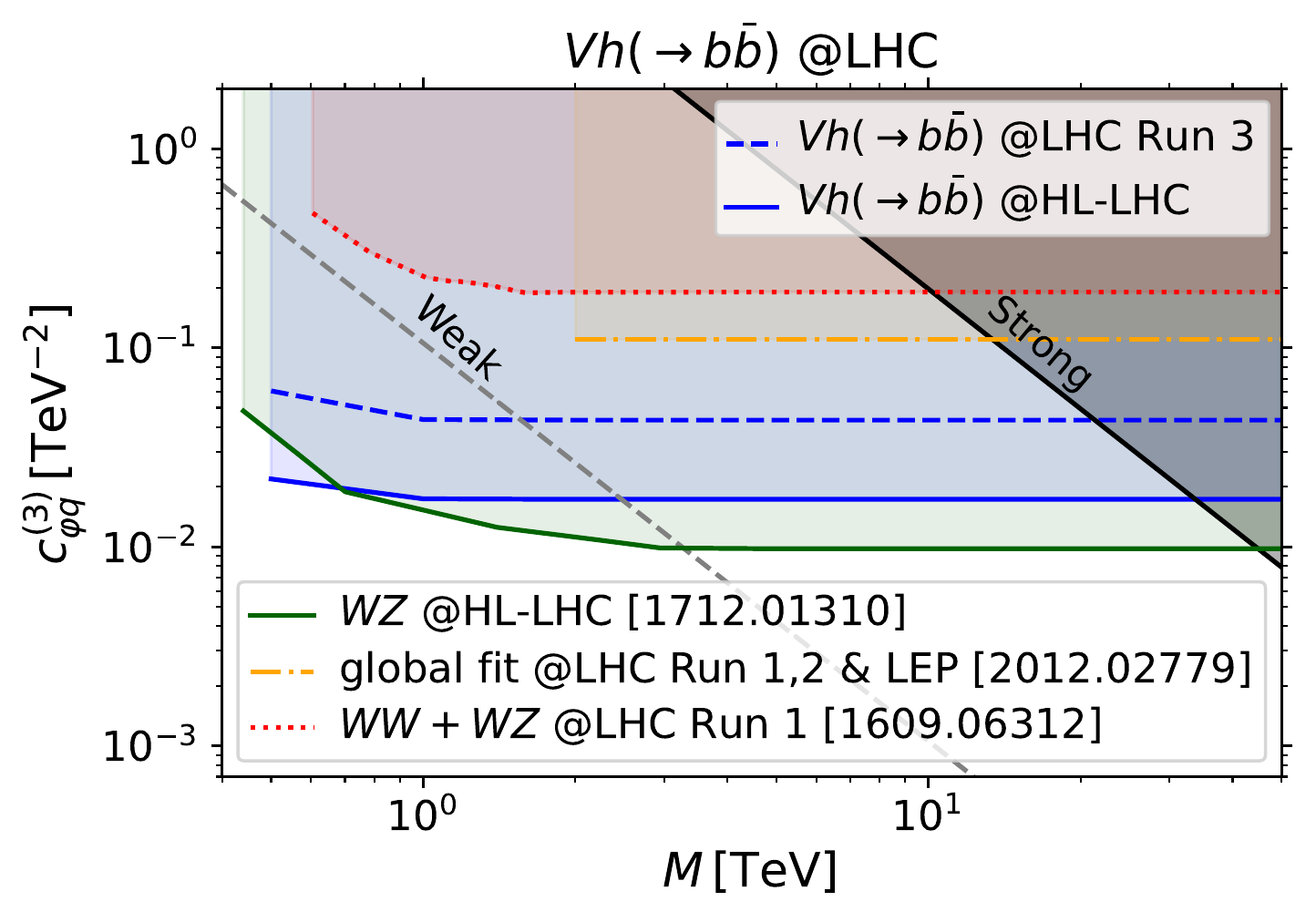} \hfill
	\includegraphics[width=0.485\linewidth]{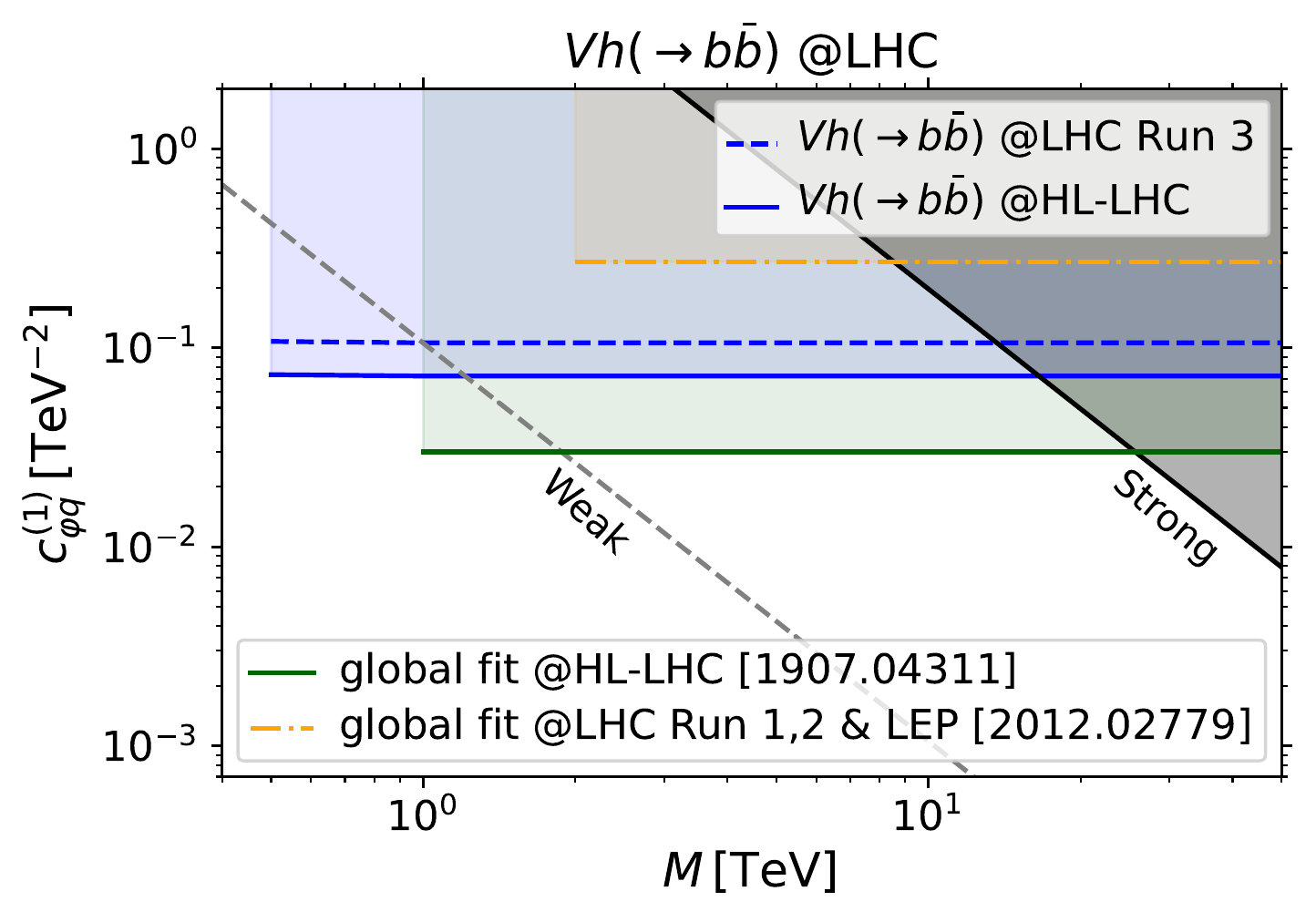} \\
	\vspace{0.3cm}
    \includegraphics[width=0.485\linewidth]{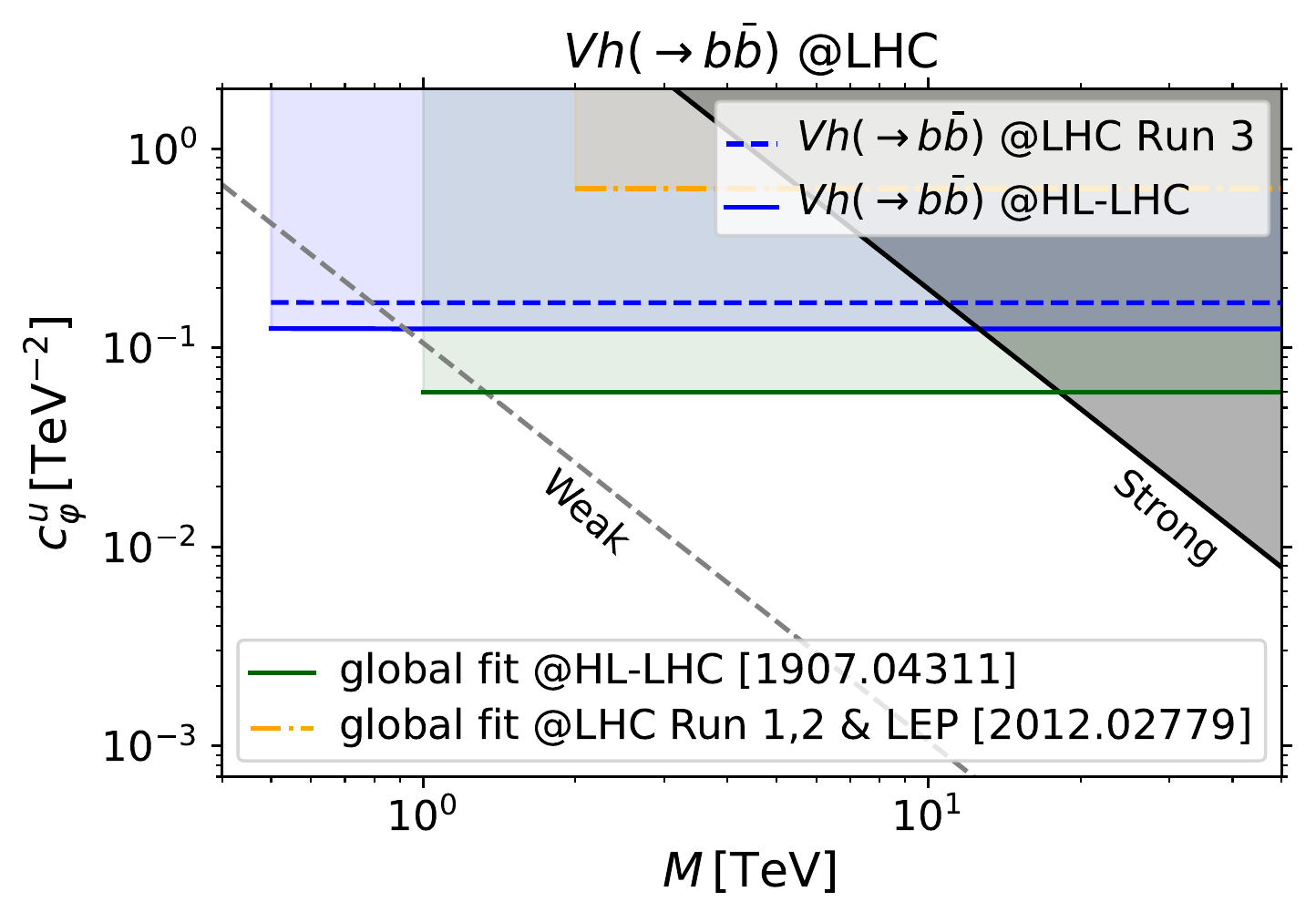}\hfill
	\includegraphics[width=0.485\linewidth]{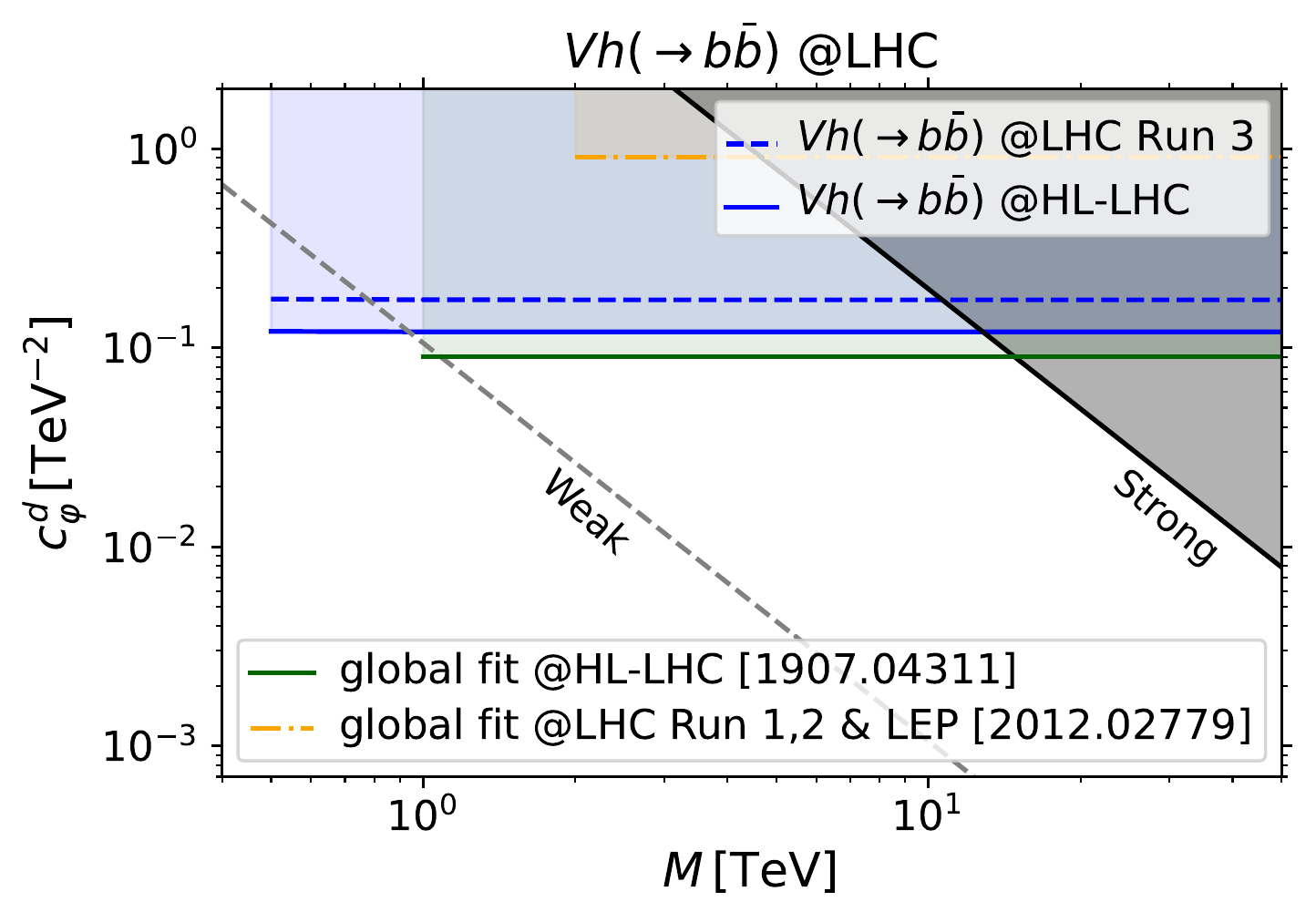}
 	\caption{Projected $95\%$ C.L.~bounds on $c_{\varphi q}^{(3)}$, $\chq$, $\chu$ and $\chd$ from one-operator fits, assuming 5\% systematics at LHC Run 3 (dashed blue lines) and HL-LHC (solid blue lines)
	as functions of the maximal-invariant-mass cut $M$.
	Current bounds and projections for some future hadron colliders are also shown. In all plots we show the marginalised global LHC data fit bound from ref.~\cite{Ellis:2020unq} (orange dot-dashed lines).
	For $\chqt$, we show the LHC Run 1 bound from ref.~\cite{Falkowski:2016cxu} (red dotted line) and the projections from the $WZ$ channel at HL-LHC from ref.~\cite{Franceschini:2017xkh} (dark green solid line). For the rest of operators we also show the 1-operator fit at HL-LHC from ref.~\cite{deBlas:2019wgy} (dark green solid lines). In all cases, if the negative and positive bound differ, we plot the maximum absolute value.
	The dashed gray and solid black lines show the values of the Wilson coefficient expected in weakly-coupled ($c \sim g^2/(4 M^2)$),
	and strongly-coupled ($c \sim (2\pi)^2/M^2$) new physics models~\cite{Franceschini:2017xkh}.}
	\label{fig:fit_cphiq_HL_LHC}
\end{figure}

In any study that uses an EFT framework, it is important to check the limits of validity of such EFT.
To this end, we present in figure~\ref{fig:fit_cphiq_HL_LHC} our single-operator bounds for the LHC after Run 3 and HL-LHC as function of the cut $M$ on the maximal center-of-mass energy of the $Vh$ system of the events included in the fit.
Since the bound interval is not symmetric, the plot shows the weaker bound in absolute value.
The plots allow us to obtain a quick evaluation of the applicability of our bounds to new-physics models.
By interpreting the maximum allowed invariant mass $M$ as a proxy for the cutoff, one can figure out which classes of models our results apply to. As can be seen from the plots, the bounds on $\chqt$ at HL-LHC can test weakly coupled theories up to a cut-off scale of order $3$\;TeV, while for strongly coupled scenarios the reach extends up to $\sim 30$\;TeV. The reach in cut-off for the other effective operators we considered is roughly a factor of $2$
weaker.

Finally, we compare our projected bounds with the ones coming from global fits at LHC and future lepton colliders.
The results from recent global fits to LEP and LHC data~\cite{Ethier:2021bye,Ellis:2020unq} show that our $Vh$ analysis at LHC Run 3 could significantly improve the bounds on $\chqt$ and $\chu$, while does not seem competitive on the determination of $\chq$ and $\chd$. The relevance of the $Vh$ channel extends also to HL-LHC, where our profiled and one-operator bounds are competitive with the projected global fits of ref.~\cite{deBlas:2019rxi}. When comparing our results with the projections for global fits at LHC Run 3 and HL-LHC it must be taken into account that the way in which these fits are performed can significantly affect the bounds. In particular, the inclusion of the LEP constraints differs in the various fits, leading to sizeable discrepancies on the estimate of the projected bounds.

On the other hand, future lepton colliders will generically improve the bounds obtained at HL-LHC by an order of magnitude. As an example we report the expected precision from a global fit at FCC-ee at $365$ GeV (${\cal L} = 1.5$~ab$^{-1}$) combined with lower-energy runs~\cite{deBlas:2019wgy}:
\begin{equation}\label{eq:bounds_fcc-ee}
\begin{array}{l@{\hspace{.8em}}l@{\hspace{3.em}}l}
\rule{0pt}{1.25em}\chqt\in[-6.3, 6.3] \times 10^{-3}\;\TeV^{-2}& ([-4.8, 4.8] \times 10^{-4}\; \TeV^{-2})\;, \\
\rule{0pt}{1.25em}\chq\in[-0.018, 0.018]\;\TeV^{-2}& \left([-0.0017, 0.0017]\;\TeV^{-2}\right)\,,\\
\rule{0pt}{1.25em}\chu\in[-0.04, 0.04]\;\TeV^{-2}& \left([-0.003, 0.003]\;\TeV^{-2}\right)\,,\\
\rule{0pt}{1.25em}\chd\in[-0.095, 0.095]\;\TeV^{-2}& \left([-0.004, 0.004]\;\TeV^{-2}\right)\,,
\end{array}
\end{equation}
where we also quoted in parentheses the bounds from one-operator fits.

\subsection{FCC-hh}\label{sec:FCChh}

\begin{table}[t]
\begin{centering}
\begin{tabular}{c|c|c}
\toprule
Coefficient & Profiled Fit & One-Operator Fit \tabularnewline
\midrule
$c_{\varphi q}^{(3)}\,$[TeV$^{-2}$] &
\begin{tabular}{ll}
\rule{0pt}{1.25em}$[-2.0,\,2.1]\times10^{-3}$ & $1\%$ syst.\\
\rule{0pt}{1.25em}$[-4.9,\,3.7]\times10^{-3}$ & $5\%$ syst.\\
\rule[-.65em]{0pt}{1.9em}$[-7.6,\,5.1]\times10^{-3}$ & $10\%$ syst.
\end{tabular}
&
\begin{tabular}{ll}
\rule{0pt}{1.25em}$[-1.1,\,1.1]\times10^{-3}$ & $1\%$ syst.\\
\rule{0pt}{1.25em}$[-2.5,\,2.4]\times10^{-3}$ & $5\%$ syst.\\
\rule[-.65em]{0pt}{1.9em}$[-4.0,\,3.6]\times10^{-3}$ & $10\%$ syst.
\end{tabular}
\tabularnewline

\hline
$c_{\varphi q}^{(1)}\,$[TeV$^{-2}$] &
\begin{tabular}{ll}
\rule{0pt}{1.25em}$[-10.6,\,9.0]\times10^{-3}$ & $1\%$ syst.\\
\rule{0pt}{1.25em}$[-14.8,\,13.6]\times10^{-3}$ & $5\%$ syst.\\
\rule[-.65em]{0pt}{1.9em}$[-17.2,\,16.4]\times10^{-3}$ & $10\%$ syst.
\end{tabular}
&
\begin{tabular}{ll}
\rule{0pt}{1.25em}$[-8.2,\,8.1]\times10^{-3}$ & $1\%$ syst.\\
\rule{0pt}{1.25em}$[-11.3,\,11.5]\times10^{-3}$ & $5\%$ syst.\\
\rule[-.65em]{0pt}{1.9em}$[-13.1,\,13.3]\times10^{-3}$ & $10\%$ syst.
\end{tabular}
\tabularnewline

\hline
$c_{\varphi u}\,$[TeV$^{-2}$] &
\begin{tabular}{ll}
\rule{0pt}{1.25em}$[-15.9,\,9.0]\times10^{-3}$ & $1\%$ syst.\\
\rule{0pt}{1.25em}$[-27.0,\,13.5]\times10^{-3}$ & $5\%$ syst.\\
\rule[-.65em]{0pt}{1.9em}$[-30.4,\,16.4]\times10^{-3}$ & $10\%$ syst.
\end{tabular}
&
\begin{tabular}{ll}
\rule{0pt}{1.25em}$[-6.2,\,4.9]\times10^{-3}$ & $1\%$ syst.\\
\rule{0pt}{1.25em}$[-24.9,\,8.2]\times10^{-3}$ & $5\%$ syst.\\
\rule[-.65em]{0pt}{1.9em}$[-30.2,\,10.4]\times10^{-3}$ & $10\%$ syst.\\
\end{tabular}
\tabularnewline

\hline
$c_{\varphi d}\,$[TeV$^{-2}$] &
\begin{tabular}{ll}
\rule{0pt}{1.25em}$[-17.9,\,23.6]\times10^{-3}$ & $1\%$ syst.\\
\rule{0pt}{1.25em}$[-22.0,\,26.5]\times10^{-3}$ & $5\%$ syst.\\
\rule[-.65em]{0pt}{1.9em}$[-25.1,\,29.5]\times10^{-3}$ & $10\%$ syst.
\end{tabular}
&
\begin{tabular}{ll}
\rule{0pt}{1.25em}$[-9.8,\,23.0]\times10^{-3}$ & $1\%$ syst.\\
\rule{0pt}{1.25em}$[-14.0,\,24.5]\times10^{-3}$ & $5\%$ syst.\\
\rule[-.65em]{0pt}{1.9em}$[-16.9,\,26.4]\times10^{-3}$ & $10\%$ syst.\\
\end{tabular} \\
\bottomrule

\end{tabular}
\par\end{centering}
\caption[caption]{ Bounds at $95\%$ C.L.~on the coefficients of the $\Ohqt$, $\Ohq$, $\Ohu$ and $\Ohd$ operators for FCC-hh with integrated luminosity of $30\,\mathrm{ab}^{-1}$.
{\bf Left column:} Bounds from the global fit, profiled over the other coefficients. {\bf Right column:} Bounds from a one-operator fit (i.e. setting the other coefficients to zero).
}
\label{tab:bounds_summary_FCC}
\end{table}

Our projected bounds at FCC-hh ($100$\;TeV with $\lag=30\,$ab$^{-1}$) are collected in table~\ref{tab:bounds_summary_FCC}. They are presented for three benchmark choices of systematic uncertainties: $1\%$, $5\%$ and $10\%$. We believe that the middle value represents the more probable scenario, while the others should be considered as limiting optimistic and pessimistic choices. The comparison with our HL-LHC projections, taking the $5\%$ systematic scenario, reveals an improvement of all the bounds by a factor $\sim 5$. This is a consequence of the much higher statistics especially in the tails of the kinematic distributions, due to the increase in cross section at higher center-of-mass energy and to the larger integrated luminosity. Since the statistical error proves to be quite small, our projected bounds at FCC-hh show strong dependence on the systematic uncertainty. %, which is a consequence of the low statistical uncertainty and the high backgrounds.
Doubling the integrated luminosity provides only a mild improvement of the bounds, of order $10 - 15\%$.

Our results can be directly compared with the ones obtained from $Vh$ production with the Higgs decaying in a pair of photons~\cite{Bishara:2020pfx}. Both channels provide quite similar sensitivity to the four effective operators for $5\%$ systematic uncertainty, while $h\to b\bar b$ provides slightly stronger bounds than $h\to\gamma\gamma$ for $1\%$ systematic and slightly worse for $10\%$. This behavior reflects the fact that the $h\to b \bar b$ channel has high statistics but, at the same time, high background, so that its sensitivity is mostly determined by the amount of systematic uncertainties. On the contrary, the $h\to \gamma\gamma$ channel is almost background-free and its precision is mainly limited by the low statistics and not by the systematic uncertainties. The small background also explains why the $h\to \gamma\gamma$ channel is less sensitive to a change in the systematic uncertainty, whose precise determination at FCC-hh could ultimately decide which Higgs decay channel has the best reach. 

%Our results can be directly compared with the ones obtained from $Vh$ production with the Higgs decaying in a pair of photons~\cite{Bishara:2020pfx}. Both channels provide quite similar sensitivity to the four effective operators for $5\%$ systematic uncertainty, while $h\to b\bar b$ provides slightly stronger bounds than $h\to\gamma\gamma$ for $1\%$ systematic and slightly worse for $10\%$. This behavior reflects the fact that the $h\to b \bar b$ channel has high statistics but, at the same time, high background, so that its sensitivity is mostly determined by the amount of systematic uncertainties. On the contrary, the $h\to \gamma\gamma$ channel is almost background-free and its precision is mainly limited by the low statistics and not by the systematic uncertainties.
\begin{figure}[t]
	\centering
	\includegraphics[width=0.6\linewidth]{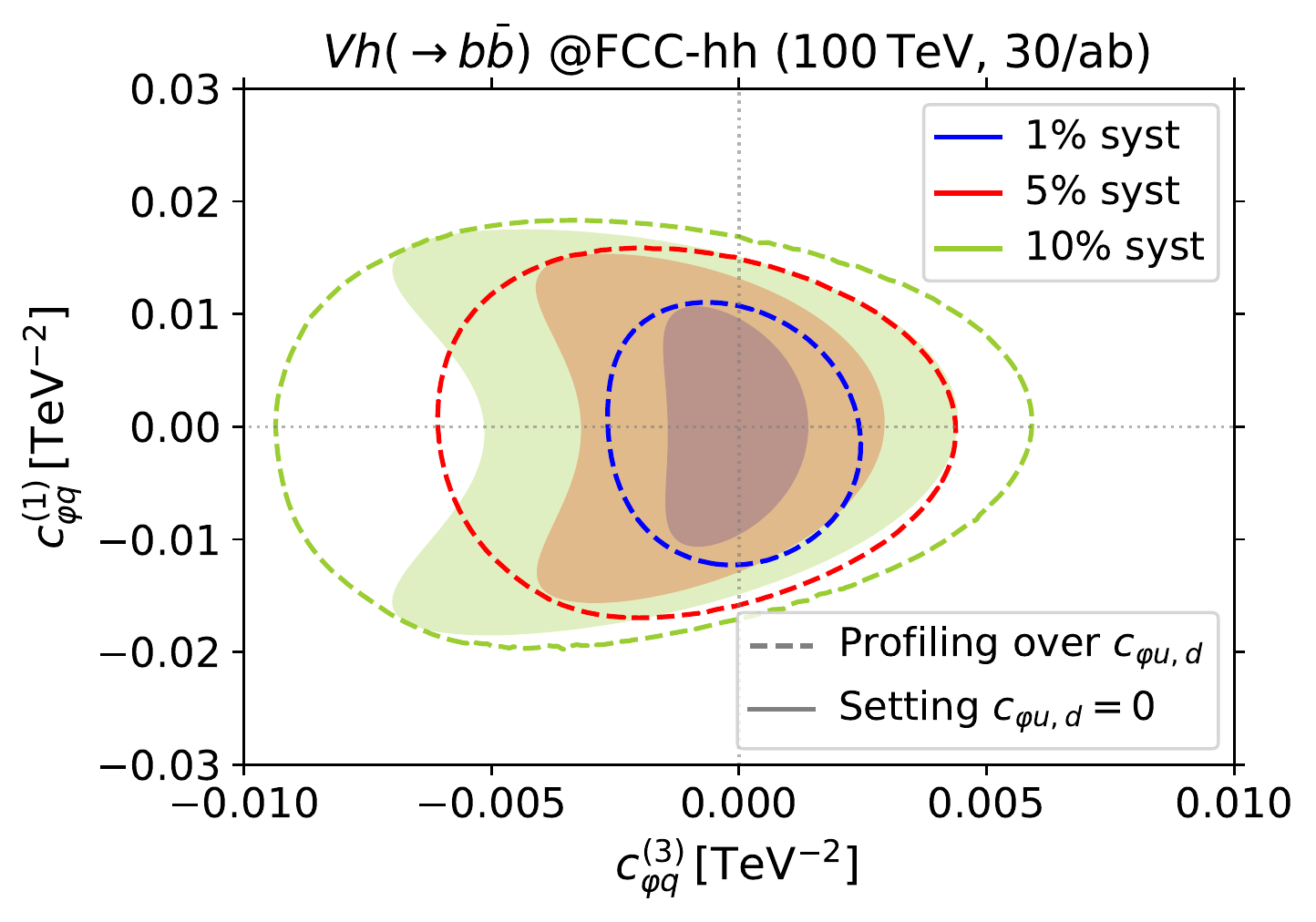}
	\caption{Projected $95\%$ C.L.~bounds on $\chq$ and $\chqt$ at FCC-hh.
	Solid (dashed) lines correspond to the bounds when profiling over (setting to zero) the Wilson coefficients not appearing in the plot.
	}
	\label{fig:2D_bounds_WhZh_FCC}
\end{figure}

%This is a clear reflection of the fact that both channels achieve similar results in opposite different ways. On the one hand, $h\to b \bar b$ seizes on higher cross sections and has a low statistical uncertainty, and on the other hand, $h\to \gamma\gamma$ suffers from higher statistical uncertainties but profits from almost negligible backgrounds. The small backgrounds also explains why the $h\to \gamma\gamma$ is less sensitive to a change in the systematic uncertainty, whose precise determination at FCC-hh could ultimately decide which Higgs decay channel has the best reach.

In figure~\ref{fig:2D_bounds_WhZh_FCC}, we show the $95\%$ C.L. bounds in the $\chq$-$\chqt$ plane for the different levels of systematic uncertainties and either profiling over (dashed lines) or setting to zero (full lines) the $\chu$ and $\chd$ coefficients. The plot shows that our analysis is much more sensitive to $\chqt$ than to $\chq$ and confirms that the sensitivity depends heavily on the systematic uncertainty.~\footnote{For a comparison with the $h\to\gamma\gamma$ channel, see the right panel of figure~3 in ref.~\cite{Bishara:2020pfx}.}
A certain degree of correlation between $\chq$ and $\chqt$, which almost disappears in the full fit profiled over $\chu$ and $\chd$, can also be seen.

\begin{figure}[t]
	\centering
	\includegraphics[width=0.485\linewidth]{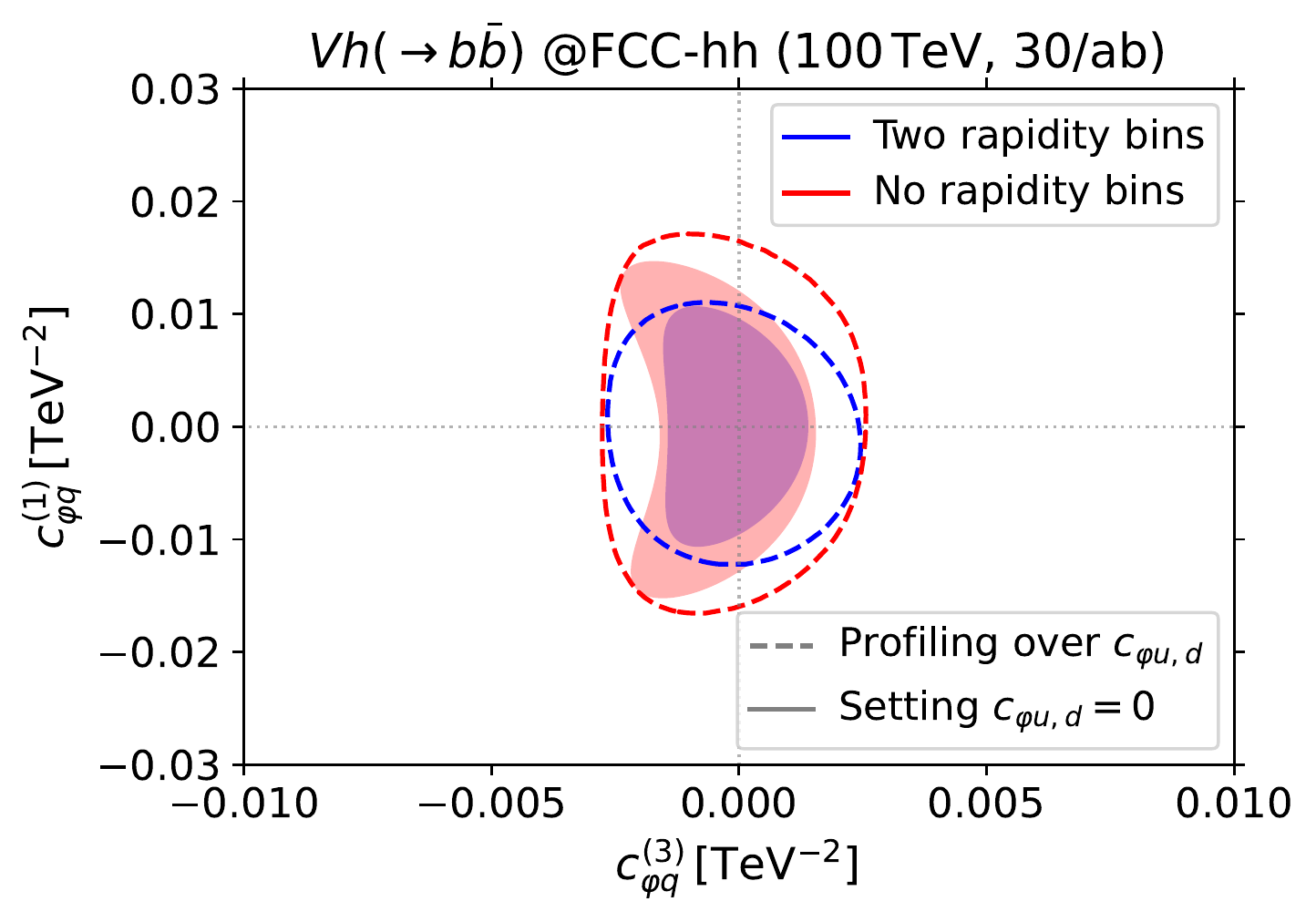}\hfill
	\includegraphics[width=0.485\linewidth]{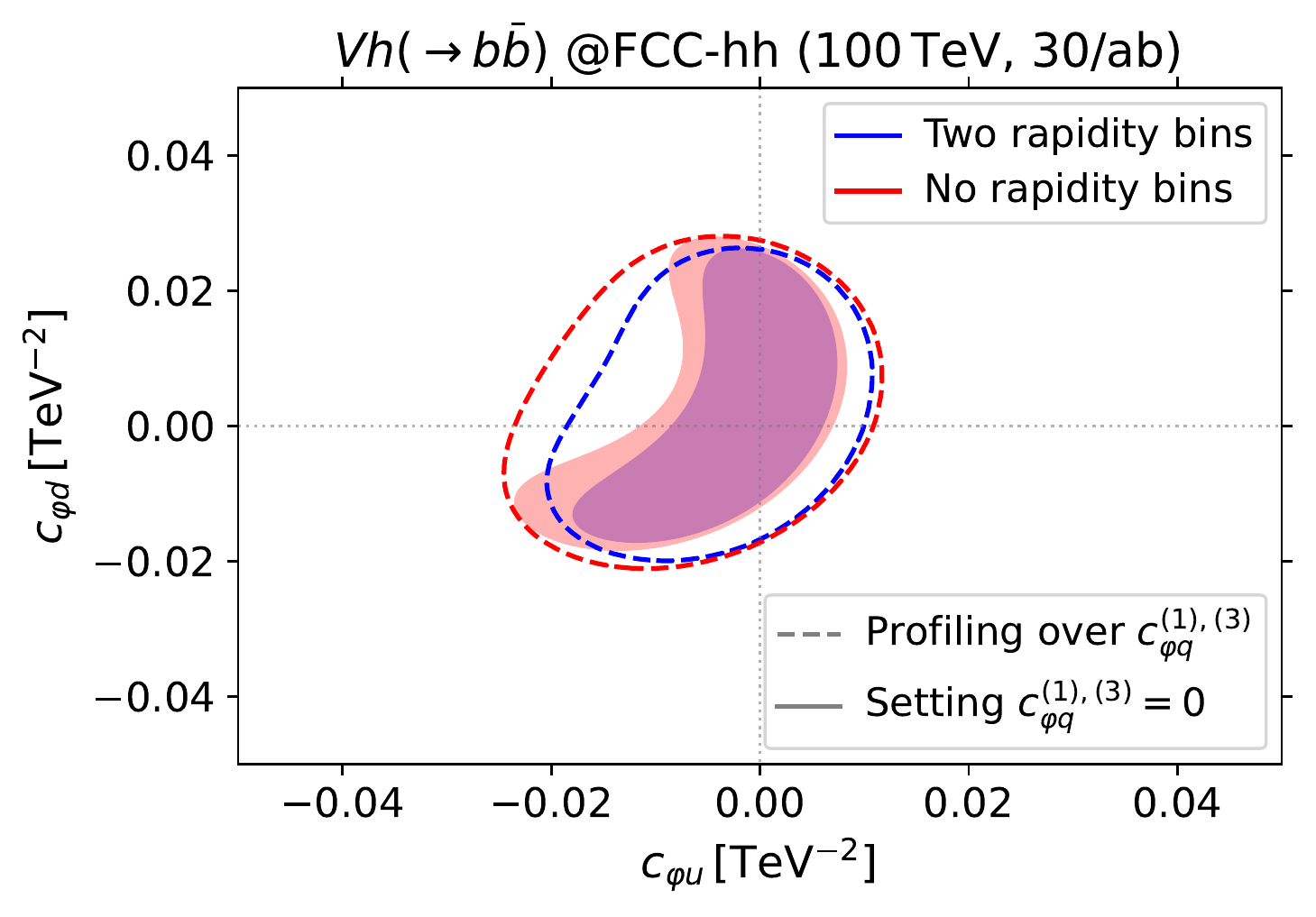}\\
	\caption{Projected $95\%$ C.L. bounds in the $\chq$-$\chqt$ (left panel) and $\chu$-$\chd$ (right panel) planes
	at the FCC-hh 
	(for $1\%$ systematic uncertainty).
	Blue regions and lines show the bound obtained by using a double binning in $p_{T}$ and rapidity, as explained in section~\ref{sec:binning}. The red regions and lines show the bound obtained by using only the $p_{T}$ binning. Solid (dashed) lines correspond to the bounds when setting to zero (profiling over) the Wilson coefficients not included in the plot.
	}
	\label{fig:2D_bounds_WhZh_FCC_rapidity binning}
\end{figure}

As explained in sections~\ref{sec:Interference} and \ref{sec:binning}, the analysis of the 0- and 2-lepton channels for FCC-hh includes a double binning strategy, adding a binning in the rapidity of the Higgs boson or of the $Zh$ system. Figure~\ref{fig:2D_bounds_WhZh_FCC_rapidity binning} shows the effect of this second binning on the bounds in the planes $\chq$-$\chqt$ and $\chu$-$\chd$. In the figure, we plot the results assuming $1\%$ systematics to highlight the effect of exploiting the rapidity distribution. %, since the addition of bins reduces the effect of systematics even with flat underlying distributions.
From the left panel of Figure~\ref{fig:2D_bounds_WhZh_FCC_rapidity binning}, one can see clearly that the rapidity binning tightens the bound on $\chq$ much more than on $\chqt$ thanks to the partial uplifting of the cancellation between up- and down-type quark contributions (see discussion in section~\ref{sec:Interference}). The right panel shows that this second binning also helps to decorrelate  $\chu$ and $\chd$ thanks to the different rapidity distribution of the up and down-type quarks that contribute to the interference of each of them. Due to the shape of the $\chi^2$ in the $\chu$-$\chd$ plane, this smaller correlation translates mainly onto a better bound on the negative value of $\chu$. Numerically, we checked that, under the $1\%$ systematics assumption, the rapidity binning can improve the bound on $\chq$ by $\sim30-40\%$ and the negative bound $\chu$ by up to $30\%$. The other bounds improve by $\sim10\%$.

\begin{figure}[t]
	\centering
	\includegraphics[width=0.485\linewidth]{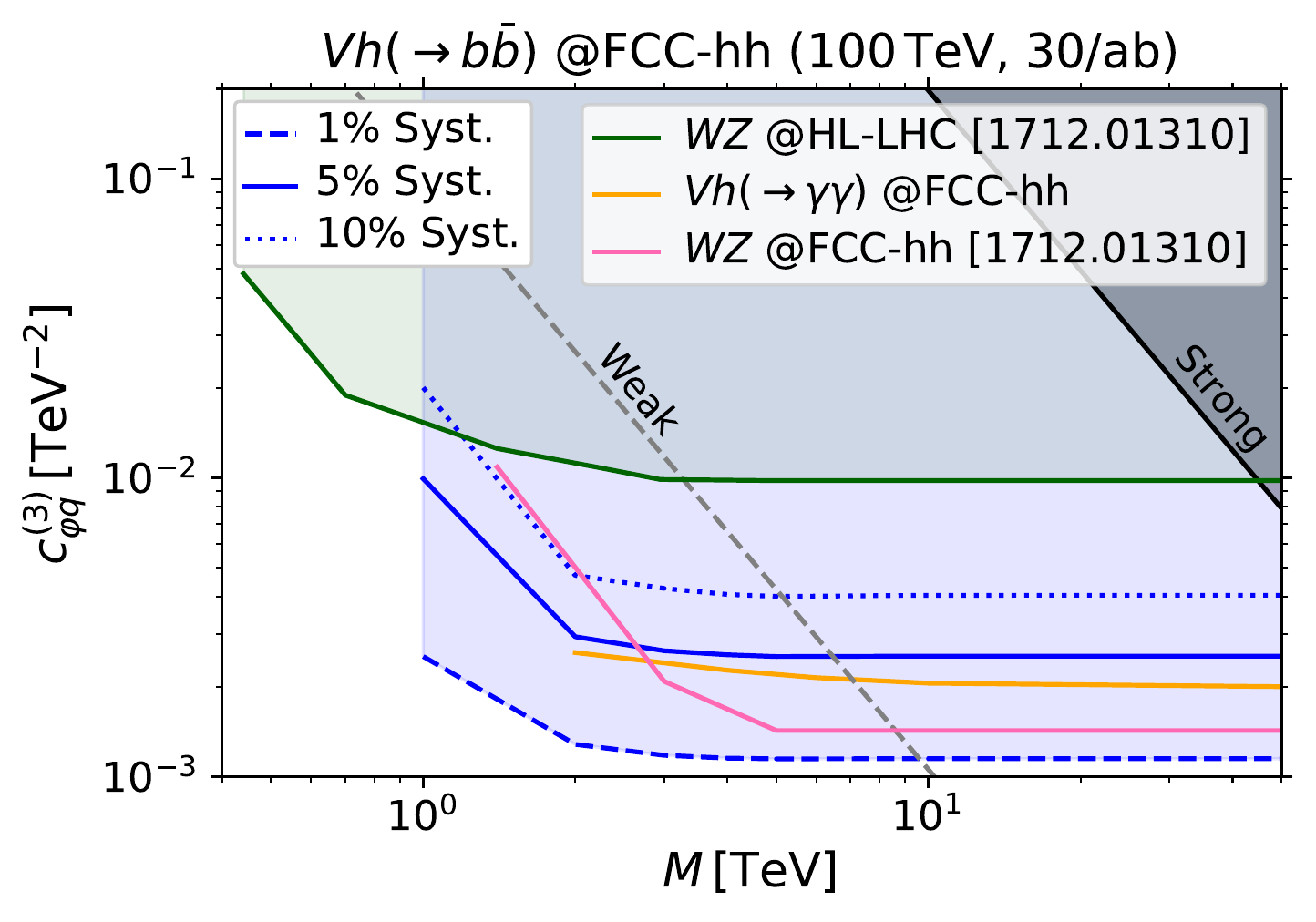} \hfill
	\includegraphics[width=0.485\linewidth]{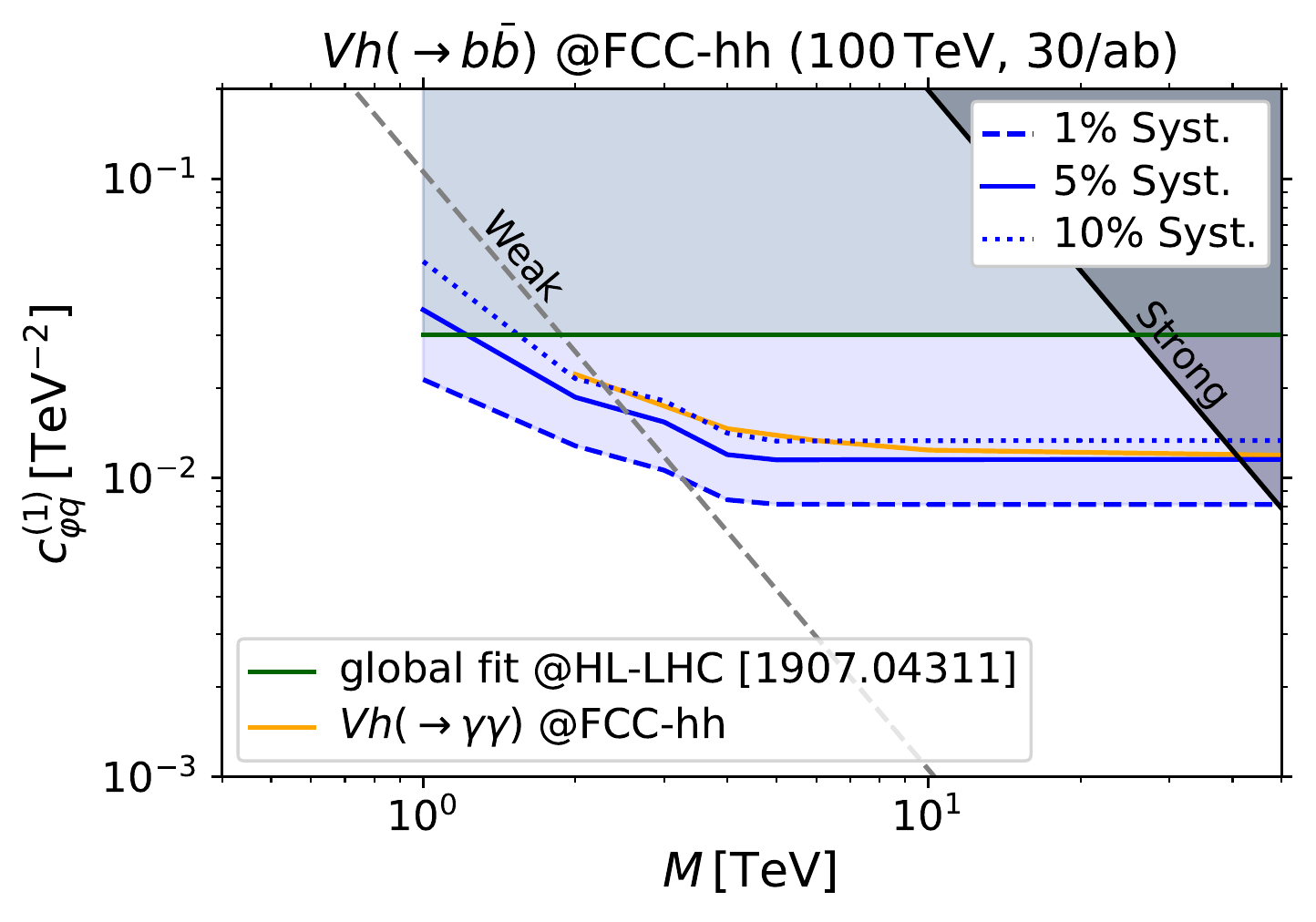} \\
    \includegraphics[width=0.485\linewidth]{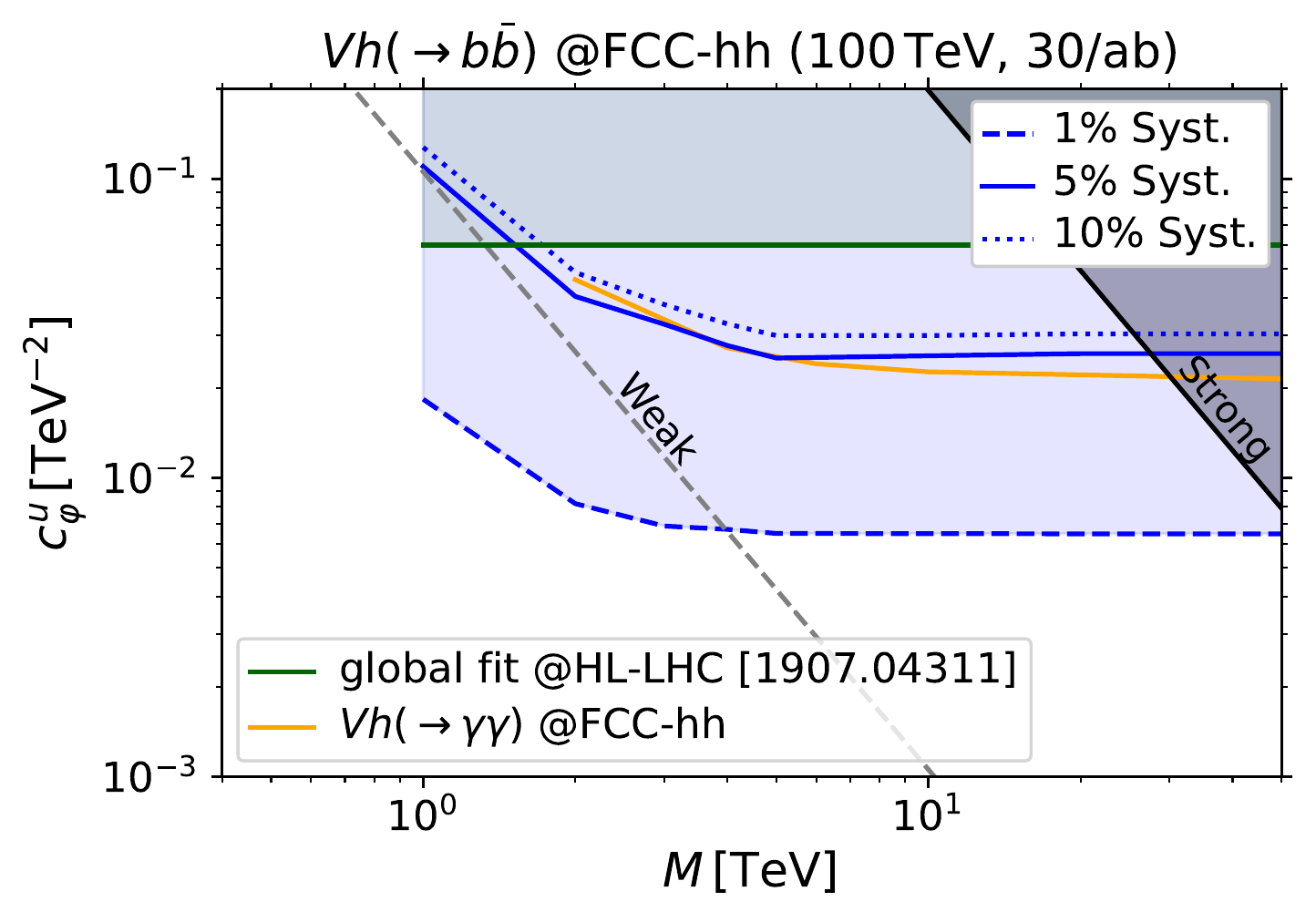}\hfill
	\includegraphics[width=0.485\linewidth]{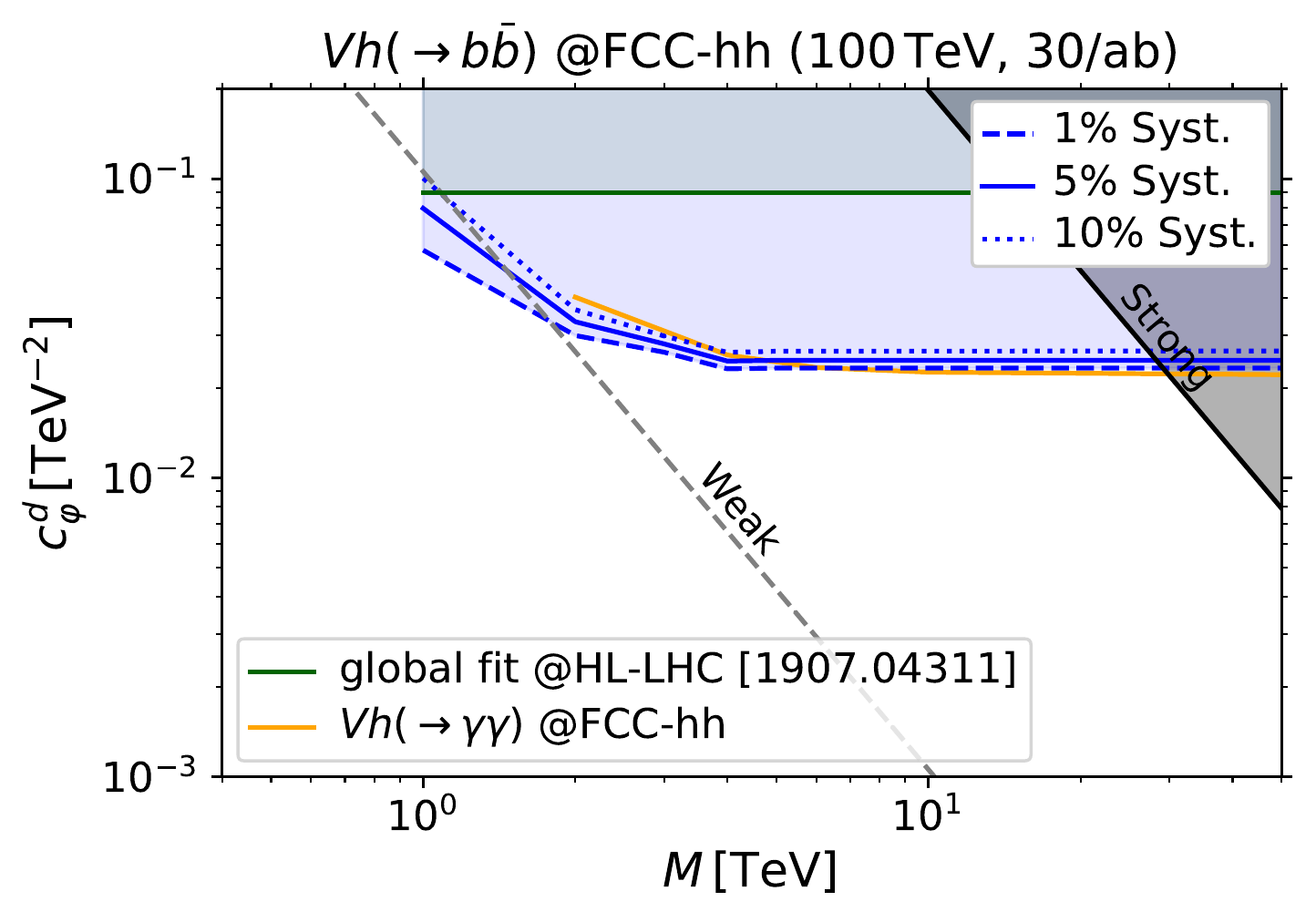}
 	\caption{Projected $95\%$ C.L.~bounds on $c_{\varphi q}^{(3)}$, $\chq$, $\chu$ and $\chd$ from one-operator fits at FCC-hh
	as function of the maximal-invariant-mass cut $M$.
	The dashed, solid and dotted blue lines show the bounds for $1\%$, $5\%$ and $10\%$ systematic errors. 
	Projections for some future hadron colliders are also shown.
	For $\chqt$, we show
	the projections from the $WZ$ channel at the HL-LHC (dark green solid line) and at the FCC-hh (pink solid line) from ref.~\cite{Franceschini:2017xkh}, assuming 5\% systematic uncertainty. Note that in ref.~\cite{Franceschini:2017xkh}, the bounds for the FCC-hh assume an integrated luminosity of $20\,\mathrm{{ab}^{-1}}$, whereas our bounds correspond to ${\cal L} = 30\,\mathrm{{ab}^{-1}}$. For the other operators, we show the 1-operator fit at HL-LHC from ref.~\cite{deBlas:2019wgy} (dark green solid lines). For all the operators we compare to the corresponding bounds from $Vh(\rightarrow \gamma\gamma)$ at the FCC-hh with $5\%$ systematics~\cite{Bishara:2020pfx} (orange solid lines).
	}
	\label{fig:fit_cphiq_FCC}
\end{figure}

Finally, figure~\ref{fig:fit_cphiq_FCC} shows our projected bounds for FCC-hh as a function of the maximal invariant mass $M$ of the events included in the fits. Notice that, when the bounds are asymmetric, we show the weaker one (in absolute value). The bounds degrade significantly only for invariant masses $M\lesssim 5$~TeV, which indicates that all our bounds can be used safely for EFTs with cutoffs above that value. 
This value is much smaller than the centre of mass energy of the hadronic collisions, phenomenon related to the PDF energy suppression (see ref.~\cite{Cohen:2021gdw} for a recent analytical study of this). 

In the plot corresponding to $\chqt$, upper-left panel, we also plot the bound from the fully-leptonic $WZ$ channel obtained in ref.~\cite{Franceschini:2017xkh} at HL-LHC and FCC-hh with a $5\%$ syst. uncertainty. The $WZ$ bound at FCC-hh is marginally better than the ones provided by either $Vh(\to b\bar b)$ or $Vh(\to \gamma \gamma)$. However, the $WZ$ analysis assumed a lower luminosity than ours, $\mathcal{L}=20$ ab$^{-1}$ instead of $30$ ab$^{-1}$, and the complete absence of backgrounds.

Figure~\ref{fig:fit_cphiq_FCC} also shows that the bounds on $\chu$ are strongly affected by the level of systematic uncertainties, in particular they degrade by a factor of $\sim 4$ going from $1\%$ to $5\%$. This stems from the fact that the likelihood has a highly non-quadratic shape due to cancellations in the amplitude between the SM-BSM interference term and the BSM squared contribution.
A much milder dependence of the bound on the systematic uncertainties is found for positive values of $\chu$ (see table~\ref{tab:bounds_summary_FCC}).

\subsection{Diboson impact on anomalous Triple Gauge Couplings}\label{sec:Res_aTGCs}

The impact of diboson precision measurements can also be appreciated when they are interpreted in terms of bounds on anomalous Triple Gauge Couplings (aTGCs). By adopting the Higgs basis, one can see that the Wilson coefficients considered in this work are related to vertex corrections and aTGCs as
\begin{eqnarray}
	\begin{aligned}
		\chqt = & +   \frac{1}{4 m_W^2} g^2 \left(\delta g_L^{Zu} - \delta g_L^{Zd} - c_{\textsc w}^2 \, \delta g_{1z}\right) \\
		\chq = & - \frac{1}{4 m_W^2} g^2\left(\delta g_L^{Zu} + \delta g_L^{Zd} +\frac{1}{3}\left(t_{\textsc w}^2 \delta\kappa_{\gamma}-s_{\textsc w}^2 \delta g_{1z}\right)\right) \\
		\chu = & - \frac{1}{2 m_W^2} g^2 \left(\delta g_{R}^{Zu}+\frac{2}{3}\left(t_{\textsc w}^2 \delta\kappa_{\gamma}-s_{\textsc w}^2 \delta g_{1z}\right)\right) \\
		\chd = & - \frac{1}{2 m_W^2} g^2 \left(\delta g_{R}^{Zd}-\frac{1}{3}\left(t_{\textsc w}^2 \delta\kappa_{\gamma}-s_{\textsc w}^2 \delta g_{1z}\right)\right)
	\end{aligned}
	\label{eq:Higgsbasis}
\end{eqnarray}
where $c_{\textsc w}$, $s_{\textsc w}$ and $t_{\textsc w}$  are the cosine, sine and tangent of the weak mixing angle respectively. 

Our results can not be translated directly to the Higgs basis without the appearance of 2 flat directions. However, if we assume that the UV theory belongs to the class of Universal Theories, the Wilson coefficients must fulfill the relation~\cite{Wells:2015uba,Franceschini:2017xkh}
\begin{equation}
	\chq = \frac{1}{4}\chu = -\frac{1}{2}\chd\,.
\end{equation}
Additionally, for this class of UV models, the vertex corrections are fully determined by the Peskin--Takeuchi oblique parameters, which are heavily constrained by EW precision observables and will be measured with even further precision at future lepton colliders. Hence, we can assume $\delta g_{L}^{Zq}\sim 0$ and express our results as bounds on the aTGCs $\delta\kappa_{\gamma}$ and $\delta g_{1z}$. 

We summarize our bounds on the aforementioned aTGCs in table~\ref{tab:Bounds_aTGCs}. We include the bounds that can be achieved at LHC Run 3, HL-LHC and FCC-hh from both profiled and one-operator fits. HL-LHC can improve the reach of LHC Run 3 on $\delta g_{1z}$ by a factor $\sim2$ but only tightens the bound $\delta\kappa_\gamma$ by $\sim30\%$. FCC-hh can easily improve the bound on both aTGCs by an order of magnitude with respect to HL-LHC.

\begin{table}[t]
\begin{centering}

\begin{tabular}{c|c|c|c}
\toprule
aTGC & Collider & Profiled Fit & One-Operator Fit\tabularnewline
\midrule 
\multirow{3}{*}{$\delta g_{1z}$} & \rule{0pt}{1.25em} LHC Run 3 & $[-2.8,\,6.5]\times10^{-3}$ & $[-2.7,\,4.3]\times10^{-3}$ \tabularnewline
 & \rule[-.65em]{0pt}{1.9em} HL-LHC & $[-1.5,\,3.2]\times10^{-3}$ & $[-1.3,\,1.7]\times10^{-3}$ \tabularnewline
 & \rule[-.65em]{0pt}{1.9em} FCC-hh & $[-2.9,\,3.5]\times10^{-4}$ & $[-2.1,\,2.3]\times10^{-4}$ \tabularnewline
\hline 
\multirow{3}{*}{$\delta\kappa_{\gamma}$} & \rule{0pt}{1.25em} LHC Run 3 & $[-1.4,\,2.8]\times10^{-2}$  & $[-1.3,\,2.2]\times10^{-2}$ \tabularnewline
 & \rule[-.65em]{0pt}{1.9em} HL-LHC & $[-9.3,\,19.8]\times10^{-3}$ & $[-7.1,\,16.4]\times10^{-3}$ \tabularnewline
 & \rule[-.65em]{0pt}{1.9em} FCC-hh & $[-1.4,\,3.5]\times10^{-3}$  & $[-1.1,\,3.3]\times10^{-3}$ \tabularnewline
\bottomrule
\end{tabular}

\par\end{centering}
\caption[caption]{Bounds at $95\%$ C.L.~on the aTGCs $\delta\kappa_{\gamma}$ and $\delta g_{1z}$ from $Vh(\to b\bar b)$ at present and future hadron colliders with $5\%$ syst. uncertainty. These results are valid under the assumption of Universal Theories and with vanishing vertex corrections. 
{\bf Left column:} Bounds from the global fit, profiled over the other coefficients. {\bf Right column:} Bounds from a one-operator fit (i.e. setting the other coefficients to zero).}
\label{tab:Bounds_aTGCs}
\end{table}

The bound on $\delta g_{1z}$ can be further improved if the $Vh$ and $WZ$ channels are combined. Taking the analysis of $WZ$ from ref.~\cite{Franceschini:2017xkh}, we obtain the following bounds:
\begin{equation}
\label{eq:bounds_aTGC_Vh_WZ}
\def\arraystretch{1.3}
\begin{array}{l@{\hspace{.8em}}l@{\hspace{3.em}}l@{\hspace{2em}}l}
\textrm{LHC Run 3} & {(300\;{\rm fb}^{-1})} &  \delta g_{1z} \in[-1.6, 2.6]\times10^{-3}\,\left([-1.6, 2.5]\times 10^{-3}\right)\;, \\
\textrm{HL-LHC} & {(3\;{\rm ab}^{-1})} &  \delta g_{1z} \in[-7.3, 9.5]\times 10^{-4}\,\left([-7.1, 8.9]\times 10^{-4}\right)\; ,\\
\textrm{FCC-hh} & {(30\;{\rm ab}^{-1})} &  \delta g_{1z} \in[-1.5, 1.7]\times 10^{-4}\,\left([-1.0, 1.3]\times 10^{-4}\right)\; ,
\end{array}
\end{equation}
where we report the bounds obtained by profiling over $\delta\kappa_\gamma$ and by setting it to zero, the latter in parenthesis. The power of $WZ$ in constraining $\chqt$, and hence $\delta g_{1z}$, is shown by the fact that the bound improves in at least a factor of $2$ in all cases. %The power of $WZ$ in constraining $\chqt$ is shown by the fact that the bound improves in at least a factor of $2$ in all cases.

\begin{figure}[t]
	\centering
	\includegraphics[width=0.47\linewidth]{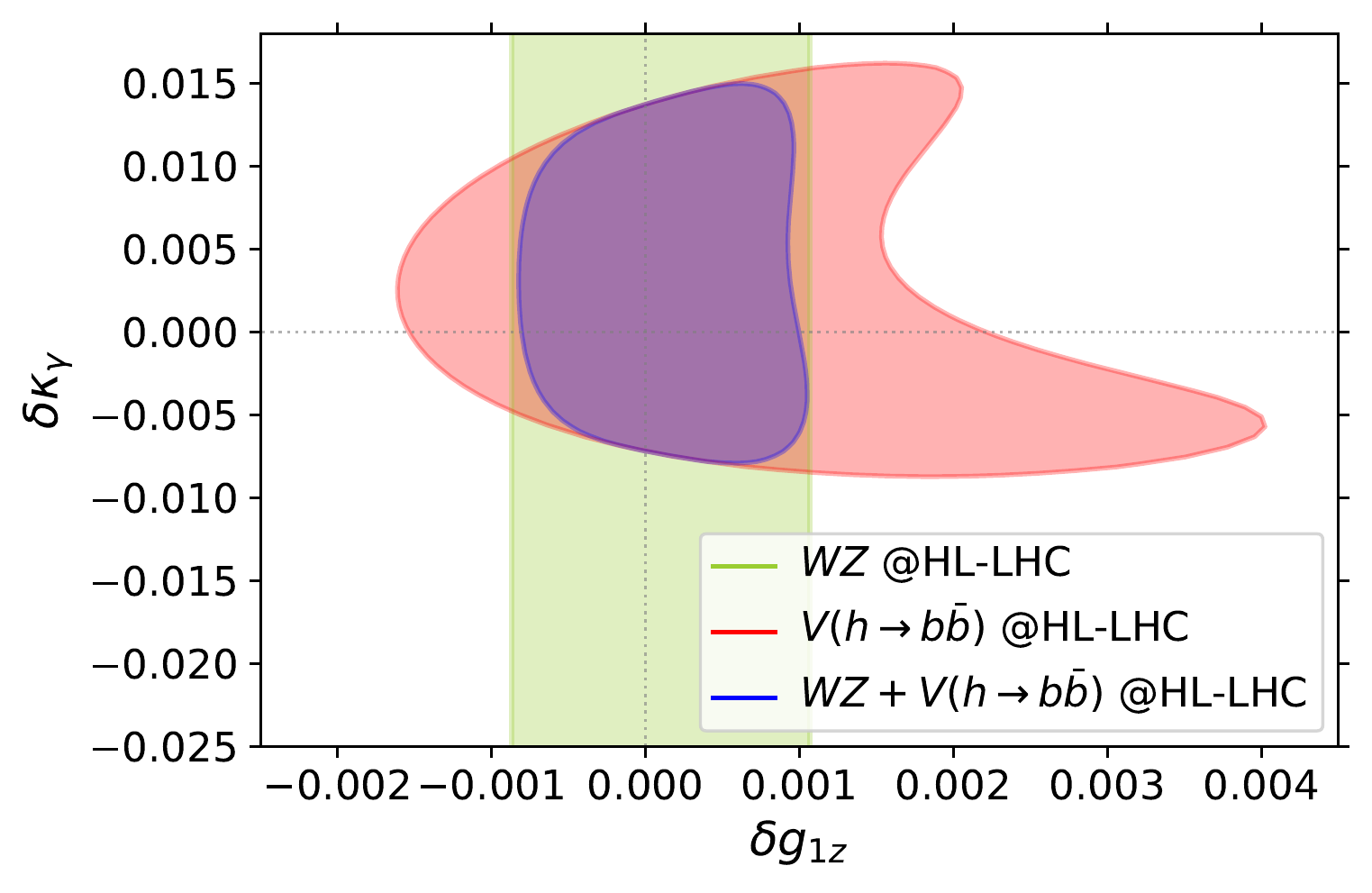}
	\hfill
	\includegraphics[width=0.47\linewidth]{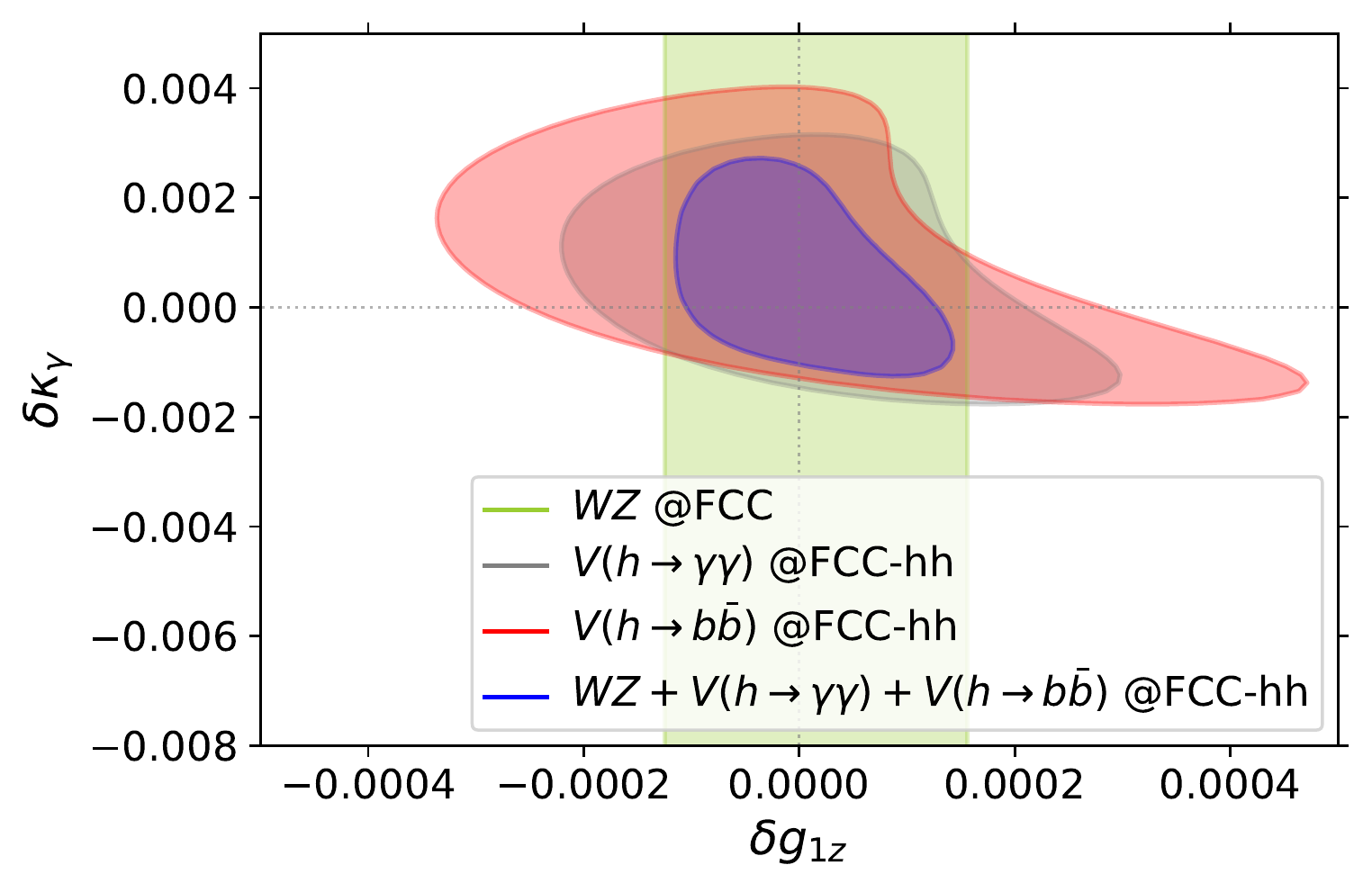}
	\caption{$95\%$ C.L. bounds on the anomalous Triple Gauge Couplings $\delta g_{1z}$ and $\delta\kappa_{\gamma}$ for Universal Theories. We show the bounds obtained from our analysis of $Vh(\to b \bar b)$ at the HL-LHC and the FCC-hh, and compare them to the bounds obtained from different studies. Additionally, we present the results of combining the bounds from all the analyses we are comparing for each of the two colliders, respectively.
	\textbf{Left panel:} Bounds at the HL-LHC. We compare our results from $Vh(\to b \bar b)$ with the bounds from the leptonic $WZ$ channel~\cite{Franceschini:2017xkh}.
	\textbf{Right panel:} Bounds at the FCC-hh. We compare our results from $Vh(\to b \bar b)$ with the bounds from the leptonic $WZ$ channel~\cite{Franceschini:2017xkh} and from $Vh(\to\gamma\gamma)$~\cite{Bishara:2020pfx}.
	}
	\label{fig:atgc_FCC}
\end{figure}

We also plot our results on the $\delta\kappa_{\gamma}$-$\delta g_{1z}$ plane in figure~\ref{fig:atgc_FCC}. Here, we also compare the impact of combining the results of this paper with the ones from ref.~\cite{Bishara:2020pfx} at FCC-hh. This figure shows how the inclusion of the $Vh$ process is essential to constrain $\delta\kappa_\gamma$, whereas the bound on $\delta g_{1z}$ is mostly determined by the $WZ$ channel.
We note that, at FCC-hh, the $\gamma\gamma$ or $b\bar b$ Higgs decay channels in the $Vh$ process have similar constraining power. 

As found in ref.~\cite{Bishara:2020pfx}, the diboson channels $WZ$ and $Vh(\to\gamma\gamma)$ at FCC-hh could be combined with global fits at the future CEPC and FCC-ee to tighten (by a factor $\sim 2$) the bounds on $\delta\kappa_{\gamma}$ and $\delta g_{1z}$ in Flavour Universal scenarios. The above results show that the inclusion of the $Vh(\to b\bar b)$ channel could improve further the global fit.

\section{Summary and conclusions} \label{sec:conclusions}

In this paper we continued the study of the leptonic $Vh$ production processes with the aim of assessing their potential for testing new physics through precision measurements. In particular, we focused on the $h \to b\bar b$ decay channel, which provides a sizeable cross section and can be exploited not only at future hadron colliders, but also at current and near-future LHC runs.
The challenge with the processes we considered in this work is the fact that they suffer from large QCD-induced backgrounds because the final state contains two $b$-quarks. This aspect markedly differs from many other precision diboson studies, which focused on clean channels with relatively small backgrounds (for instance fully leptonic $WZ$~\cite{Franceschini:2017xkh}, leptonic $W\gamma$~\cite{Panico:2017frx}, or leptonic $Vh$ with $h \to \gamma\gamma$~\cite{Bishara:2020pfx,Bishara:2020vix}). As a consequence, a tailored analysis strategy was needed to achieve good sensitivity.
Specifically, we considered new physics effects parametrized by four dimension-6 EFT operators,
namely $\Ohqt$, $\Ohq$, $\Ohu$ and $\Ohd$ in the Warsaw basis, which induce energy-growing corrections to the SM amplitudes.

Since the main new-physics effects are expected in the high-energy tails of kinematic distributions, we found it convenient to isolate energetic events by exploiting boosted-Higgs identification techniques. In our analysis we split the events in two categories, depending on whether a boosted Higgs candidate or two resolved $b$-jets were present; see appendix~\ref{sec:btagger_details} for more details. Moreover we classified the events depending on the number of charged leptons ($0$, $1$, or $2$) in the final state. For each class, we devised optimized cuts to improve the sensitivity to new physics (the selection cuts are reported in appendix~\ref{sec:AppCuts}).

The combined analysis of boosted and resolved events provides a significant improvement in sensitivity. With respect to an analysis exploiting only boosted events, the combination of the boosted and resolved categories yields a 17\% improvement on the most strongly bounded Wilson coefficient at the LHC Run 3 and a 7\% improvement at FCC-hh.

We found that at LHC Run 3 our analysis provides bounds competitive with the ones derived from the $WZ$ diboson channel. The main limitation at this stage is low statistics which results in uncertainties larger than the expected systematic ones.
The HL-LHC program, thanks to the tenfold increase in integrated luminosity, allows for a significant improvement in the bounds; see table~\ref{tab:bounds_summary_LHC}.
In this case the statistical error becomes of order $5\%$, which is most likely comparable to the expected systematic uncertainty. We found that at the end of the HL-LHC, the $Vh(\to b\bar b)$ processes could have an important impact on bounding $\chqt$ and $\chu$, even when included in a global EFT fit.

An additional strong improvement in sensitivity could be achieved at FCC-hh, thanks to the much higher integrated luminosity and, especially, the wider energy range. The possibility to access very energetic events ($p_T \gtrsim 600$\,GeV) gives a substantial bonus for the analysis. In fact, in the high-$p_T$ bins in the boosted analysis for the 0- and 2-lepton categories, the signal-to-background ratio is found to be of order one or better, while the number of expected events is still sizeable (see figs.~\ref{fig:histograms_vv_FCC} and \ref{fig:histograms_ll_FCC}). This should be contrasted with the LHC case, in which the background dominates over the signal in all bins (figs.~\ref{fig:histograms_vv}-\ref{fig:histograms_ll}).
Furthermore, the significantly larger number of events at FCC-hh enables us  to additionally bin with respect to the rapidity of the $Vh$ system.
This helps in reducing the correlation among the new-physics effects and enhances the sensitivity to $\Ohq$ and $\Ohu$.
Taking these aspects into account, we found that FCC-hh could improve the bounds on the four EFT operators we considered by nearly one order of magnitude with respect to HL-LHC.
The $Vh(\to b\bar b)$ process is thus expected to remain competitive with other precision probes as a test of $\Ohqt$ and $\Ohq$. In particular, the sensitivity to $\Ohqt$ was found to be comparable to the one expected from the analysis of the fully-leptonic $WZ$ process (see upper left panel of fig.~\ref{fig:fit_cphiq_FCC}).

For FCC-hh, it is also interesting to compare the sensitivity of the $Vh(\to b \bar b)$ analysis with the one expected from $Vh(\to \gamma \gamma)$ which was studied in refs.~\cite{Bishara:2020pfx,Bishara:2020vix}.
We found that the relative importance of the two channels crucially depends on the amount of systematic uncertainty that characterizes the $h \to b \bar b$ final state.
For $5\%$ systematic uncertainty, both channels provide similar bounds, but thanks to harnessing different features.
Lower systematic uncertainties favor the $Vh(\to b \bar b)$ process, while larger uncertainties benefit the essentially background-free $h \to \gamma \gamma$ analysis.

For both the LHC and FCC-hh, we also studied the validity of the EFT expansion by deriving the bounds as a function of a cut on the maximum energy of any event included in the fits (which can be interpreted as a proxy for the EFT cut-off scale).
We found that our analysis could probe weakly coupled new physics scenarios with a new-physics scale up to $1.6$ TeV at LHC Run 3, $2.5$ TeV at HL-LHC, and $6.5$ TeV at FCC-hh (see figs.~\ref{fig:fit_cphiq_HL_LHC} and \ref{fig:fit_cphiq_FCC}). In the case of strongly-coupled new physics, the sensitivity extends well above $40$ TeV.

Finally, we assessed the relevance of our studies for testing aTGCs in Universal Theories. The $Vh$ processes (in particular $Zh$) are useful to tighten the constraint on $\delta \kappa_\gamma$, while a more limited impact is expected on $\delta g_{1z}$, which can be better determined through the $WZ$ process (see fig.~\ref{fig:atgc_FCC}).
At FCC-hh, the $Vh(\to b\bar b)$ channel shows a sensitivity comparable with the one of $Vh(\to \gamma\gamma)$.
The results obtained in the present work clearly highlight the relevance of the $Vh$ processes in the context of precision measurements at present and future colliders. 

We presented a detailed analysis of the main new-physics effects testable in these processes, however there are other aspects that could deserve further investigation.
First, subleading new physics effects due to CP-odd operators could be considered. In this case, as shown in ref.~\cite{Bishara:2020vix}, an extended analysis including angular distributions could be used to enhance the sensitivity.
A second aspect worth investigation is the final state in which the $W$ and $Z$ bosons decay hadronically with the Higgs decaying to two photons (this requires FCC-hh) or to a pair of $b$-quarks. The latter final state, $jjbb$, would be accessible at the LHC but is extremely challenging with regards to background suppression. 
This all hadronic channel is perhaps better suited to machine learning techniques rather than cut and count analyses if it is to be at all tractable.

\section*{Acknowledgments}
We thank K. Tackmann, G. Magni, and R. S. Gupta for useful discussions. We thank Marc Montull for collaboration in the early stages of this project. The work of F.B., P.E., C.G. and A.R. was partially supported by the Deutsche Forschungsgemeinschaft (DFG, German Research Foundation) under grant 491245950 and under Germany’s Excellence Strategy — EXC 2121 “Quantum Universe” — 390833306. The work of C.G. and A.R. was also partially supported by the International Helmholtz-Weizmann Research School for Multimessenger Astronomy, largely funded through the Initiative and Networking Fund of the Helmholtz Association. A.R. has received funding from the European Research Council (ERC) under the European Union’s Horizon 2020 research and innovation programme (Grant agreement No. 949451). G.P. was supported in part by the MIUR under contract 2017FMJFMW (PRIN2017).
This work was performed in part at the Aspen Center for Physics, which is supported by National Science Foundation grant PHY-1607611.

\appendix
\section{Details on the event simulation}
\label{sec:AppMCdetails}

For the simulations of LHC events,
we generated samples assuming a center-of-mass energy of $13\,\mathrm{TeV}$.
Although the actual LHC energy could slightly differ from this value, we expect the analysis not to be affected in a significant way.
For FCC-hh, we assume a center-of-mass energy of $100\,\mathrm{TeV}$.

The three signal processes, $pp\rightarrow Z(\rightarrow \nu\bar\nu)h$, $pp\rightarrow W(\rightarrow \ell\nu)h$ and $pp\rightarrow Z(\rightarrow \ell^+\ell^-)h$, were generated at NLO in the QCD coupling\footnote{The process $pp\rightarrow Zh \rightarrow \ell^+\ell^- b \bar b$ was computed in SMEFT with a subset of the operators up to NNLO in QCD ref.~\cite{Haisch:2022nwz}.}.
We accounted for QED NLO corrections by extracting the corresponding $k$-factors from ref.~\cite{Frederix:2018nkq}, where they are given as a function of the transverse momentum of the Higgs. We applied the $k$-factors by reweighting each event according to the transverse momentum of the reconstructed Higgs. The EW $k$-factors extracted from ref.~\cite{Frederix:2018nkq} are listed in table~\ref{tab:k-factors}.

\begin{table}[t]
\begin{centering}
\renewcommand{\arraystretch}{1.25}
\begin{tabular}{c|c|c}
\toprule
$p_T^h$ bin [GeV] & $k$-factor for $Zh$ & $k$-factor for $Wh$\tabularnewline
\midrule
$[0,\,230]$  & $0.96$ & $0.99$ \tabularnewline
$[230,\,330]$  & $0.94$ & $0.95$ \tabularnewline
$[330,\,500]$  & $0.90$ & $0.90$ \tabularnewline
$[500,\,700]$  & $0.87$ & $0.83$ \tabularnewline
$[700,\,1000]$  & $0.79$ & $0.75$ \tabularnewline
$[1000,\,1500]$  & $0.70$ & $0.63$ \tabularnewline
$[1500,\,\infty]$  & $0.57$ & $0.57$ \tabularnewline
\bottomrule
\end{tabular}
\par\end{centering}
\caption[caption]{NLO EW $k$-factors for the signal processes of our $Vh(\rightarrow b\bar b)$ study per bin of the transverse momentum of the Higgs boson $p_T^h$. The $k$-factors are extracted from ref.~\cite{Frederix:2018nkq}.
}
\label{tab:k-factors}
\end{table}

The EW corrections are fairly small in the low-$p_T^h$ bins, but become sizeable at high energy. Notice that the EW $k$-factors can be significantly smaller than one, and therefore tend to reduce the sensitivity to the signal.
On the other hand, we did not apply the EW $k$-factors to the background processes,
in such way to obtain more conservative bounds.

For the simulations at $13\,\mathrm{TeV}$, all the backgrounds were simulated at NLO in the QCD coupling, with the exception of the $t\bar t$ background, which was simulated at leading order but with an additional hard QCD jet. This was done in order to account for a big part of the QCD corrections while still reducing significantly the simulation time.
The high efficiency of our cuts on the $t\bar t$ channel required a large number of simulated events in order to achieve a reasonably low statistical error, hence the aforementioned compromise.

For the simulations at $100\,\mathrm{TeV}$ center-of-mass energy, we reduced the order of simulation for the $Z(\rightarrow \nu\bar \nu) b \bar b$ and $Z(\rightarrow \ell^+\ell^-) b \bar b$ backgrounds to LO because of computational costs.
The corresponding reduction in the accuracy of our simulations is most likely smaller than the uncertainty on the detector performance and on the technical details of the FCC-hh machine (eg.~acceptance regions, total integrated luminosity, etc.).

In order to improve the efficiency with which the simulated events pass our selection cuts, we applied a set of generation-level cuts and binning in the $m_{b\bar b}$, $\Delta R^{b \bar b}$ and $p_T^V$ variables. These cuts and bin boundaries are listed in table~\ref{tab:gen_cuts_LHC} for the simulations at $13\,\mathrm{TeV}$ and in table~\ref{tab:gen_cuts_FCC} for the simulations at $100\,\mathrm{TeV}$.

\begin{table}[t]
\begin{centering}
\centering\renewcommand*{\arraystretch}{1.5}
\begin{tabular}{c | m{2.5cm} | m{2.5cm} |  m{2.5cm}}
\toprule
 & \centering{$Z\rightarrow\nu\bar{\nu}$}  & \centering{$W\rightarrow\ell \nu$} & \centering{$Z\rightarrow\ell^+ \ell^-$} \tabularnewline
\midrule 
$p_{T,\min}^{j}$ {[}GeV{]} & \centering{$7^{a}$} & \centering{$7^{a}$} & \centering{-} \tabularnewline
$p_{T,\min}^{b}$ {[}GeV{]} & \centering{$7^{a}$} & \centering{$7^{a}$} & \centering{-} \tabularnewline
$p_{T,\min}^{\ell}$ {[}GeV{]} & \centering{$7$} & \centering{$7$}  & \centering{$7$}  \tabularnewline
$|\eta_{max}^{\ell}|$ & \centering{$\infty$} & \centering{$2.8$} & \centering{$2.8$}  \tabularnewline
$m_{b\bar b}$ {[}GeV{]}& \multicolumn{2}{c|}{$\{0,\,100,\,150,\,\infty\}^{a}$} & \centering{-}  \tabularnewline
$\Delta R^{b\bar b}$  & \multicolumn{2}{c|}{$\{0,\,1,\,\infty\}^{a}$} & \centering{-} \tabularnewline
$p_{T}^{V}$ & \multicolumn{3}{c}{$\{0,\,200,\,400,\,600,\,800,\,1200,\,\infty\}^{b}$} \tabularnewline
\bottomrule
\end{tabular}
\par\end{centering}
\caption{Parton level generation cuts for signal and background processes at $13\,\mathrm{TeV}$.
$p_{T}^{V}$ denotes the vector boson $p_T$ and a dash means that the cut was not used for that channel.
$^{a}$: only applied to the simulations of $t\bar t$. $^{b}$: applied to all simulations except $t\bar t$. }
\label{tab:gen_cuts_LHC}
\end{table}

\begin{table}[t]
\begin{centering}
\centering\renewcommand*{\arraystretch}{1.5}
\begin{tabular}{c | m{2.5cm} | m{2.5cm} |  m{2.5cm}}
\toprule
 & \centering{$Z\rightarrow\nu\bar{\nu}$}  & \centering{$W\rightarrow\ell \nu$} & \centering{$Z\rightarrow\ell^+ \ell^-$} \tabularnewline
\midrule 
$p_{T,\min}^{j}$ {[}GeV{]} & \centering{$10^{a}$} & \centering{$10^{a}$} & \centering{-} \tabularnewline
$p_{T,\min}^{b}$ {[}GeV{]} & \centering{$20^{c}$} & \centering{$20^{c}$} & \centering{$20^{c}$} \tabularnewline
$p_{T,\min}^{\ell}$ {[}GeV{]} & \centering{$10$} & \centering{$30$}  & \centering{$30$}  \tabularnewline
$|\eta_{max}^{\ell}|$ & \centering{$\infty$} & \centering{$6.1$} & \centering{$6.1$}  \tabularnewline
$m_{b\bar b}$ {[}GeV{]}& \multicolumn{2}{c|}{$\{0,\,100,\,150,\,\infty\}^{a}$} & \centering{-}  \tabularnewline
$\Delta R^{b\bar b}$  & \multicolumn{2}{c|}{$\{0,\,1,\,\infty\}^{a}$} & \centering{-} \tabularnewline
$p_{T}^{V}$ & \multicolumn{3}{c}{$\{0,\,200,\,400,\,600,\,800,\,1200,\,\infty\}^{b}$} \tabularnewline
\bottomrule
\end{tabular}
\par\end{centering}
\caption{Parton level generation cuts for signal and background processes at $100\,\mathrm{TeV}$.
$p_{T}^{V}$ denotes the vector boson $p_T$ and a dash means that the cut was not used for that channel.
$^{a}$: only applied to the simulations of $t\bar t$. $^{b}$: applied to all simulations except $t\bar t$. $^{c}$: only applied to LO simulations.}
\label{tab:gen_cuts_FCC}
\end{table}

\section{Details on the tagging algorithm}\label{sec:btagger_details}

In the following, we will describe in detail how the classification into boosted and resolved events was performed for LHC events.
This classification strategy is inspired by the one presented in ref.~\cite{Gouzevitch:2013qca}.
For FCC-hh, we used the same algorithm and modified certain parameters to reflect the expected wider coverage of the future detectors.
We specify those changes at the end of this appendix.

For each event, after showering, we clustered the final partons using the anti-$k_T$ algorithm \cite{Cacciari:2008gp} with radius parameter $R_\mathrm{minor}=0.4$. We denote the resulting jets ``minijets'' in the following. If the event contains a $b$-flavoured final parton that has an angular separation $\Delta R\leqslant R_\mathrm{minor}$ with respect to one of the minijets, this minijet receives a $b$-tag with a probability
\begin{equation}
    \mathrm{eff}_b^{\mathrm{LHC}} = \begin{cases} 
      0 & \mathrm{if }\,\,p_T \leq 20\mathrm{\,GeV\,\,or\,\,} |\eta| > 2.5 \\
      0.8\tanh(0.003 p_T)\frac{30}{1+0.086 p_T} & \mathrm{else} \\
   \end{cases}\, ,
\end{equation}
where $p_T$ and $\eta$ are the transverse momentum and pseudorapidity of the minijet respectively. If, instead, the event contains a $c$-parton within $\Delta R\leqslant R_\mathrm{minor}$ of the minijet, the minijet can be mistagged as a $b$-jet with a probability
\begin{equation}
    \mathrm{eff}_c^{\mathrm{LHC}} = \begin{cases} 
      0 & \mathrm{if }\,\,p_T \leq 20\mathrm{\,GeV\,\,or\,\,} |\eta| > 2.5 \\
      0.2\tanh(0.02 p_T)\frac{1}{1+0.0034 p_T} & \mathrm{else} \\
   \end{cases}\,.
\end{equation}
Finally if the event contains only light partons in the vicinity of the minijet, the mistag probability is given by
\begin{equation}
    \mathrm{eff}_j^{\mathrm{LHC}} = \begin{cases} 
      0 & \mathrm{if }\,\,p_T \leq 20\mathrm{\,GeV\,\,or\,\,} |\eta| > 2.5 \\
      0.002\tanh(7.3 \cdot 10^{-6}\cdot p_T ) & \mathrm{else} \\
   \end{cases}\,.
\end{equation}

Starting again from the final partons after parton-shower, we perform another clustering step using the anti-$k_T$ algorithm with radius parameter $R_\mathrm{major}=1.0$. We call the resulting jets ``fatjets''. For each fatjet, we recluster its constituents using the Cambridge/Aachen algorithm~\cite{Dokshitzer:1997in, Wobisch:1998wt} with radius parameter $R_\mathrm{major}$. Then, we apply the mass-drop tagging algorithm to the resulting $p_T$-leading jet. If there is a jet in the event that receives a mass-drop tag, we loop through all the previously $b$-tagged minijets and check whether one of them has an angular distance to the mass-drop tagged jet $\Delta R \leqslant R_\mathrm{flavor}=0.2$. For each such alignment, the mass-drop tagged jet receives a $b$-tag.

Events that contain at least one mass-drop-tagged jet are classified as ``boosted'' and the fatjets found inside them are stored for analysis. The rest of the events are considered as ``resolved'' and instead of the fatjets, we store the minijets. In the latter class of events, we expect the Higgs boson to appear as two $b$-tagged minijets, hence the ``resolved'' name.

The $b$-tagging algorithm was adapted for FCC-hh following the reference parameters available in the FCC-hh CDR~\cite{FCC:2018vvp} and the FCC-hh Delphes card~\cite{deFavereau:2013fsa}. In particular, the $b$-tagging efficiency is assumed to be $\mathrm{eff}_b^{\mathrm{FCC}}=0.85$ for any $b$-jet with $p_T>20$~GeV and $|\eta|<4.5$ and as $0$ outside that region. The probability of mistagging a $c$ or light jet as $b$-jet is assumed to be $\mathrm{eff}_c^{\mathrm{FCC}}=0.05$ and $\mathrm{eff}_j^{\mathrm{FCC}}=0.01$ inside the same region mentioned above and as $0$ in any other case. The parameters $R_\mathrm{major}$, $R_\mathrm{minor}$ and $R_\mathrm{flavor}$ were not modified.

\section{Selection cuts}\label{sec:AppCuts}

In this appendix, we describe the selection cuts applied for each event category.
We present in details the cuts used in the LHC analyses, and at the end of each section, we comment on the changes applied for FCC-hh. Finally, we present the cut efficiencies for each of the different parts of our analysis.

\subsection{Zero-lepton category}
\label{sec:Sel_cuts_0lep}

In the zero-lepton category, we require the absence of charged leptons in the acceptance region that is defined as $p_T > 7\,$GeV and $|\eta| < 2.5$ for electrons or $|\eta| < 2.7$ for muons. This acceptance region for leptons will be called loose. We furthermore require the absence of light non-$b$-tagged jets in the acceptance region $\left(p_T > 20\,\mathrm{GeV} \land |\eta| <2.5\right)$~$\lor$\\$\left(p_T > 30 \,\mathrm{GeV}\land 2.5 \leq |\eta| < 4.5\right)$. The acceptance regions for the different particles, both at (HL-)LHC and FCC-hh, are summarised in table~\ref{tab:accept_cuts}.

\begin{table}[t]
	\centering{
		\renewcommand{\arraystretch}{1.0}
    \begin{tabular}{ >{\centering\arraybackslash} m {0.09\textwidth} | >{\centering\arraybackslash} m {0.09\textwidth} |  >{\centering\arraybackslash}m{0.08\textwidth} | >{\centering\arraybackslash}m{0.08\textwidth} | >{\centering\arraybackslash}m{0.08\textwidth} | >{\centering\arraybackslash}m{0.08\textwidth} | >{\centering\arraybackslash} m {0.15\textwidth} | >{\centering\arraybackslash} m {0.1\textwidth} }
			\toprule
		\multirow{2}*{}	& \multirow{2}*{Collider} & \multicolumn{2}{c|}{ Electrons} & \multicolumn{2}{c|}{ Muons} & \multirow{2}*{Light Jets} & \multirow{2}*{$b$-jets} \\ 
	& & Loose & Tight & Loose & Tight & & \\
	\midrule
	\multirow{2}*{$p_T$~[GeV]}& LHC & $>7$ & $>27$ & $>7$ & $>25$ & $>30\,(>20)$ & \multirow{2}*{$>20$}\\
	& FCC-hh & \multicolumn{2}{c|}{$>30$} & \multicolumn{2}{c|}{$>30$} & $>30$ & \\
	\midrule
	\multirow{2}*{$|\eta|$}& LHC & \multicolumn{2}{c|}{$<2.5$} & \multicolumn{2}{c|}{$<2.7$}  &$<4.5\,(<2.5)$&$<2.5$\\
	& FCC-hh & \multicolumn{2}{c|}{$<6.0$} & \multicolumn{2}{c|}{$<6.0$} & $<6.0$ & $<4.5$ \\\bottomrule
		\end{tabular}}
		\caption{Acceptance regions for charged leptons, light non-$b$-tagged jets and $b$-tagged jets used in our analysis for LHC, HL-LHC and FCC-hh. The acceptance regions of LHC and HL-LHC are equal. In the case of light jets at LHC, the minimum $p_T$ outside (between) the parenthesis applies to the jets that fulfill the $|\eta|$ condition outside (between) the parenthesis, see text for details. All the values were chosen following refs.~\cite{ATLAS:2020fcp,ATLAS:2020jwz,FCC:2018vvp}}
		\label{tab:accept_cuts}
	\end{table}

In the resolved category of the (HL-)LHC analysis, in order to reconstruct the Higgs, we look for $b$-tagged jets within the region defined by $p_{T}^{b} \geqslant 20\,$GeV and $|\eta^{b}| \leqslant 2.5$ and ask the leading $b$-quark to have $p_{T}^{b,\mathrm{leading}} \geqslant 45$~GeV. 
The angular distance between the two $b$-jets must be $\Delta R_{bb} \leqslant 1.5$. Furthermore, the azimuthal angle between the $b$-quarks, $\Delta \phi (b_1, b_2)$, defined such that it lies in the range $[0,180]^\circ$, must be in the interval $[0,140]^\circ$. The $p_T$ of the $b$-quarks is further constrained by requiring $H_T=\sum_{b} p_{T}>H_{T,\min}=130$~GeV. We demand the azimuthal angle between the missing transverse momentum and the reconstructed Higgs to be $\Delta \phi (E_T^\mathrm{miss}, h_\mathrm{cand}) \in [120^\circ,\,240^\circ]$ and the azimuthal angles between the missing transverse momentum and the $b$-jets to be $\Delta \phi (E_T^\mathrm{miss}, b\mathrm{-jets}) \in [20^\circ,\,340^\circ]$. Lastly, we require the missing transverse momentum to be at least $E_{T,\min}^{\mathrm{miss}} = 150\,$GeV.

\begin{table}[t]
	\centering{
		\renewcommand{\arraystretch}{1.25}
		\begin{tabular}{ c @{\hspace{.5em}} |  >{\centering\arraybackslash}m{0.15\textwidth} | >{\centering\arraybackslash}m{0.15\textwidth}  |  >{\centering\arraybackslash}m{0.15\textwidth} | >{\centering\arraybackslash}m{0.15\textwidth} }
			\toprule
			\multirow{2}*{Selection cuts} & \multicolumn{2}{ c|}{Boosted category}  & \multicolumn{2}{ c}{Resolved category} \\
			\cline{2-5} & (HL-)LHC & FCC-hh & (HL-)LHC & FCC-hh  \\ 
			\midrule
			$p_{T,\min}^{b}$ [GeV] & \multicolumn{2}{c|}{-} & \multicolumn{2}{c}{ 20 } \\
			$p_{T,\min}^{b,\mathrm{leading}}$ [GeV] & \multicolumn{2}{c|}{-} & 45 & - \\
			$\eta_{\max}^{b}$ & \multicolumn{2}{c|}{-}& 2.5 & 4.5 \\
			$\eta_{\max}^{h_\mathrm{cand}}$ & 2.0 & 4.5 & \multicolumn{2}{c}{-}  \\
			$\Delta R_{bb}^\mathrm{max}$ & \multicolumn{2}{c|}{-} & \multicolumn{2}{c}{$1.5$} \\
			$H_{T,\min}$ [GeV] & \multicolumn{2}{c|}{-} & $130$ & - \\
			$\Delta \phi (E_T^\mathrm{miss}, h_\mathrm{cand})$ & \multicolumn{2}{c|}{$[120^\circ,\,240^\circ]$} & \multicolumn{2}{c}{$[120^\circ,\,240^\circ]$} \\
			$\Delta \phi (b_1, b_2)$& \multicolumn{2}{c|}{-} &$[0, 140]^\circ $ & $[0, 110]^\circ $ \\
			$\Delta \phi (E_T^\mathrm{miss}, b\mathrm{-jets})$ & \multicolumn{2}{c|}{-} &  \multicolumn{2}{c}{$[20^\circ,\,340^\circ]$} \\
			$E_{T,\min}^{\mathrm{miss}}$ [GeV] & $270$ & - & $150$ & - \\
			$m_{h_\mathrm{cand}}$ [GeV] & \multicolumn{4}{c}{ $[90,\,120]$} \\\bottomrule
		\end{tabular}}
		\caption{Summary of the selection cuts in the 0-lepton category for both LHC and FCC-hh analyses.}
		\label{tab:sel_cuts_0}
	\end{table}

In the boosted category, we require the Higgs candidate to fulfill $|\eta^{h_\mathrm{cand}}| \leqslant  2.0$. The cut on the azimuthal angle between the missing transverse momentum and the Higgs candidate is the same as in the resolved category and the minimum missing transverse momentum is $E_{T,\min}^{\mathrm{miss}} = 270\,$GeV.

For both the resolved and the boosted events, we select events where the mass of the reconstructed Higgs fulfills $m_{h_\mathrm{cand}} \in [90,\,120]\,$GeV. The selection cuts of the (HL-)LHC analysis are summarized in table \ref{tab:sel_cuts_0}.

The analysis described above was adapted to the FCC-hh scenario with minimal changes. The acceptance regions were extended as can be seen in table~\ref{tab:accept_cuts}, in particular with a much wider angular coverage but higher minimum $p_T$. In both boosted and resolved categories, we eliminated the $E_{T,\min}^{\mathrm{miss}}$ cut due to its poor discriminating power and being based on the ATLAS trigger specification at LHC. The boosted analysis suffered one further modification with respect to LHC: the increase in $\eta_{\max}^{h_\mathrm{cand}}$ to $4.5$. In the resolved analysis, we removed the $p_{T,\min}^{b,\mathrm{leading}}$ and $H_{T,\min}$ cuts and reduced the allowed region for $\Delta \phi (b_1, b_2)$ by $30^\circ$, to $[0, 110]^\circ $.

\subsection{One-lepton category}
\label{sec:Sel_cuts_1lep}

For the one-lepton category, we require exactly one charged lepton in the tight acceptance region defined as $p_T > 27\,$ GeV and $|\eta| < 2.5$ for electrons or $p_T > 25\,$ GeV and  $|\eta| < 2.7$ for muons. We veto events with additional $b$-tagged jets in the acceptance region, which is defined like in the zero-lepton category. All the acceptance regions are defined in table~\ref{tab:accept_cuts}.

At (HL-)LHC and for resolved events, we impose the same cuts on the $p_T$ and the pseudorapidity of $b$-jets as in the zero-lepton category. However, we simplify the angular-separation cuts and only require $\Delta R_{bb}\leqslant 2.0$. Lastly, we reject events with missing transverse momentum below $E_{T,\min}^{\mathrm{miss}} = 30\,$GeV if the charged lepton is an electron. In the muon sub-channel, we require $\textrm{min}\lbrace p_{T,\mathrm{min}}^{\mu, E_{T}^\mathrm{miss}}\rbrace \geqslant 90$~ GeV to replicate the ATLAS analysis~\cite{ATLAS:2020fcp}. In the summary table \ref{tab:sel_cuts_1}, we abbreviate this cut as $E_T^{\mathrm{miss}}$ for conciseness.

\begin{table}[t]
	\centering{
		\renewcommand{\arraystretch}{1.25}
\begin{tabular}{ >{\centering\arraybackslash} m {0.2\textwidth} |  >{\centering\arraybackslash}m{0.15\textwidth} | >{\centering\arraybackslash}m{0.15\textwidth}  |  >{\centering\arraybackslash}m{0.15\textwidth} | >{\centering\arraybackslash}m{0.15\textwidth} }
			\toprule
			\multirow{2}*{Selection cuts} & \multicolumn{2}{ c|}{Boosted category}  & \multicolumn{2}{ c}{Resolved category} \\
			\cline{2-5} & (HL-)LHC & FCC-hh & (HL-)LHC & FCC-hh  \\ \midrule
			$p_{T,\min}^{b}$ [GeV] &  \multicolumn{2}{c|}{-} & \multicolumn{2}{c}{20} \\
			$p_{T,\min}^{b,\mathrm{leading}}$ [GeV] &  \multicolumn{2}{c|}{-} & $45$ & -  \\
			$\eta_{\max}^{b}$ & \multicolumn{2}{c|}{-} & 2.5 & 4.5\\
			$\eta_{\max}^{h_\mathrm{cand}}$ & 2.0 & 4.5 & \multicolumn{2}{c}{-} \\
			$\Delta R_{bb}^\mathrm{max}$ & \multicolumn{2}{c|}{-} & \multicolumn{2}{c}{$2.0$} \\
			$E_{T,\min}^{\mathrm{miss}}$ [GeV] & $\Big\lbrace$\scriptsize\begin{tabular}{c}
			     $50$ if $\ell=e$ \\
			     $90$ if $\ell=\mu$
			\end{tabular}  & - & $\Big\lbrace$\scriptsize\begin{tabular}{c}
			     $30$ if $\ell=e$ \\
			     $90$ if $\ell=\mu$
			\end{tabular} & - \\
			$|\Delta y (W, h_\mathrm{cand})|_\mathrm{max}$ & $1.4$ & $1.2$ & \multicolumn{2}{c}{-} \\
			$m_{h_\mathrm{cand}}$ [GeV] & \multicolumn{4}{c}{ $[90,\,120]$}  \\\bottomrule
		\end{tabular}}
		\caption{Summary of the selection cuts in the 1-lepton category for LHC and FCC-hh analyses.}
		\label{tab:sel_cuts_1}
	\end{table}

In the boosted category, we impose the same maximum cut on the $|\eta|$ of the Higgs candidate as in the 0-lepton category. The cut on $E_{T,\min}^{\mathrm{miss}}$ is exactly like in the resolved category for muons but in the case of electrons, the minimum value is raised to $50$~GeV.
Finally, we only accept events where the maximum rapidity difference between the reconstructed $W$ and the Higgs candidate is $|\Delta y (W, h_\mathrm{cand})|_\mathrm{max} = 1.4$.

For both the resolved and the boosted events, we impose the same Higgs mass window cut as in the zero-lepton category, i.e. $m_{h_\mathrm{cand}} \in [90,\,120]$\,GeV.

For FCC-hh, we modify the acceptance regions of leptons and jets as described in table~\ref{tab:accept_cuts}. As in the 0-lepton category, we modify the resolved analysis by eliminating the  $E_{T,\textrm{min}}^{\textrm{miss}}$ and $p_{T,\min}^{b,\mathrm{leading}}$ cuts. In the boosted analysis, we also eliminate $E_{T,\textrm{min}}^{\textrm{miss}}$, increase $\eta_{\max}^{h_\mathrm{cand}}$ up to $4.5$ and reduce $|\Delta y (W, h_\mathrm{cand})|_\mathrm{max}$ to $1.2$. All the selection cuts for both colliders are summarised in table~\ref{tab:sel_cuts_1}.

\subsection{Two-lepton category}

In the two-lepton category at LHC, we require exactly two same-flavour charged leptons in the loose acceptance region, see table~\ref{tab:accept_cuts} for its definition. For events without a boosted Higgs candidate, we require the absence of non-$b$-tagged jets in the acceptance region. The cuts $p_{T,\mathrm{min}}^b$, $p_{T,\min}^{b,\mathrm{leading}}$, $\eta_{\max}^{b}$ and $\Delta R_{bb}$ in the resolved category and $\eta_{\max}^{h_\mathrm{cand}}$ in the boosted category are the same as in the zero-lepton category. 

Additionally, in the resolved category, we employ a minimum cut on the leading charged lepton of $p_{T,\min}^{\ell, \mathrm{leading}} = 27\,$ GeV. In order to select events where the dilepton pair comes from a $Z$ boson, we only accept events with invariant mass of the pair of charged leptons fulfilling $m_{\ell\ell}\in[81,101]\,$GeV.

For boosted events, there must be at least one charged lepton with $p_T>27\,$GeV and $|\eta|<2.5$. Furthermore, we require the difference in rapidity between the reconstructed $Z$ and the Higgs candidate to be at most $|\Delta y (Z, h_\mathrm{cand} )|_\mathrm{max} = 1.0$. The invariant mass of the lepton pair has to be $m_{\ell\ell}\in[66,116]\,$GeV. We also impose a maximum cut on the $p_{T}^{\ell}$ imbalance, defined as $|p_{T}^{\ell_1}-p_{T}^{\ell_2}|/p_{T}^{Z}\leqslant 0.8$. Finally, if the charged leptons are muons, we reject the event if the transverse momentum of the $Z$-$E_{T}^\mathrm{miss}$ system is not above $p_{T,\mathrm{min}}^{Z,E_T^\mathrm{miss}} = 90\,$GeV.

As in the zero- and one-lepton category, we impose the same Higgs mass window cut to the Higgs candidate. However, in this category we add a cut on the $p_T$ of the $Z$ boson, $p_T^{Z}>200$~GeV. This helps us to reduce Monte Carlo uncertainties in the backgrounds without worsening significantly our final results. This additional cut also forces us to use, at LHC, only 1 bin in the boosted category and 2 in the resolved one, as it will be explained in the next subsection. 

In the resolved category for FCC-hh, we removed the cut on $p_{T,\min}^{\ell, \mathrm{leading}}$ and raised $\Delta R_{bb}^{\mathrm{max}}$ and $|\eta_{\max}^{b}|$ to $2.0$ and $4.5$ respectively. The changes for the boosted category were broader since we raised $p_{T,\mathrm{min}}^{Z,E_T^\mathrm{miss}}$ for muons to $200$~GeV, increased the maximum allowed $|\eta^{h_\mathrm{cand}}|$ to $4.5$ and reduced the maximum allowed $p_T^{\ell}$ imbalance to $0.5$. Furthermore, we removed the cuts requiring at least one lepton with $p_T>27\,$GeV and $|\eta|<2.5$ and $p_{T,\mathrm{min}}^{Z}$ of $200$~GeV. 
A summary of the cuts for both resolved and boosted categories at FCC-hh and LHC is in table \ref{tab:sel_cuts_2}.

\begin{table}[t]
	\centering{
		\renewcommand{\arraystretch}{1.37}
    \begin{tabular}{ >{\centering\arraybackslash} m {0.215\textwidth} |  >{\centering\arraybackslash}m{0.16\textwidth} | >{\centering\arraybackslash}m{0.16\textwidth}  |  >{\centering\arraybackslash}m{0.16\textwidth} | >{\centering\arraybackslash}m{0.16\textwidth} }
			\toprule
			\multirow{2}*{Selection cuts} & \multicolumn{2}{ c|}{Boosted category}  & \multicolumn{2}{ c}{Resolved category} \\
			\cline{2-5} & (HL-)LHC & FCC-hh & (HL-)LHC & FCC-hh  \\ \midrule
			$p_{T,\min}^{b}$ [GeV] & \multicolumn{2}{c|}{-} &\multicolumn{2}{c}{$20$}\\
			$p_{T,\min}^{b,\mathrm{leading}}$ [GeV] & \multicolumn{2}{c|}{-} & $45$ & - \\
			$\eta_{\max}^{b}$ & \multicolumn{2}{c|}{-} & $2.5$ & $4.5$\\
			$\eta_{\max}^{h_\mathrm{cand}}$ & $2.0$ & $4.5$ & \multicolumn{2}{c}{-} \\
			$\Delta R_{bb}^\mathrm{max}$ & \multicolumn{2}{c|}{-} & $1.5$ & $2.0$ \\
			Leptons &  \scriptsize{\begin{tabular}{c}
			     $ \exists\, \ell$ with \\
			     $p_T > 27\,\mathrm{GeV}$ \\
			     $\textrm{and } |\eta| < 2.5$
			\end{tabular} }  & - & $\scriptstyle p_{T,\min}^{\ell, \mathrm{lead}} = 27\,\mathrm{GeV}$ & -\\
			$\Delta y (Z, h_\mathrm{cand})_\mathrm{max}$ & \multicolumn{2}{c|}{$1.0$}  & \multicolumn{2}{c}{-} \\
			$m_{\ell \ell}$ [GeV] &\multicolumn{2}{c|}{$[66,\,116]$}  & \multicolumn{2}{c}{$[81,\,101]$}  \\
			max. $p_{T}^{\ell}$ imbalance & $0.8$ & $0.5$ & \multicolumn{2}{c}{-} \\
			$p_{T,\mathrm{min}}^{Z,E_T^\mathrm{miss}}$ [GeV] & $90$ if $\ell=\mu$ & 200 & \multicolumn{2}{c}{-} \\
			$m_{h_\mathrm{cand}}$ [GeV] &\multicolumn{4}{c}{$[90,\,120]$}  \\
			$p_{T,\mathrm{min}}^{Z}$ [GeV] & $200$ & -  & - & - \\\bottomrule
		\end{tabular}}
		\caption{Summary of the selection cuts in the 2-lepton category for the LHC and FCC-hh analyses.}
		\label{tab:sel_cuts_2}
	\end{table}

\subsection{Cut efficiencies}\label{sec:cut_efficiency}

Tables \ref{tab:Cutflow_NuNu_Boosted} to \ref{tab:Cutflow_EllEll_Boosted} display the cutflows of the 0-, 1- and 2-lepton categories for the boosted and the resolved events. Notice that the first two rows (the first row in the 2-lepton category) are identical for the boosted and resolved categories, since we differentiate between the two categories in the subsequent step, requiring either a mass-drop tagged jet with 2 $b$-tags or 2 resolved $b$-jets,
as explained before.

\begin{table}[t]
\centering
\renewcommand{\arraystretch}{1.15}
\resizebox{\textwidth}{!}{
\begin{tabular}{c|c|c|c|c|c|c|c|c|c|c}
\toprule
\multicolumn{1}{c|}{\multirow{2}{*}{Cuts / Eff.}} & \multicolumn{2}{c|}{$Zh$}& \multicolumn{2}{c|}{$Wh$} & \multicolumn{2}{c|}{$Wb\bar b$} & \multicolumn{2}{c|}{$Zb\bar b$} & \multicolumn{2}{c}{$t\bar t$} \\
& LHC & FCC &LHC & FCC & LHC & FCC & LHC & FCC & LHC & FCC \\
\midrule
$0$ $\ell^{\pm}$                                    & $1$    & $1$    & $0.32$  & $0.41$  & $0.34$   & $0.40$  & $0.78$  & $1$     & $0.98$   & $1$ \tabularnewline
0 UT jets                                           & $0.37$ & $0.22$ & $0.036$ & $0.019$ & $0.02$   & $0.009$ & $0.12$  & $0.061$ & $0.011$  & $0.022$ \tabularnewline
1 MDT DBT jet                                       & $0.29$ & $0.19$ & $0.026$ & $0.012$ & $0.014$  & $0.005$ & $0.048$ & $0.018$ & $0.0018$ & $0.0012$ \tabularnewline
$\eta_{\max}^{h_\mathrm{cand}}$                     & $0.26$ & $0.19$ & $0.022$ & $0.012$ & $0.012$  & $0.005$ & $0.044$ & $0.018$ & $0.0016$ & $0.0012$ \tabularnewline
$\Delta \phi (E_T^\mathrm{miss},h_\mathrm{cand})$   & $0.26$ & $0.19$ & $0.022$ & $0.012$ & $0.012$  & $0.005$ & $0.044$ & $0.018$ & $0.0016$ & $0.0012$ \tabularnewline
$E_{T}^\mathrm{miss}$                               & $0.12$ & -      & $0.007$ & -       & $0.003$  & -       & $0.013$ & -       & $0.0005$ & - \tabularnewline
$m_{h_\mathrm{cand}}$                               & $0.12$ & $0.19$ & $0.007$ & $0.012$ & $0.0008$ & $0.001$ & $0.003$ & $0.003$ & $4\cdot10^{-5}$ & $0.0001$ \tabularnewline
\bottomrule         
\end{tabular}
}
\caption{Cutflow for the boosted events in the 0-lepton category at LHC and FCC-hh. The acceptance regions for charged leptons and jets at the different colliders are defined in the text. A dash means that the particular cut was not applied. UT, MDT and BDT stand for untagged, mass-drop-tagged and doubly-$b$-tagged respectively.}
\label{tab:Cutflow_NuNu_Boosted}
\end{table}

\begin{table}[t]
\centering
\renewcommand{\arraystretch}{1.15}
\resizebox{\textwidth}{!}{
\begin{tabular}{c|c|c|c|c|c|c|c|c|c|c}
\toprule
\multicolumn{1}{c|}{\multirow{2}{*}{Cuts / Eff.}} & \multicolumn{2}{c|}{$Zh$}& \multicolumn{2}{c|}{$Wh$} & \multicolumn{2}{c|}{$Wb\bar b$} & \multicolumn{2}{c|}{$Zb\bar b$} & \multicolumn{2}{c}{$t\bar t$} \\
& LHC & FCC &LHC & FCC & LHC & FCC & LHC & FCC & LHC & FCC \\
\midrule
$0$ $\ell^{\pm}$ & $1$     & $1$   & $0.32$ & $0.40$  & $0.34$ & $0.4$ & $0.78$ & $1$ & $0.98 $ & $1$ \tabularnewline
0 UT jets        & $0.37$ & $0.22$ & $0.036$ & $0.019$ & $0.020$ & $0.092$ & $0.12 $ & $0.061$ & $0.011 $ & $0.022$ \tabularnewline
2 res. $b$-jets  &  $0.028$  &  $0.0037$  &  $0.0027$  &  $0.003$  &  $0.0016$  &  $0.0061$  &   $0.015 $  &  $0.025$  &  $ 6\cdot10^{-5}$  & $1\cdot 10^{-5}$  \tabularnewline
$\Delta R_{bb}$ & $0.027$ & $0.003$ & $0.0024$ & $0.0003$ & $0.0006$ & $0.0002$ & $0.0035 $ & $0.0034$ & $ 1\cdot 10^{-5}$ & $4\cdot10^{-8}$ \tabularnewline
$H_T$ & $0.027$ & - & $0.0024$ & - & $0.0006$ & - & $0.0035 $ & - & $ 1\cdot 10^{-5}$ & - \tabularnewline
$p_{T,\min}^{b,\mathrm{leading}}$ & $0.027$ & - & $0.0024$ & - & $0.0006$ & - & $0.0035 $ & - & $1\cdot10^{-5} $ & - \tabularnewline
$\Delta \phi (E_T^\mathrm{miss}, h_\mathrm{cand})$ & $0.027$ & $0.0030$ & $0.0024$ & $0.0003$ & $0.0006$ & $0.0002$ & $0.0035 $ & $0.0034$ & $1\cdot 10^{-5} $ & $4\cdot10^{-8}$ \tabularnewline
$\Delta \phi (b_1, b_2)$ & $0.027$ & $0.0026$ & $0.0024$ & $0.0003$ & $0.0006$ & $ 0.0002$ & $0.0035 $ & $0.0029$ & $1\cdot 10^{-5}$ &$4\cdot 10^{-8}$ \tabularnewline
$\Delta \phi (E_T^\mathrm{miss}, b\mathrm{-jets})$ & $0.027$ & $0.0026$ & $0.0024$ & $0.0003$ & $0.0006$ & $0.0002$ & $0.0035 $ & $0.0003$ & $1\cdot 10^{-5}$ & $4\cdot 10^{-8}$ \tabularnewline
$E_{T}^\mathrm{miss}$ & $0.027$ & - & $0.0024$ & - & $0.0006$ & - & $0.0035 $ & - & $1\cdot 10^{-5}$ & - \tabularnewline
$m_{h_\mathrm{cand}}$ & $0.027$ & $0.0026$ & $0.0024$ & $0.0003$ & $ 3\cdot 10^{-5}$ & $2\cdot10^{-5}$ & $ 10^{-4}$ & $0.0001$ & $<10^{-5}$ &$ <4 \cdot 10^{-8}$ \tabularnewline
\bottomrule         
\end{tabular}
}
\caption{Cutflow for the resolved events in the 0-lepton category at the LHC and FCC-hh. 
A dash means that the particular cut was not applied and UT stands for untagged.
}
\label{tab:Cutflow_NuNu_Resolved}
\end{table}

\begin{table}[t]
\begin{centering}
\begin{scriptsize}
\renewcommand{\arraystretch}{1.1}
\begin{tabular}{c|c|c|c|c|c|c}
\toprule
\multicolumn{1}{c|}{\multirow{2}{*}{Cuts / Eff.}} & \multicolumn{2}{c|}{$Wh$} & \multicolumn{2}{c|}{$Wb\bar b$} & \multicolumn{2}{c}{$t\bar t$} \\
& LHC & FCC &  LHC & FCC & LHC & FCC \\
\midrule
$1$ $\ell^{\pm}$                                    & $0.66$ & $0.59$ & $0.59$  & $0.59$  & $0.88$   & $0.87$  \tabularnewline
0 UT jets                                           & $0.25$ & $0.14$ & $0.075$ & $0.031$ & $0.021$  & $0.012$ \tabularnewline
1 MDT DBT jet                                       & $0.18$ & $0.12$ & $0.051$ & $0.021$ & $0.010$  & $0.005$ \tabularnewline
$E_{T}^\mathrm{miss}$                               & $0.16$ &  -     & $0.043$ & -       & $0.0097$ & -       \tabularnewline
$\eta_{\max}^{h_\mathrm{cand}}$                     & $0.14$ & $0.12$ & $0.038$ & $0.021$ & $0.0089$ & $0.005$ \tabularnewline
$\Delta y (W, h_\mathrm{cand})_\mathrm{max}$        & $0.13$ & $0.10$ & $0.030$ & $0.014$ & $0.0072$ & $0.003$ \tabularnewline
$m_{h_\mathrm{cand}}$                               & $0.13$ & $0.10$ & $0.007$ & $0.003$ & $0.0005$ & $0.0001$ \tabularnewline
\bottomrule         
\end{tabular}
\end{scriptsize}
\par\end{centering}
\caption{Cutflow for the boosted events in the 1-lepton category at the LHC and FCC-hh.
A dash means that the particular cut was not applied. UT, MDT and DBT stand for untagged, mass-drop-tagged and doubly-$b$-tagged respectively.}
\label{tab:Cutflow_EllNu_Boosted}
\end{table}

\begin{table}[htb!]
\begin{centering}
\begin{scriptsize}
\renewcommand{\arraystretch}{1.1}
\begin{tabular}{c|c|c|c|c|c|c}
\toprule
\multicolumn{1}{c|}{\multirow{2}{*}{Cuts / Eff.}} & \multicolumn{2}{c|}{$Wh$} & \multicolumn{2}{c|}{$Wb\bar b$} & \multicolumn{2}{c}{$t\bar t$} \\
& LHC & FCC &  LHC & FCC & LHC & FCC \\
\midrule
$1$ $\ell^{\pm}$                                    &           $0.66$      &           $0.59$      &       $0.59$          &           $0.60$       &           $0.88$      &         $0.87$        \tabularnewline
0 UT jets                                           &           $0.25$      &           $0.14$      &       $0.075$         &           $0.031$      &           $0.021$     &      $0.012$          \tabularnewline
2 res. b-jets                                       & $0.025$ & $0.0023$  &  $0.006$  &  $0.002$  &  $0.002$  &  $<10^{-4}$  \tabularnewline
$\Delta R_{bb}$                                     &           $0.025$     &           $0.0019$    &       $0.004$         &        $0.001$        &        $0.0017$       &           $<10^{-4}$     \tabularnewline
$E_{T}^\mathrm{miss}$                               &           $0.024$     &           -           &       $0.003$         &           -           &        $0.0016$       &           -           \tabularnewline
$p_{T,\min}^{b,\mathrm{leading}}$                   &           $0.024$     &           -           &       $0.003$         &           -           &        $0.0016$       &           -           \tabularnewline
$m_{h_\mathrm{cand}}$                               &           $0.024$     &           $0.0019$    &   $7\cdot10^{-5}$     &  $5\cdot10^{-5}$      &        $<5\cdot 10^{-6}$         &           $<10^{-4}$     \tabularnewline
\bottomrule         
\end{tabular}
\end{scriptsize}
\par\end{centering}
\caption{Cutflow for the resolved events in the 1-lepton category at the LHC and FCC-hh.
A dash means that the particular cut was not applied and UT stands for untagged.}
\label{tab:Cutflow_EllNu_Resolved}
\end{table}

\begin{table}[t]
\begin{centering}
\begin{scriptsize}
\renewcommand{\arraystretch}{1.1}
\begin{tabular}{@{}c|c|c|c|c@{}}
\toprule
\multicolumn{1}{c|}{\multirow{2}{*}{Cuts / Eff.}} & \multicolumn{2}{c|}{$Zh$} & \multicolumn{2}{c}{$Zb\bar b$} \\
& LHC & FCC &  LHC & FCC \\
\midrule
$2$ $\ell^{\pm}$                             & $0.48$  & $0.44$ & $0.71$  & $0.62$ \tabularnewline
1 MDT DBT jet                                & $ 0.21$ & $0.30$ & $0.18 $ & $0.23$\tabularnewline
Leptons                                      & $ 0.21$ & -      & $0.18$  & - \tabularnewline
$\eta_{\max}^{h_\mathrm{cand}}$              & $ 0.19$ & $0.30$ & $0.17 $ & $0.23$\tabularnewline
$\Delta y (Z, h_\mathrm{cand})_\mathrm{max}$ & $0.16$  & $0.24$ & $0.10 $ & $0.12$\tabularnewline
$m_{\ell\ell}$                               & $0.15$  & $0.23$ & $0.10$  & $0.12$\tabularnewline
max. $p_{T}^{\ell}$ imbalance                & $0.14$  & $0.15$ & $0.078$ & $0.053$\tabularnewline
$p_{T,\mathrm{min}}^{Z,E_T^\mathrm{miss}}$   & $0.14$  & $0.15$ & $0.078$ & $0.053$\tabularnewline
$p_{T,\mathrm{min}}^Z$                       & $0.14$  & -      & $0.074$ & - \tabularnewline
$m_{h_\mathrm{cand}}$                        & $0.14 $ & $0.15$ & $0.020$ & $0.011$\tabularnewline
\bottomrule         
\end{tabular}
\hfill
\begin{tabular}{@{}c|c|c|c|c@{}}
\toprule
\multicolumn{1}{c|}{\multirow{2}{*}{Cuts / Eff.}} & \multicolumn{2}{c|}{$Zh$} & \multicolumn{2}{c}{$Zb\bar b$} \\
& LHC & FCC &  LHC & FCC \\
\midrule
$2$ $\ell^{\pm}$                  & $0.48$                 &                  $0.44$ &                  $0.71$  &                 $0.62$ \tabularnewline
2 res. $b$-jets                     & $0.061$ & $0.011$  &  $0.057$  &  $0.10$  \tabularnewline
$\Delta R_{bb}$                   & $0.052 $               &                 $0.008$ & $ 0.014$                 & $0.030$                  \tabularnewline
0 UT jets                         & $0.033 $               & $0.0019$                & $0.0051 $                & $0.0081$ \tabularnewline
Leptons                           & $0.033 $               & -                       & $0.0051 $                & - \tabularnewline
$p_{T,\min}^{b,\mathrm{leading}}$ & $0.033 $               & $0.0019$                & $0.0051$                 & $0.0081$ \tabularnewline
$m_{\ell\ell}$                    & $0.031 $               & $0.0017$                & $0.0047 $                & $0.0076$ \tabularnewline
$m_{h_\mathrm{cand}}$             & $ 0.031$               & $0.0017$                & $0.0002 $               & $0.0016$ \tabularnewline
\bottomrule         
\end{tabular}
\end{scriptsize}
\par\end{centering}
\caption{Cutflow for the boosted (left) and resolved (right) events in the 2-lepton category at LHC and FCC-hh.
A dash means that the particular cut was not used in the analysis for that collider. MDT, DBT and UT stand for mass-drop-tagged, doubly-$b$-tagged, and untagged respectively.}
\label{tab:Cutflow_EllEll_Boosted}
\end{table}

For the cutflows at LHC (and HL-LHC), we only took into account events where the transverse momentum of the reconstructed vector boson satisfies $p_T^V > 200\,\mathrm{GeV}$, whereas the cutflows at FCC-hh are restricted to $p_T^V > 400\,\mathrm{GeV}$. With this choice, the cutflows reflect the behaviour of high-energy events, which provide the highest sensitivity to New Physics effects.
We stress that the classification of an event as boosted or resolved is mutually exclusive.

\section{Full results}
\label{sec:AppFullResults}

\begin{table}[t]
\begin{centering}
\begin{footnotesize}
\begin{tabular}{c|c|c}
\toprule
Coefficient & Profiled Fit & One-Operator Fit \tabularnewline
\midrule
$c_{\varphi q}^{(3)}\,$[TeV$^{-2}$] &
\begin{tabular}{ll}
\rule{0pt}{1.25em}$[-9.2,\,4.4]\times10^{-2}$ & $1\%$ syst.\\
\rule{0pt}{1.25em}$[-11.1,\,4.6]\times10^{-2}$ & $5\%$ syst.\\
\rule[-.65em]{0pt}{1.9em}$[-14.5,\,4.9]\times10^{-2}$ & $10\%$ syst.
\end{tabular}
&
\begin{tabular}{ll}
\rule{0pt}{1.25em}$[-5.9,\,4.0]\times10^{-2}$ & $1\%$ syst.\\
\rule{0pt}{1.25em}$[-6.8,\,4.3]\times10^{-2}$ & $5\%$ syst.\\
\rule[-.65em]{0pt}{1.9em}$[-8.3,\,4.6]\times10^{-2}$ & $10\%$ syst.
\end{tabular}
\tabularnewline
\hline
$c_{\varphi q}^{(1)}\,$[TeV$^{-2}$] &
\begin{tabular}{ll}
\rule{0pt}{1.25em}$[-1.4,\,1.4]\times10^{-1}$ & $1\%$ syst.\\
\rule{0pt}{1.25em}$[-1.5,\,1.4]\times10^{-1}$ & $5\%$ syst.\\
\rule[-.65em]{0pt}{1.9em}$[-1.5,\,1.5]\times10^{-1}$ & $10\%$ syst.
\end{tabular}
&
\begin{tabular}{ll}
\rule{0pt}{1.25em}$[-1.1,\,1.2]\times10^{-1}$ & $1\%$ syst.\\
\rule{0pt}{1.25em}$[-1.1,\,1.2]\times10^{-1}$ & $5\%$ syst.\\
\rule[-.65em]{0pt}{1.9em}$[-1.1,\,1.3]\times10^{-1}$ & $10\%$ syst.
\end{tabular}
\tabularnewline
\hline
$c_{\varphi u}\,$[TeV$^{-2}$] &
\begin{tabular}{ll}
\rule{0pt}{1.25em}$[-2.1,\,1.4]\times10^{-1}$ & $1\%$ syst.\\
\rule{0pt}{1.25em}$[-2.1,\,1.4]\times10^{-1}$ & $5\%$ syst.\\
\rule[-.65em]{0pt}{1.9em}$[-2.2,\,1.5]\times10^{-1}$ & $10\%$ syst.
\end{tabular}
&
\begin{tabular}{ll}
\rule{0pt}{1.25em}$[-1.9,\,1.1]\times10^{-1}$ & $1\%$ syst.\\
\rule{0pt}{1.25em}$[-1.9,\,1.1]\times10^{-1}$ & $5\%$ syst.\\
\rule[-.65em]{0pt}{1.9em}$[-2.0,\,1.2]\times10^{-1}$ & $10\%$ syst.\\
\end{tabular}
\tabularnewline

\hline
$c_{\varphi d}\,$[TeV$^{-2}$] &
\begin{tabular}{ll}
\rule{0pt}{1.25em}$[-2.0,\,2.6]\times10^{-1}$ & $1\%$ syst.\\
\rule{0pt}{1.25em}$[-2.1,\,2.4]\times10^{-1}$ & $5\%$ syst.\\
\rule[-.65em]{0pt}{1.9em}$[-2.2,\,2.5]\times10^{-1}$ & $10\%$ syst.
\end{tabular}
&
\begin{tabular}{ll}
\rule{0pt}{1.25em}$[-1.6,\,2.0]\times10^{-1}$ & $1\%$ syst.\\
\rule{0pt}{1.25em}$[-1.6,\,2.0]\times10^{-1}$ & $5\%$ syst.\\
\rule[-.65em]{0pt}{1.9em}$[-1.7,\,2.1]\times10^{-1}$ & $10\%$ syst.\\
\end{tabular} \\
\bottomrule
\end{tabular}
\end{footnotesize}
\par\end{centering}
\caption[caption]{ Bounds at $95\%$ C.L.~on the coefficients of the $\Ohqt$, $\Ohq$, $\Ohu$ and $\Ohd$ operators for 13 TeV LHC with integrated luminosity of $139\,\mathrm{fb}^{-1}$.
{\bf Left column:} Global fit, profiled over the other coefficients. {\bf Right column:} One-operator fit (i.e. setting the other coefficients to zero).
}
\label{tab:bounds_summary}
\end{table}

\begin{table}[t]
\begin{centering}
\begin{footnotesize}
\begin{tabular}{c|c|c}
\toprule
Coefficient & Profiled Fit & One-Operator Fit \tabularnewline
\midrule
$c_{\varphi q}^{(3)}\,$[TeV$^{-2}$] &
\begin{tabular}{ll}
\rule{0pt}{1.25em}$[-8.5,\,4.8]\times10^{-2}$ & $1\%$ syst.\\
\rule{0pt}{1.25em}$[-11.7,\,5.2]\times10^{-2}$ & $5\%$ syst.\\
\rule[-.65em]{0pt}{1.9em}$[-18.1,\,5.9]\times10^{-2}$ & $10\%$ syst.
\end{tabular}
&
\begin{tabular}{ll}
\rule{0pt}{1.25em}$[-6.7,\,4.5]\times10^{-2}$ & $1\%$ syst.\\
\rule{0pt}{1.25em}$[-8.5,\,5.0]\times10^{-2}$ & $5\%$ syst.\\
\rule[-.65em]{0pt}{1.9em}$[-14.4,\,5.7]\times10^{-2}$ & $10\%$ syst.
\end{tabular}
\tabularnewline

\hline
$c_{\varphi q}^{(1)}\,$[TeV$^{-2}$] &
\begin{tabular}{ll}
\rule{0pt}{1.25em}$[-1.5,\,1.5]\times10^{-1}$ & $1\%$ syst.\\
\rule{0pt}{1.25em}$[-1.6,\,1.6]\times10^{-1}$ & $5\%$ syst.\\
\rule[-.65em]{0pt}{1.9em}$[-1.7,\,1.6]\times10^{-1}$ & $10\%$ syst.
\end{tabular}
&
\begin{tabular}{ll}
\rule{0pt}{1.25em}$[-1.2,\,1.4]\times10^{-1}$ & $1\%$ syst.\\
\rule{0pt}{1.25em}$[-1.2,\,1.4]\times10^{-1}$ & $5\%$ syst.\\
\rule[-.65em]{0pt}{1.9em}$[-1.3,\,1.5]\times10^{-1}$ & $10\%$ syst.
\end{tabular}
\tabularnewline

\hline
$c_{\varphi u}\,$[TeV$^{-2}$] &
\begin{tabular}{ll}
\rule{0pt}{1.25em}$[-2.3,\,1.6]\times10^{-1}$ & $1\%$ syst.\\
\rule{0pt}{1.25em}$[-2.4,\,1.7]\times10^{-1}$ & $5\%$ syst.\\
\rule[-.65em]{0pt}{1.9em}$[-2.5,\,1.8]\times10^{-1}$ & $10\%$ syst.
\end{tabular}
&
\begin{tabular}{ll}
\rule{0pt}{1.25em}$[-2.1,\,1.3]\times10^{-1}$ & $1\%$ syst.\\
\rule{0pt}{1.25em}$[-2.2,\,1.4]\times10^{-1}$ & $5\%$ syst.\\
\rule[-.65em]{0pt}{1.9em}$[-2.2,\,1.5]\times10^{-1}$ & $10\%$ syst.\\
\end{tabular}
\tabularnewline

\hline
$c_{\varphi d}\,$[TeV$^{-2}$] &
\begin{tabular}{ll}
\rule{0pt}{1.25em}$[-2.2,\,2.6]\times10^{-1}$ & $1\%$ syst.\\
\rule{0pt}{1.25em}$[-2.3,\,2.7]\times10^{-1}$ & $5\%$ syst.\\
\rule[-.65em]{0pt}{1.9em}$[-2.4,\,2.8]\times10^{-1}$ & $10\%$ syst.
\end{tabular}
&
\begin{tabular}{ll}
\rule{0pt}{1.25em}$[-1.9,\,2.3]\times10^{-1}$ & $1\%$ syst.\\
\rule{0pt}{1.25em}$[-1.9,\,2.3]\times10^{-1}$ & $5\%$ syst.\\
\rule[-.65em]{0pt}{1.9em}$[-2.0,\,2.4]\times10^{-1}$ & $10\%$ syst.\\
\end{tabular} \\
\bottomrule
\end{tabular}
\end{footnotesize}
\par\end{centering}
\caption[caption]{Alternative bounds at $95\%$ C.L.~on the coefficients of the $\Ohqt$, $\Ohq$, $\Ohu$ and $\Ohd$ operators for 13 TeV LHC with integrated luminosity of $139\,\mathrm{fb}^{-1}$. Background and $Wh$-signal cross-sections have been rescaled to match the ATLAS $m_{b\bar b}$ distributions in the Higgs window~\cite{ATLAS:2020fcp,ATLAS:2020jwz}.
}
\label{tab:alternative_bounds_summary}
\end{table}

In this appendix, we collect our projected bounds for different collider settings. Table~\ref{tab:bounds_summary} presents our $95\%$ C.L. bounds on $\chqt$, $\chq$, $\chu$ and $\chd$ for a collider with a c.o.m. energy of $13$~TeV and $139$ fb$^{-1}$ of integrated luminosity, i.e. like LHC Run 2. The middle column presents the bounds obtained after profiling a fit on the 4 Wilson coefficients, while the last column presents the results from a one-operator fit. The bounds are presented for 3 different assumptions on the size of systematic uncertainty: $1\%$, $5\%$ and $10\%$.

\begin{table}[t]
\begin{centering}
\begin{footnotesize}
\begin{tabular}{c|c|c}
\toprule
Coefficient & Profiled Fit & One-Operator Fit \tabularnewline
\midrule
$c_{\varphi q}^{(3)}\,$[TeV$^{-2}$] &
\begin{tabular}{ll}
\rule{0pt}{1.25em}$[-5.9,\,3.2]\times10^{-2}$ & $1\%$ syst.\\
\rule{0pt}{1.25em}$[-7.9,\,3.5]\times10^{-2}$ & $5\%$ syst.\\
\rule[-.65em]{0pt}{1.9em}$[-10.6,\,4.0]\times10^{-2}$ & $10\%$ syst.
\end{tabular}
&
\begin{tabular}{ll}
\rule{0pt}{1.25em}$[-3.7,\,2.9]\times10^{-2}$ & $1\%$ syst.\\
\rule{0pt}{1.25em}$[-4.3,\,3.2]\times10^{-2}$ & $5\%$ syst.\\
\rule[-.65em]{0pt}{1.9em}$[-5.4,\,3.6]\times10^{-2}$ & $10\%$ syst.
\end{tabular}
\tabularnewline

\hline
$c_{\varphi q}^{(1)}\,$[TeV$^{-2}$] &
\begin{tabular}{ll}
\rule{0pt}{1.25em}$[-1.2,\,1.2]\times10^{-1}$ & $1\%$ syst.\\
\rule{0pt}{1.25em}$[-1.2,\,1.3]\times10^{-1}$ & $5\%$ syst.\\
\rule[-.65em]{0pt}{1.9em}$[-1.4,\,1.3]\times10^{-1}$ & $10\%$ syst.
\end{tabular}
&
\begin{tabular}{ll}
\rule{0pt}{1.25em}$[-8.5,\,10.4]\times10^{-2}$ & $1\%$ syst.\\
\rule{0pt}{1.25em}$[-8.7,\,10.6]\times10^{-2}$ & $5\%$ syst.\\
\rule[-.65em]{0pt}{1.9em}$[-9.3,\,11.1]\times10^{-2}$ & $10\%$ syst.
\end{tabular}
\tabularnewline

\hline
$c_{\varphi u}\,$[TeV$^{-2}$] &
\begin{tabular}{ll}
\rule{0pt}{1.25em}$[-1.8,\,1.2]\times10^{-1}$ & $1\%$ syst.\\
\rule{0pt}{1.25em}$[-1.9,\,1.2]\times10^{-1}$ & $5\%$ syst.\\
\rule[-.65em]{0pt}{1.9em}$[-2.0,\,1.4]\times10^{-1}$ & $10\%$ syst.
\end{tabular}
&
\begin{tabular}{ll}
\rule{0pt}{1.25em}$[-16.6,\,8.7]\times10^{-2}$ & $1\%$ syst.\\
\rule{0pt}{1.25em}$[-16.8,\,9.0]\times10^{-2}$ & $5\%$ syst.\\
\rule[-.65em]{0pt}{1.9em}$[-17.5,\,9.7]\times10^{-2}$ & $10\%$ syst.\\
\end{tabular}
\tabularnewline

\hline
$c_{\varphi d}\,$[TeV$^{-2}$] &
\begin{tabular}{ll}
\rule{0pt}{1.25em}$[-1.7,\,2.0]\times10^{-1}$ & $1\%$ syst.\\
\rule{0pt}{1.25em}$[-1.8,\,2.1]\times10^{-1}$ & $5\%$ syst.\\
\rule[-.65em]{0pt}{1.9em}$[-1.9,\,2.2]\times10^{-1}$ & $10\%$ syst.
\end{tabular}
&
\begin{tabular}{ll}
\rule{0pt}{1.25em}$[-1.3,\,1.7]\times10^{-1}$ & $1\%$ syst.\\
\rule{0pt}{1.25em}$[-1.3,\,1.7]\times10^{-1}$ & $5\%$ syst.\\
\rule[-.65em]{0pt}{1.9em}$[-1.4,\,1.8]\times10^{-1}$ & $10\%$ syst.\\
\end{tabular} \\
\bottomrule

\end{tabular}
\end{footnotesize}
\par\end{centering}
\caption[caption]{ Bounds at $95\%$ C.L.~on the coefficients of the $\Ohqt$, $\Ohq$, $\Ohu$ and $\Ohd$ operators for 13 TeV LHC with integrated luminosity of $300\,\mathrm{fb}^{-1}$.
}
\label{tab:bounds_summary_300fb-1}
\end{table}

\begin{table}[t]
\begin{centering}
\begin{footnotesize}
\begin{tabular}{c|c|c}
\toprule
Coefficient & Profiled Fit & One-Operator Fit \tabularnewline
\midrule
$c_{\varphi q}^{(3)}\,$[TeV$^{-2}$] &
\begin{tabular}{ll}
\rule{0pt}{1.25em}$[-5.5,\,3.4]\times10^{-2}$ & $1\%$ syst.\\
\rule{0pt}{1.25em}$[-8.2,\,4.0]\times10^{-2}$ & $5\%$ syst.\\
\rule[-.65em]{0pt}{1.9em}$[-12.7,\,4.9]\times10^{-2}$ & $10\%$ syst.
\end{tabular}
&
\begin{tabular}{ll}
\rule{0pt}{1.25em}$[-4.2,\,3.2]\times10^{-2}$ & $1\%$ syst.\\
\rule{0pt}{1.25em}$[-5.5,\,3.8]\times10^{-2}$ & $5\%$ syst.\\
\rule[-.65em]{0pt}{1.9em}$[-8.1,\,4.6]\times10^{-2}$ & $10\%$ syst.
\end{tabular}
\tabularnewline

\hline
$c_{\varphi q}^{(1)}\,$[TeV$^{-2}$] &
\begin{tabular}{ll}
\rule{0pt}{1.25em}$[-1.3,\,1.3]\times10^{-1}$ & $1\%$ syst.\\
\rule{0pt}{1.25em}$[-1.4,\,1.4]\times10^{-1}$ & $5\%$ syst.\\
\rule[-.65em]{0pt}{1.9em}$[-1.5,\,1.5]\times10^{-1}$ & $10\%$ syst.
\end{tabular}
&
\begin{tabular}{ll}
\rule{0pt}{1.25em}$[-1.0,\,1.2]\times10^{-1}$ & $1\%$ syst.\\
\rule{0pt}{1.25em}$[-1.0,\,1.2]\times10^{-1}$ & $5\%$ syst.\\
\rule[-.65em]{0pt}{1.9em}$[-1.1,\,1.3]\times10^{-1}$ & $10\%$ syst.
\end{tabular}
\tabularnewline

\hline
$c_{\varphi u}\,$[TeV$^{-2}$] &
\begin{tabular}{ll}
\rule{0pt}{1.25em}$[-2.0,\,1.3]\times10^{-1}$ & $1\%$ syst.\\
\rule{0pt}{1.25em}$[-2.1,\,1.4]\times10^{-1}$ & $5\%$ syst.\\
\rule[-.65em]{0pt}{1.9em}$[-2.2,\,1.5]\times10^{-1}$ & $10\%$ syst.
\end{tabular}
&
\begin{tabular}{ll}
\rule{0pt}{1.25em}$[-1.8,\,1.1]\times10^{-1}$ & $1\%$ syst.\\
\rule{0pt}{1.25em}$[-1.9,\,1.1]\times10^{-1}$ & $5\%$ syst.\\
\rule[-.65em]{0pt}{1.9em}$[-2.0,\,1.2]\times10^{-1}$ & $10\%$ syst.\\
\end{tabular}
\tabularnewline

\hline
$c_{\varphi d}\,$[TeV$^{-2}$] &
\begin{tabular}{ll}
\rule{0pt}{1.25em}$[-1.8,\,2.2]\times10^{-1}$ & $1\%$ syst.\\
\rule{0pt}{1.25em}$[-1.9,\,2.3]\times10^{-1}$ & $5\%$ syst.\\
\rule[-.65em]{0pt}{1.9em}$[-2.1,\,2.5]\times10^{-1}$ & $10\%$ syst.
\end{tabular}
&
\begin{tabular}{ll}
\rule{0pt}{1.25em}$[-1.5,\,1.9]\times10^{-1}$ & $1\%$ syst.\\
\rule{0pt}{1.25em}$[-1.6,\,2.0]\times10^{-1}$ & $5\%$ syst.\\
\rule[-.65em]{0pt}{1.9em}$[-1.8,\,2.2]\times10^{-1}$ & $10\%$ syst.\\
\end{tabular} \\
\bottomrule
\end{tabular}
\end{footnotesize}
\par\end{centering}
\caption[caption]{ Alternative bounds at $95\%$ C.L.~on the coefficients of the $\Ohqt$, $\Ohq$, $\Ohu$ and $\Ohd$ operators for 13 TeV LHC with integrated luminosity of $300\,\mathrm{fb}^{-1}$. Background and $Wh$-signal cross-sections have been rescaled to match the ATLAS $m_{b\bar b}$ distributions in the Higgs window.
}
\label{tab:bounds_summary_300fb-1_alternative}
\end{table}

Table~\ref{tab:alternative_bounds_summary} shows the bounds on the WCs with the same energy and luminosity than the previous table, but in this case the $Wh$ and background cross-sections were rescaled to match the ones reported by the ATLAS collaboration in refs.~\cite{ATLAS:2020fcp,ATLAS:2020jwz}.

\begin{table}[t]
\begin{centering}
\begin{footnotesize}
\begin{tabular}{c|c|c}
\toprule
Coefficient & Profiled Fit & One-Operator Fit \tabularnewline
\midrule
$c_{\varphi q}^{(3)}\,$[TeV$^{-2}$] &
\begin{tabular}{ll}
\rule{0pt}{1.25em}$[-2.1,\,1.4]\times10^{-2}$ & $1\%$ syst.\\
\rule{0pt}{1.25em}$[-3.9,\,1.9]\times10^{-2}$ & $5\%$ syst.\\
\rule[-.65em]{0pt}{1.9em}$[-6.7,\,2.7]\times10^{-2}$ & $10\%$ syst.
\end{tabular}
&
\begin{tabular}{ll}
\rule{0pt}{1.25em}$[-1.1,\,1.0]\times10^{-2}$ & $1\%$ syst.\\
\rule{0pt}{1.25em}$[-1.7,\,1.5]\times10^{-2}$ & $5\%$ syst.\\
\rule[-.65em]{0pt}{1.9em}$[-2.9,\,2.3]\times10^{-2}$ & $10\%$ syst.
\end{tabular}
\tabularnewline

\hline
$c_{\varphi q}^{(1)}\,$[TeV$^{-2}$] &
\begin{tabular}{ll}
\rule{0pt}{1.25em}$[-5.8,\,6.9]\times10^{-2}$ & $1\%$ syst.\\
\rule{0pt}{1.25em}$[-7.9,\,8.7]\times10^{-2}$ & $5\%$ syst.\\
\rule[-.65em]{0pt}{1.9em}$[-10.4,\,10.8]\times10^{-2}$ & $10\%$ syst.
\end{tabular}
&
\begin{tabular}{ll}
\rule{0pt}{1.25em}$[-4.5,\,6.3]\times10^{-2}$ & $1\%$ syst.\\
\rule{0pt}{1.25em}$[-5.4,\,7.2]\times10^{-2}$ & $5\%$ syst.\\
\rule[-.65em]{0pt}{1.9em}$[-6.9,\,8.7]\times10^{-2}$ & $10\%$ syst.
\end{tabular}
\tabularnewline

\hline
$c_{\varphi u}\,$[TeV$^{-2}$] &
\begin{tabular}{ll}
\rule{0pt}{1.25em}$[-11.5,\,5.9]\times10^{-2}$ & $1\%$ syst.\\
\rule{0pt}{1.25em}$[-13.5,\,8.2]\times10^{-2}$ & $5\%$ syst.\\
\rule[-.65em]{0pt}{1.9em}$[-16.1,\,10.7]\times10^{-2}$ & $10\%$ syst.
\end{tabular}
&
\begin{tabular}{ll}
\rule{0pt}{1.25em}$[-11.1,\,3.9]\times10^{-2}$ & $1\%$ syst.\\
\rule{0pt}{1.25em}$[-12.4,\,4.9]\times10^{-2}$ & $5\%$ syst.\\
\rule[-.65em]{0pt}{1.9em}$[-14.4,\,6.8]\times10^{-2}$ & $10\%$ syst.\\
\end{tabular}
\tabularnewline

\hline
$c_{\varphi d}\,$[TeV$^{-2}$] &
\begin{tabular}{ll}
\rule{0pt}{1.25em}$[-1.0,\,1.2]\times10^{-1}$ & $1\%$ syst.\\
\rule{0pt}{1.25em}$[-1.3,\,1.5]\times10^{-1}$ & $5\%$ syst.\\
\rule[-.65em]{0pt}{1.9em}$[-1.6,\,1.8]\times10^{-1}$ & $10\%$ syst.
\end{tabular}
&
\begin{tabular}{ll}
\rule{0pt}{1.25em}$[-6.6,\,10.6]\times10^{-2}$ & $1\%$ syst.\\
\rule{0pt}{1.25em}$[-8.0,\,12.0]\times10^{-2}$ & $5\%$ syst.\\
\rule[-.65em]{0pt}{1.9em}$[-10.4,\,14.4]\times10^{-2}$ & $10\%$ syst.\\
\end{tabular} \\
\bottomrule

\end{tabular}
\end{footnotesize}
\par\end{centering}
\caption[caption]{ Bounds at $95\%$ C.L.~on the coefficients of the $\Ohqt$, $\Ohq$, $\Ohu$ and $\Ohd$ operators for 14 TeV HL-LHC with integrated luminosity of $3\,\mathrm{ab}^{-1}$.
}
\label{tab:bounds_summary_HL_LHC}
\end{table}

\begin{table}[t]
\begin{centering}
\begin{footnotesize}
\begin{tabular}{c|c|c}
\toprule
Coefficient & Profiled Fit & One-Operator Fit \tabularnewline
\midrule
$c_{\varphi q}^{(3)}\,$[TeV$^{-2}$] &
\begin{tabular}{ll}
\rule{0pt}{1.25em}$[-2.0,\,1.5]\times10^{-2}$ & $1\%$ syst.\\
\rule{0pt}{1.25em}$[-4.4,\,2.4]\times10^{-2}$ & $5\%$ syst.\\
\rule[-.65em]{0pt}{1.9em}$[-8.6,\,3.5]\times10^{-2}$ & $10\%$ syst.
\end{tabular}
&
\begin{tabular}{ll}
\rule{0pt}{1.25em}$[-1.3,\,1.2]\times10^{-2}$ & $1\%$ syst.\\
\rule{0pt}{1.25em}$[-2.5,\,2.1]\times10^{-2}$ & $5\%$ syst.\\
\rule[-.65em]{0pt}{1.9em}$[-4.8,\,3.4]\times10^{-2}$ & $10\%$ syst.
\end{tabular}
\tabularnewline

\hline
$c_{\varphi q}^{(1)}\,$[TeV$^{-2}$] &
\begin{tabular}{ll}
\rule{0pt}{1.25em}$[-7.3,\,8.5]\times10^{-2}$ & $1\%$ syst.\\
\rule{0pt}{1.25em}$[-9.5,\,10.4]\times10^{-2}$ & $5\%$ syst.\\
\rule[-.65em]{0pt}{1.9em}$[-12.7,\,12.6]\times10^{-2}$ & $10\%$ syst.
\end{tabular}
&
\begin{tabular}{ll}
\rule{0pt}{1.25em}$[-5.3,\,7.2]\times10^{-2}$ & $1\%$ syst.\\
\rule{0pt}{1.25em}$[-6.8,\,8.7]\times10^{-2}$ & $5\%$ syst.\\
\rule[-.65em]{0pt}{1.9em}$[-9.1,\,11.0]\times10^{-2}$ & $10\%$ syst.
\end{tabular}
\tabularnewline

\hline
$c_{\varphi u}\,$[TeV$^{-2}$] &
\begin{tabular}{ll}
\rule{0pt}{1.25em}$[-12.8,\,6.3]\times10^{-2}$ & $1\%$ syst.\\
\rule{0pt}{1.25em}$[-15.6,\,9.2]\times10^{-2}$ & $5\%$ syst.\\
\rule[-.65em]{0pt}{1.9em}$[-19.0,\,12.6]\times10^{-2}$ & $10\%$ syst.\\
\end{tabular}
&
\begin{tabular}{ll}
\rule{0pt}{1.25em}$[-12.4,\,4.9]\times10^{-2}$ & $1\%$ syst.\\
\rule{0pt}{1.25em}$[-14.4,\,6.7]\times10^{-2}$ & $5\%$ syst.\\
\rule[-.65em]{0pt}{1.9em}$[-17.3,\,9.5]\times10^{-2}$ & $10\%$ syst.\\
\end{tabular}
\tabularnewline

\hline
$c_{\varphi d}\,$[TeV$^{-2}$] &
\begin{tabular}{ll}
\rule{0pt}{1.25em}$[-1.0,\,1.4]\times10^{-1}$ & $1\%$ syst.\\
\rule{0pt}{1.25em}$[-1.4,\,1.7]\times10^{-1}$ & $5\%$ syst.\\
\rule[-.65em]{0pt}{1.9em}$[-1.8,\,2.1]\times10^{-1}$ & $10\%$ syst.
\end{tabular}
&
\begin{tabular}{ll}
\rule{0pt}{1.25em}$[-7.9,\,12.0]\times10^{-2}$ & $1\%$ syst.\\
\rule{0pt}{1.25em}$[-10.4,\,14.4]\times10^{-2}$ & $5\%$ syst.\\
\rule[-.65em]{0pt}{1.9em}$[-14.0,\,18.0]\times10^{-2}$ & $10\%$ syst.\\
\end{tabular} \\
\bottomrule

\end{tabular}
\end{footnotesize}
\par\end{centering}
\caption[caption]{Alternative bounds at $95\%$ C.L.~on the coefficients of the $\Ohqt$, $\Ohq$, $\Ohu$ and $\Ohd$ operators for 14 TeV HL-LHC with integrated luminosity of $3\,\mathrm{ab}^{-1}$. Background and $Wh$-signal cross-sections have been rescaled to match the ATLAS $m_{b\bar b}$ distributions in the Higgs window.
}
\label{tab:alternative_bounds_summary_HL_LHC}
\end{table}

We present our projected bounds also for 13 TeV LHC, with the full Run 3 integrated luminosity, $300$~fb$^{-1}$, and for HL-LHC in tables~\ref{tab:bounds_summary_300fb-1} and~\ref{tab:bounds_summary_HL_LHC}.
Finally, as we did for the LHC Run 2, in tables~\ref{tab:bounds_summary_300fb-1_alternative} and~\ref{tab:alternative_bounds_summary_HL_LHC}, we also give the bounds obtained including the rescaling of the signal and background cross sections to match the distributions.

\section{Fitting the $\mathbf{t\bar t}$ background at FCC-hh}\label{sec:fit_tt}

\begin{figure}[t]\centering
	\includegraphics[scale=1]{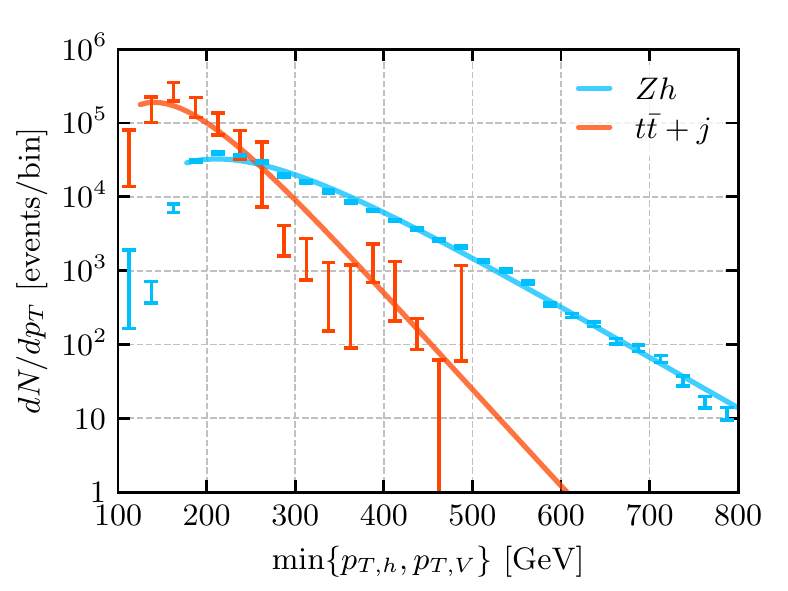}
	\caption{The $p_T$ spectrum for the signal and the $t\bar{t}+j$ background at FCC-hh along with the corresponding fits of the functional form in eq.~\eqref{eq:fit_function} with the best fit parameters given in eq.~\eqref{eq:best_fit_parameters}.}
	\label{fig:min_pt_fit}
\end{figure}

To improve the modelling to the $t\bar{t}$ background where few Monte Carlo events survive the analysis cuts, we perform a fit of the $\min\{p_{T,h},p_{T,V}\}$ spectrum.
The functional form
\begin{equation}
\frac{dN}{dx} = \left(1-x^{1/3}\right)^b\,x^{a_0+a_1\,\log x}\,,
\label{eq:fit_function}
\end{equation}
which was taken from ref.~\cite{ATLAS:2016gzy}, models the spectrum well beyond the peak as shown in fig.~\ref{fig:min_pt_fit}.
Here, $N$ is the number of events and $x=p_T/\rho$ is the normalized $p_T$ and $\rho$ is an arbitrary mass scale. In principle, one should normalize by the center-of-mass energy, i.e. take $\rho=\sqrt{s}$, but numerically it is better (and sufficient for our purpose) to set it to a smaller value, $\rho=3$ TeV.
Note that taking the $\log$ of both sides of eq.~\eqref{eq:fit_function} makes the fit function linear in the unknown coefficients, $\{a_0,a_1,b\}$, which results in a better fit.
The best fit parameters for the $Zh$ (signal) and $t\bar{t}+j$ (background) are given by
\begingroup
\setlength{\tabcolsep}{12pt}
\begin{equation}
\begin{tabular}{rccc}\toprule[1pt]
 & $a_0$ & $a_1$ & $b$ \\\midrule[0.5pt]
$Zh$ & -14.70 & -3.248 & 10.78\\
$t\bar{t}+j$ &-19.34 & -3.888 & 23.72 \\\bottomrule[1pt]
\end{tabular}
\label{eq:best_fit_parameters}
\end{equation}
\endgroup
 
\section{Signal and background cross-sections in $Vh$ diboson processes}
\label{app:EvtNumbersVH}

In this appendix, we collect the tables reporting the number of signal and backgorund events after the selection cuts per bin for each $pp\to Vh$ channel. We present the results for HL-LHC ($14$~TeV and $3$~ab$^{-1}$) and  FCC-hh  ($100$~TeV and  $30$~ab$^{-1}$).
The number of signal events is given as a quadratic function of the Wilson coefficients studied in each channel. The SM value of the number of signal events agrees with figs.~\ref{fig:histograms_vv}-\ref{fig:histograms_ll_FCC} and summing over the background contributions in the figures one obtains the numbers in these tables. 

\subsection{The 0-lepton channel}
\label{app:EvtNumbers_Zh_vv}

\begin{table}[t]
	\centering
		\begin{scriptsize}
			\begin{tabular}{|c|c|c|c@{\hspace{.25em}}|}
				\hline
				\multicolumn{3}{|c|}{0-lepton channel, resolved, HL-LHC} \tabularnewline
				\hline
				\multirow{2}{*}{
				\hspace{-2.5em}
				\begin{tabular}{c}
				$p_{T,\mathrm{min}}$ bin\\
				$[$GeV$]$
				\end{tabular}
				\hspace{-2.5em}} & \multicolumn{2}{c|}{Number of expected events}\tabularnewline
				\cline{2-3} &  \rule{0pt}{1.15em}Signal & Background \tabularnewline
				\hline
				$[0-160]$ & 
				$\begin{aligned} \phSpa 300 \,& + 1120 \,\chqt
					+ (-39 \pm 47)\,\chq + (155 \pm 39 ) \,\chu - (80\pm30) \,\chd
					 + 1250 \,\left(\chqt\right)^{2}\\
					\rule[-1.em]{0pt}{1em}& + 1010 \,\left(\chq\right)^{2}
					 + (550 \pm 75)\,\left(\chu\right)^{2} +  (400\pm47)\,\left(\chd\right)^{2} - (190 \pm 250)\,\chqt\,\chq
				\end{aligned}$ & $8000\pm3700$ \tabularnewline \hline
	            $[160-200]$ & $\begin{aligned} \phSpa 708 \,& + 3230 \,\chqt
					+ (-160 \pm 60)\,\chq + ( 596 \pm 49 ) \,\chu - ( 263 \pm 39 ) \,\chd + 4460 \,\left(\chqt\right)^{2} \\
					\rule[-1.em]{0pt}{1em}& + 3340 \,\left(\chq\right)^{2} + 1920 \,\left(\chu\right)^{2} +  1510 \,\left(\chd\right)^{2} - (1400 \pm 340)\,\chqt\,\chq
				\end{aligned}$ & $5350\pm1400$ \tabularnewline
				\hline 
				 $[200-250]$ &  $\begin{aligned} \phSpa 195 \,& + 1160 \,\chqt
					- (55\pm22)\,\chq + (223\pm15) \,\chu - (90\pm13) \,\chd + 2075 \,\left(\chqt\right)^{2}\\
					\rule[-1.em]{0pt}{1em}& + 1750 \,\left(\chq\right)^{2} + 955 \,\left(\chu\right)^{2} +  698\,\left(\chd\right)^{2} - (430 \pm 150)\,\chqt\,\chq
				\end{aligned}$ & $1310\pm90$ \tabularnewline
				\hline 
				 $[250-\infty]$ &  $\begin{aligned} \phSpa 33 \,& + 312 \,\chqt
					+ (-32\pm10)\,\chq + (66\pm7) \,\chu - (26\pm6) \,\chd + 1020 \,\left(\chqt\right)^{2}\\
					\rule[-1.em]{0pt}{1em}& + 907 \,\left(\chq\right)^{2} + 517\,\left(\chu\right)^{2} +  351\,\left(\chd\right)^{2} - (360 \pm 85)\,\chqt\,\chq
				\end{aligned}$ & $265\pm37$ \tabularnewline
				\hline 
\end{tabular}
\end{scriptsize}
	\caption{Number of expected signal events as a function of the Wilson coefficients (in units of TeV$^{-2}$) and of total background events in the $Zh \rightarrow \nu \bar{\nu} b\bar{b}$ channel, resolved category, at HL-LHC.
	The Monte Carlo errors on the fitted coefficients, when not explicitly specified, are $\lesssim \textit{few}$ \%.
	}
	\label{tab:App_sigma_full_Zh_neut_HL_LHC_res}
\end{table}
\begin{table}[t]
	\centering
			\begin{scriptsize}
			\begin{tabular}{|c|c|c|c@{\hspace{.25em}}|}
				\hline
				\multicolumn{3}{|c|}{0-lepton channel, boosted, HL-LHC} \tabularnewline
				\hline
				\multirow{2}{*}{
				\hspace{-2.5em}
				\begin{tabular}{c}
				$p_{T,\mathrm{min}}$ bin\\
				$[$GeV$]$
				\end{tabular}
				\hspace{-2.5em}} &
				\multicolumn{2}{c|}{Number of expected events}\tabularnewline
				\cline{2-3} &  \rule{0pt}{1.15em}Signal & Background \tabularnewline
				\hline
				$[0-300]$ & 
				$\begin{aligned} \phSpa
				118 \,& + 1175 \,\chqt
					- (17 \pm 20) \,\chq + 248 \,\chu - (123\pm13) \,\chd + 3479 \,\left(\chqt\right)^{2}\\
					& + 3064 \,\left(\chq\right)^{2}
					\rule[-1.em]{0pt}{1em}
					+ 1724 \,\left(\chu\right)^{2} +  1300 \,\left(\chd\right)^{2} - (1190 \pm 171)\,\chqt\,\chq
				\end{aligned}$ & $492\pm50$\tabularnewline
				\hline 
				 $[300-350]$ &  $\begin{aligned} \phSpa 117 \,& + 1423 \,\chqt
					- (123 \pm 21)\,\chq + 272 \,\chu - (77\pm13) \,\chd + 5222 \,\left(\chqt\right)^{2}\\
					\rule[-1.em]{0pt}{1em}& + 4643 \,\left(\chq\right)^{2} + 2670\,\left(\chu\right)^{2} +  1810 \,\left(\chd\right)^{2} - (1670 \pm 191)\,\chqt\,\chq
				\end{aligned}$ & $492\pm43$ \tabularnewline
				\hline
				 $[350-\infty]$ &  $\begin{aligned} \phSpa 111 \,& + 2115 \,\chqt
					- (217\pm15)\,\chq + 489 \,\chu - (162\pm9) \,\chd + 12820 \,\left(\chqt\right)^{2}\\
					\rule[-1.em]{0pt}{1em}& + 11790 \,\left(\chq\right)^{2} + 7060\,\left(\chu\right)^{2} +  4650\,\left(\chd\right)^{2} - 4700\,\chqt\,\chq
				\end{aligned}$ & $243\pm16$ \tabularnewline
				\hline
\end{tabular}
\end{scriptsize}
	\caption{Number of expected signal and background events in the $Zh \rightarrow \nu \bar{\nu} b\bar{b}$ channel, boosted category, at HL-LHC.
	}
	\label{tab:App_sigma_full_Zh_neut_HL_LHC_boos}
\end{table}

We report the number of signal and background events in the 0-lepton channel at HL-LHC in tables~\ref{tab:App_sigma_full_Zh_neut_HL_LHC_res} and \ref{tab:App_sigma_full_Zh_neut_HL_LHC_boos} for the resolved and boosted channels respectively. The corresponding results for FCC-hh are shown in tables~\ref{tab:App_sigma_full_Zh_neut_FCC_res} and~\ref{tab:App_sigma_full_Zh_neut_FCC_boos}. Notice that in this channel the number of signal events includes the contributions from both $Zh\to\nu\bar\nu b\bar b$ and $Wh\to \nu\ell b \bar b$ with a missing lepton.

In the last bin in table~\ref{tab:App_sigma_full_Zh_neut_FCC_boos} we found no background events in the Monte Carlo simulation. We verified however that our bounds are almost unaffected if the bin is excluded from the analysis, so that the exact determination of the background is not essential.

\begin{table}[t]
\centering
\begin{scriptsize}
\begin{tabular}{|@{\;}c@{\hspace{.45em}}|c|c@{\;}|@{\;}c@{\hspace{.5em}}|}
				\hline
				\multicolumn{4}{|c|}{0-lepton channel, resolved, FCC-hh} \tabularnewline
				\hline
			    \multirow{2}{*}{
				\hspace{-2.5em}
				\begin{tabular}{c}
				$p_{T,\mathrm{min}}$ bin\\
				$[$GeV$]$
				\end{tabular}
				\hspace{-2.5em}} &\multirow{2}{*}{$|y_{h}|$ bin} & \multicolumn{2}{c|}{Number of expected events}\tabularnewline
				\cline{3-4} & & \rule{0pt}{1.15em}Signal & Background \tabularnewline
				\hline
				\multirow{2}{*}[-22pt]{$[0-200]$} & $[0-2]$ & 
				$\begin{aligned} & \rule{0pt}{1.25em} 2.63\times\!10^5 + 9.8\times\!10^5 \,\chqt
					+ (1.3\pm0.14)\times\!10^5\,\chq + (1.3\pm0.1) \times\!10^5 \,\chu \\
					& \hspace{1.5em} - (6\pm1)\times\!10^4 \,\chd + 1.11\times\!10^6 \left(\chqt\right)^{2}+ 8.04\times\!10^5 \left(\chq\right)^{2}\\
					\rule[-.65em]{0pt}{1em}& \hspace{1.5em} + (3.8\pm0.2) \times10^5\left(\chu\right)^{2} + 4.1 \times\!10^5 \left(\chd\right)^{2} + ( 1.0 \pm 0.8) \times\!10^5 \,\chqt\,\chq
				\end{aligned}$ & $1.48\times\!10^7 $\tabularnewline
				\cline{2-4}
				& $[2-6]$ & $\begin{aligned} &  \rule{0pt}{1.25em} 2.08\times\!10^5 + 7.75\times\!10^5 \,\chqt
					+ (1.2\pm 1.3)\times\!10^4 \,\chq + (1.3\pm 0.1) \times\!10^5 \,\chu\\
					& \hspace{1.5em}- (4.4\pm0.9)\times\!10^4 \,\chd+ 8.72\times\!10^5 \left(\chqt\right)^{2}+ 5.97\times\!10^5 \left(\chq\right)^{2}\\
					\rule[-.65em]{0pt}{1em}& \hspace{1.5em}+ (3.4\pm0.2)\times\!10^5\left(\chu\right)^{2} +  (2.53\pm0.13)\times\!10^5\left(\chd\right)^{2}\\
					\rule[-.75em]{0pt}{1em}& \hspace{1.5em} - (1.1 \pm 0.7) \times\!10^5 \,\chqt\,\chq
				\end{aligned}$ & $6.6\times10^6$\tabularnewline
				\hline 
				 \multirow{2}{*}[-22pt]{$[200-400]$} & $[0-2]$ &  $\begin{aligned} & \rule{0pt}{1.25em} 3.05\times\!10^4 \, + 1.99\times\!10^5 \,\chqt
					+(2.3\pm0.4)\times\!10^4 \,\chq + (3.4\pm0.3)\times\!10^4 \,\chu \\
					& \hspace{1.5em}- (1.8\pm0.3)\times\!10^4 \,\chd + 4.14\times\!10^5 \left(\chqt\right)^{2} + 3.3\times\!10^5 \left(\chq\right)^{2}\\
					\rule[-.65em]{0pt}{1em}& \hspace{1.5em}+ 1.65\times\!10^5 \left(\chu\right)^{2} +  1.74\times\!10^5 \left(\chd\right)^{2}
					+ ( 4.2 \pm 2.8)\times\!10^4\,\chqt\,\chq
				\end{aligned}$ &
				\hspace{-1.5em}
				\begin{tabular}{c}
				$5.6\,\times 10^5$\\
				$\pm\,0.4\times\!10^5$
				\end{tabular}
				\hspace{-1.5em}
				\tabularnewline
				\cline{2-4}
				& $[2-6]$ & $\begin{aligned} & \rule{0pt}{1.25em} 2.3\times\!10^4 + 1.50\times\!10^5 \,\chqt
					- (4200 \pm 3300)\,\chq + (2.6 \pm 0.2)\times\!10^4 \,\chu \\
					& \hspace{1.5em}- ( 9500 \pm 2100 ) \,\chd + 3.13\times\!10^5 \left(\chqt\right)^{2}+ 2.38\times\!10^5 \left(\chq\right)^{2}\\
					\rule[-.65em]{0pt}{1em}
					& \hspace{1.5em}+ 1.32\times\!10^5 \left(\chu\right)^{2} +  9.65\times\!10^4 \left(\chd\right)^{2} - ( 7.3 \pm 2.4 )\times\!10^4\,\chqt\,\chq
				\end{aligned}$ & $ 2.53\times10^5 $\tabularnewline
				\hline 
				\multirow{2}{*}[-22pt]{$[400-600]$} & $[0-2]$ &  $\begin{aligned} & \rule{0pt}{1.25em} (283\pm24) + (5900\pm400) \,\chqt
					+ (763\pm340) \,\chq  + (860\pm260)  \,\chu \\
					& \hspace{1.5em}- (520\pm250) \,\chd + 4.1\times\!10^4 \left(\chqt\right)^{2}+ (3.6\pm0.2)\times\!10^4 \left(\chq\right)^{2}\\
					\rule[-.65em]{0pt}{1em}& \hspace{1.5em} + (1.8\pm0.1)\times\!10^4 \left(\chu\right)^{2} + (1.9\pm0.1)\times\!10^4 \left(\chd\right)^{2} \\
					\rule[-.75em]{0pt}{1em} & \hspace{1.5em} - ( 9200 \pm 5700 )\,\chqt\,\chq
				\end{aligned}$ & $ 1700\pm400 $ \tabularnewline
				\cline{2-4}
				& $[2-6]$ & $\begin{aligned} &  \rule{0pt}{1.25em} (178\pm19) + (3500\pm310) \,\chqt
					- (330\pm260) \,\chq + (770\pm220) \,\chu \\
					& \hspace{1.5em}- (400\pm180) \,\chd + 2.4\times\!10^4 \left(\chqt\right)^{2} + (2.1\pm0.16)\times\!10^4 \left(\chq\right)^{2}\\
					\rule[-.65em]{0pt}{1em}& \hspace{1.5em}+ (1.33\pm0.09)\times\!10^4 \left(\chu\right)^{2} + (9500\pm700) \left(\chd\right)^{2}\\
					\rule[-.75em]{0pt}{1em} & \hspace{1.5em}- (1.5 \pm 0.5)\times\!10^4 \,\chqt\,\chq
				\end{aligned}$ & $ 850\pm260 $\tabularnewline
				\hline 
				\multirow{2}{*}[-22pt]{$[600-800]$} & $[0-2]$ &  $\begin{aligned} & \rule{0pt}{1.25em} (61\pm 5) + (1700\pm400) \,\chqt
					+ (220\pm210) \,\chq + (540\pm130) \,\chu \\
					& \hspace{1.5em}- (400\pm130) \,\chd + 3.4\times\!10^4 \left(\chqt\right)^{2}+ 3.4\times\!10^4 \left(\chq\right)^{2}\\
					\rule[-.65em]{0pt}{1em}& \hspace{1.5em}+ 1.71\times\!10^4 \left(\chu\right)^{2} +  1.77\times\!10^4 \left(\chd\right)^{2} - ( 1200 \pm 5000) \,\chqt\,\chq
				\end{aligned}$ & $ 330\pm100 $ \tabularnewline
				\cline{2-4}
				& $[2-6]$ & $\begin{aligned} & \rule{0pt}{1.25em} (27\pm4) + (1400\pm320) \,\chqt
					+ (32\pm 140)\,\chq + (480\pm93) \,\chu \\
					&  \hspace{1.em}- (110\pm84) \,\chd + (1.9\pm0.1)\times\!10^4 \left(\chqt\right)^{2}+ (1.6\pm0.1)\times\!10^4 \left(\chq\right)^{2}\\
					\rule[-.65em]{0pt}{1em}& \hspace{1.em}+ (1.13\pm0.07)\!\times\!10^4 \!\left(\chu\right)^{2}\! +\! (6300\pm520)\left(\chd\right)^{2}\!-\! (3500 \pm 3500)\,\chqt\,\chq
				\end{aligned}$ & $ 150\pm50 $\tabularnewline
				\hline 
				\multirow{2}{*}[-22pt]{$[800-\infty]$} & $[0-2]$ &  $\begin{aligned} & \rule{0pt}{1.25em} (26\pm 3) + (480\pm850) \,\chqt
					+ (120\pm290) \,\chq  + (380\pm150) \,\chu \\
					& \hspace{1.5em}+ (1.7\pm 180) \,\chd + 4.6\times\!10^4 \left(\chqt\right)^{2}+ (4.4\pm0.3)\times\!10^4 \left(\chq\right)^{2}\\
					\rule[-.65em]{0pt}{1em}& \hspace{1.5em}+ (2.5\pm0.2)\times\!10^4 \left(\chu\right)^{2} +  (2.3\pm0.1)\times\!10^4 \left(\chd\right)^{2} \\
					\rule[-.75em]{0pt}{1em} & \hspace{1.5em}- (4400 \pm 8000)\,\chqt\,\chq
				\end{aligned}$ & $ 36\pm 23 $ \tabularnewline
				\cline{2-4}
				& $[2-6]$ & $\begin{aligned} & \rule{0pt}{1.25em} (7\pm2) + (240\pm560) \,\chqt
					- (35 \pm 160) \,\chq + (160\pm83) \,\chu \\
					& \hspace{1.em}- (100\pm70) \,\chd + (1.6\pm0.1)\times\!10^4 \left(\chqt\right)^{2}+ (1.9\pm0.2)\times\!10^4 \left(\chq\right)^{2}\\
					\rule[-.65em]{0pt}{1em}& \hspace{1.em}+ (1.03\pm0.09)\!\times\!10^4\!\left(\chu\right)^{2}\! +\!  (5100\pm530) \!\left(\chd\right)^{2}\! -\! (8000 \pm 5100)\,\chqt\,\chq
				\end{aligned}$ & $ 3\pm19$\tabularnewline
				\hline 
\end{tabular}
\end{scriptsize}
\caption{Number of expected signal and background events in the $Zh \rightarrow \nu \bar{\nu} b\bar{b}$ channel, resolved category, at FCC-hh.}
        \label{tab:App_sigma_full_Zh_neut_FCC_res}			
\end{table}

\begin{table}[t]
\centering
		\begin{scriptsize}
           \begin{tabular}{|@{\hspace{.25em}}c|c|c|@{\hspace{.35em}}c@{\hspace{.35em}}|}
				\hline
				\multicolumn{4}{|c|}{0-lepton channel, boosted, FCC-hh} \tabularnewline
				\hline
				\multirow{2}{*}{
				\hspace{-2.5em}
				\begin{tabular}{c}
				$p_{T,\mathrm{min}}$ bin\\
				$[$GeV$]$
				\end{tabular}
				\hspace{-2.5em}}  &\multirow{2}{*}{$|y_{h}|$ bin} & \multicolumn{2}{c|}{Number of expected events}\tabularnewline
				\cline{3-4} & & \rule{0pt}{1.15em}Signal & Background \tabularnewline
				\hline
				\multirow{2}{*}[-22pt]{$[0-200]$} & $[0-2]$ & 
				$\begin{aligned} & \rule{0pt}{1.25em} 1.9\!\times\!10^4 + 1.06\!\times\!10^5 \,\chqt
					+ (1.5\pm0.3)\!\times\!10^4 \,\chq \\
					& \hspace{1.25em} + (1.3\pm0.2)\!\times\!10^4 \,\chu - (5300 \pm 2000) \,\chd+ 1.79\!\times\!10^5 \left(\chqt\right)^{2}\\
					\rule[-.65em]{0pt}{1em}
					& \hspace{1.25em} + (1.34\pm0.07)\!\times\!10^5 \left(\chq\right)^{2} + (6.8\pm0.5)\!\times\!10^4 \left(\chu\right)^{2} \\
					\rule[-.75em]{0pt}{1em}& \hspace{1.25em} + (7.1\pm0.4)\!\times\!10^4 \left(\chd\right)^{2} + (307 \pm 2\!\times\!10^4) \,\chqt\,\chq
				\end{aligned}$ & $(1.1\pm0.2)\times10^6$\tabularnewline
				\cline{2-4}
				& $[2-6]$ & $\begin{aligned} & \rule{0pt}{1.25em} 1.62\!\times\!10^4  + 9.1\!\times\!10^4 \,\chqt
					- (480\pm2800) \,\chq \\
					& \hspace{1.25em}+ (1.3\pm0.2)\!\times\!10^4 \,\chu - (6800\pm1800) \,\chd + 1.49\!\times\!10^5 \left(\chqt\right)^{2} \\
					\rule[-.65em]{0pt}{1em}& \hspace{1.25em}+ (1.09\pm0.07)\!\times\!10^5 \left(\chq\right)^{2} + (5.6\pm0.4)\!\times\! 10^4 \left(\chu\right)^{2}\\
					\rule[-.75em]{0pt}{1em}& \hspace{1.25em}+ (4.9\pm0.3)\!\times\! 10^4 \left(\chd\right)^{2} - (5.1 \pm 1.9)\!\times\!10^4 \,\chqt\,\chq
				\end{aligned}$ & $ (4.2\pm0.6)\times10^5 $  \tabularnewline
				\hline 
				 \multirow{2}{*}[-22pt]{$[200-400]$} & $[0-2]$ &  $\begin{aligned} & \rule{0pt}{1.25em} 9.61\!\times\!10^4 + 9.17\!\times\!10^5 \,\chqt
					+ (1.12 \pm 0.07)\!\times\!10^5 \,\chq + 1.46\!\times\!10^5 \,\chu \\
					& \hspace{1.25em}- (7.4\pm0.5)\!\times\!10^4 \,\chd + 2.95\!\times\!10^6 \left(\chqt\right)^{2}+ 2.42\!\times\!10^6 \left(\chq\right)^{2}\\
					\rule[-.65em]{0pt}{1em}& \hspace{1.25em}+ 1.17\!\times\!10^6 \left(\chu\right)^{2} +  1.24\!\times\!10^6 \left(\chd\right)^{2} + (2.0 \pm 0.6)\!\times\!10^5 \,\chqt\,\chq
				\end{aligned}$ & $ (9.8\pm0.6)\times10^5 $ \tabularnewline
				\cline{2-4}
				& $[2-6]$ & $\begin{aligned} & \rule{0pt}{1.25em} 7.19\!\times\!10^4  + 6.76\!\times\!10^5 \,\chqt
					- (4.6\pm0.6)\!\times\!10^4 \,\chq + 1.33\!\times\!10^5 \,\chu \\
					& \hspace{1.25em}- (4.7\pm0.4)\!\times\!10^4 \,\chd + 2.13\!\times\!10^6 \left(\chqt\right)^{2}+ 1.73\!\times\!10^6 \left(\chq\right)^{2}\\
					\rule[-.65em]{0pt}{1em}& \hspace{1.25em}+ 1.02\!\times\!10^6 \left(\chu\right)^{2} + 7.26\!\times\!10^5 \left(\chd\right)^{2} - (6.0 \pm 0.6)\!\times\!10^5 \,\chqt\,\chq
				\end{aligned}$ & $ (5.0\pm0.3)\times10^5 $\tabularnewline
				\hline 
				\multirow{2}{*}[-22pt]{$[400-600]$} & $[0-2]$ &  $\begin{aligned} & \rule{0pt}{1.25em} 1.01\!\times\!10^4 + 2.43\!\times\!10^5 \,\chqt
					+ (2.4\pm0.2)\!\times\!10^4 \,\chq\\
					& \hspace{1.25em}+ 4.2\!\times\!10^4 \,\chu - (2.2\pm0.2)\!\times\!10^4 \,\chd + 1.78\!\times\!10^6 \left(\chqt\right)^{2}\\
					\rule[-.65em]{0pt}{1em}& \hspace{1.25em}+ 1.57\!\times\!10^6 \left(\chq\right)^{2}+ 7.87\!\times\!10^5 \left(\chu\right)^{2} +  7.89\!\times\!10^5 \left(\chd\right)^{2}\\
					\rule[-.75em]{0pt}{1em}& \hspace{1.25em} - (0.1 \pm 3.7)\!\times\!10^4 \,\chqt\,\chq
				\end{aligned}$ & $ (2.1\pm0.2)\times10^4 $ \tabularnewline
				\cline{2-4}
				& $[2-6]$ & $\begin{aligned} & \rule{0pt}{1.25em} 6400 + 1.50\!\times\!10^5 \,\chqt
					- (1.63 \pm 0.18)\!\times\!10^4 \,\chq + 3.5\!\times\!10^4 \,\chu \\
					& \hspace{1.25em} -(9600\pm1100) \,\chd + 1.10\!\times\!10^6 \left(\chqt\right)^{2}+ 9.67\!\times\!10^5 \left(\chq\right)^{2}\\
					\rule[-.65em]{0pt}{1em}& \hspace{1.25em}+ 5.87\!\times\!10^5 \left(\chu\right)^{2} +  3.63\!\times\!10^5 \left(\chd\right)^{2} - ( 4.4 \pm 0.3)\!\times\!10^5 \,\chqt\,\chq
				\end{aligned}$ & $ 9600\pm1000 $\tabularnewline
				\hline 
				\multirow{2}{*}[-22pt]{$[600-800]$} & $[0-2]$ & $\begin{aligned} & \rule{0pt}{1.25em} 510  + 2.24\!\times\!10^4 \,\chqt
					+ (1860\pm550) \,\chq + (3300\pm350) \,\chu \\
					& \hspace{1.25em} - (2100\pm350) \,\chd + 2.71\!\times\!10^5 \left(\chqt\right)^{2}+ 2.45\!\times\!10^5 \left(\chq\right)^{2}\\
					\rule[-.65em]{0pt}{1em}& \hspace{1.25em} + 1.24\!\times\!10^5 \left(\chu\right)^{2} +  1.19\!\times\!10^5 \left(\chd\right)^{2} - (8600 \pm 13000)\,\chqt\,\chq
				\end{aligned}$ & $260\pm60$ \tabularnewline
				\cline{2-4}
				& $[2-6]$ & $\begin{aligned} & \rule{0pt}{1.25em} 250 + (9400\pm800) \,\chqt
					- (2100\pm380)\,\chq + (2900\pm300) \,\chu \\
					& \hspace{1.25em}- (1000\pm200) \,\chd + 1.35\!\times\!10^5 \left(\chqt\right)^{2}+ 1.25\!\times\!10^5 \left(\chq\right)^{2}\\
					\rule[-.65em]{0pt}{1em}& \hspace{1.25em}+ 8.0\!\times\!10^4\left(\chu\right)^{2} +  4.6\!\times\!10^4\left(\chd\right)^{2} - (5.7 \pm 0.9)\!\times\!10^4 \,\chqt\,\chq
				\end{aligned}$ & $ 180 \pm 60 $\tabularnewline
				\hline 
				\multirow{2}{*}[-22pt]{$[800-\infty]$} & $[0-2]$ &  $\begin{aligned} & \rule{0pt}{1.25em} (15\pm2)  + (640\pm370) \,\chqt
					+ (18\pm140) \,\chq + (260\pm70) \,\chu \\
					& \hspace{1.25em}- (120\pm90) \,\chd + (1.8\pm0.1)\!\times\!10^4 \left(\chqt\right)^{2}\\
					\rule[-.65em]{0pt}{1em}& \hspace{1.25em}+ (1.8\pm0.1)\!\times\!10^4 \left(\chq\right)^{2} + (9000\pm700) \left(\chu\right)^{2} \\
					\rule[-.75em]{0pt}{1em}& \hspace{1.25em} +  (8800\pm600) \left(\chd\right)^{2} - (3800 \pm 4100)\,\chqt\,\chq
				\end{aligned}$ & $ 1\pm1 $ \tabularnewline
				\cline{2-4}
				& $[2-6]$ & $\begin{aligned} & \rule{0pt}{1.25em} (9\pm1) + (270\pm230) \,\chqt
					- (110\pm97)  \,\chq + (84\pm54) \,\chu \\
					& \hspace{1.25em}- (11\pm49) \,\chd+ (7300\pm600) \,\left(\chqt\right)^{2}\\
					\rule[-.65em]{0pt}{1em}& \hspace{1.25em}+ (8200\pm830) \,\left(\chq\right)^{2}+ (4400\pm500)\,\left(\chu\right)^{2} \\
					\rule[-.75em]{0pt}{1em}& \hspace{1.25em}+  (2600\pm330) \,\left(\chd\right)^{2} - (4700 \pm 2700)\,\chqt\,\chq
				\end{aligned}$ & $ <1 $\tabularnewline
				\hline 
			\end{tabular}
			\end{scriptsize}
				\caption{Number of expected signal and background events in the 0-lepton channel, boosted category, at FCC-hh.}
	\label{tab:App_sigma_full_Zh_neut_FCC_boos}
			
\end{table}

\clearpage

\subsection{The 1-lepton channel}
\label{app:EvtNumbers_Wh}

In table \ref{tab:App_sigma_full_Wh_HL_LHC_res}, we show the fit of the expected number of events at HL-LHC for signal and background in the 1-lepton channel, resolved category. The same information for the boosted category can be found in table~\ref{tab:App_sigma_full_Wh_HL_LHC_boos}. The number of expected signal and background events at FCC-hh are shown in tables~\ref{tab:App_sigma_full_Wh_FCC_res} and~\ref{tab:App_sigma_full_Wh_FCC_boos} for the resolved and boosted categories respectively.

\begin{table}[t]
		\centering
        \begin{scriptsize}
			\begin{tabular}{|@{\hspace{.35em}}c|c|c@{\hspace{.5em}}|}
				\hline
				\multicolumn{3}{|c|}{1-lepton channel, resolved, HL-LHC} \tabularnewline
				\hline
				\multirow{2}{*}{$p_{T}^{h}$ bin [GeV]} & \multicolumn{2}{c|}{Number of expected events}\tabularnewline
				\cline{2-3} &  \rule{0pt}{1.15em}Signal & Background \tabularnewline
				\hline
				$[0-175]$ & $\rule[-.85em]{0pt}{2.35em} 5100 \, + 14900 \,\chqt + 12800 \,\left(\chqt\right)^{2}$ & $144000\pm9800$\tabularnewline
				\hline 
				$[175-250]$ & $\rule[-.85em]{0pt}{2.35em} 780 \, + 4400 \,\chqt + 6600 \,\left(\chqt\right)^{2}$ & $6550$\tabularnewline
				\hline 
                $[250-\infty]$ & $\rule[-.85em]{0pt}{2.35em} 41 \, + 380 \,\chqt + 950 \,\left(\chqt\right)^{2}$ & $203\pm35$\tabularnewline
				\hline
		\end{tabular}
\end{scriptsize}
	\caption{Number of expected signal and background events in the $Wh \rightarrow \nu \ell b\bar{b}$ channel, resolved category, at HL-LHC.
	}
	\label{tab:App_sigma_full_Wh_HL_LHC_res}
\end{table}

\begin{table}[t]
	\centering
\begin{scriptsize}
			\begin{tabular}{|@{\hspace{.35em}}c|c|c@{\hspace{.5em}}|}
				\hline
				\multicolumn{3}{|c|}{1-lepton channel, boosted, HL-LHC} \tabularnewline
				\hline
				\multirow{2}{*}{$p_{T}^{h}$ bin [GeV]} & \multicolumn{2}{c|}{Number of expected events}\tabularnewline
				\cline{2-3} &  \rule{0pt}{1.15em}Signal & Background \tabularnewline
				\hline
				$[0-175]$ & $\rule[-.85em]{0pt}{2.35em} (26\pm5) \, + (154\pm16) \,\chqt + (221\pm20) \,\left(\chqt\right)^{2}$ & $1630$\tabularnewline
				\hline 
				$[175-250]$ & $\rule[-.85em]{0pt}{2.35em} 560 \, + 3770 \,\chqt + 6650 \,\left(\chqt\right)^{2}$ & $5690$\tabularnewline
				\hline 
                $[250-300]$ & $\rule[-.85em]{0pt}{2.35em} 214 \, + 1920 \,\chqt + 4530 \,\left(\chqt\right)^{2}$ & $1046$\tabularnewline
				\hline 
                $[300-\infty]$ & $\rule[-.85em]{0pt}{2.35em} (79\pm5) \, + 1150 \,\chqt + 4700 \,\left(\chqt\right)^{2}$ & $425\pm25$\tabularnewline
				\hline
			\end{tabular}
\end{scriptsize}
	\caption{Number of expected signal and background events in the $Wh \rightarrow \nu \ell b\bar{b}$ channel, boosted category, at HL-LHC.
	}
	\label{tab:App_sigma_full_Wh_HL_LHC_boos}
\end{table}

\begin{table}[t]
	\centering
		\setlength{\extrarowheight}{0mm}%
          \begin{scriptsize}
			\begin{tabular}{|@{\hspace{.35em}}c|c|c@{\hspace{.5em}}|}
				\hline
				\multicolumn{3}{|c|}{1-lepton channel, resolved, FCC-hh} \tabularnewline
				\hline
				\multirow{2}{*}{$p_{T}^{h}$ bin [GeV]} & \multicolumn{2}{c|}{Number of expected events}\tabularnewline
				\cline{2-3} &  \rule{0pt}{1.15em}Signal & Background \tabularnewline
				\hline
				$[0-200]$ & $\rule[-.85em]{0pt}{2.35em} 1.31\times10^6 \, + 3.91\times 10^6 \,\chqt + 3.58\times 10^6 \,\left(\chqt\right)^{2}$ & $2.73\times10^8$\tabularnewline
				\hline 
				$[200-400]$ & $\rule[-.85em]{0pt}{2.35em} 7.5\times10^4 \, + 5.29\times10^5 \,\chqt + 1.09\times10^6 \,\left(\chqt\right)^{2}$ & $(1.1\pm0.15)\times10^6$\tabularnewline
				\hline 
                $[400-600]$ & $\rule[-.85em]{0pt}{2.35em} (465\pm 61) \, + (10400\pm 900) \,\chqt + 7.2\times10^4 \,\left(\chqt\right)^{2}$ & $(3800\pm1200)$\tabularnewline
				\hline 
				$[600-\infty]$ & $\rule[-.85em]{0pt}{2.35em} (41\pm8) \, + (2700\pm630) \,\chqt + (27000\pm2400) \,\left(\chqt\right)^{2}$ & $(500\pm190)$ \tabularnewline
				\hline 
		\end{tabular}
		\end{scriptsize}
	\caption{Number of expected signal and background events in the $Wh \rightarrow \nu \ell b\bar{b}$ channel, resolved category, at FCC-hh.
	}
	\label{tab:App_sigma_full_Wh_FCC_res}
\end{table}

\begin{table}[t]
	\centering
\begin{scriptsize}
			\begin{tabular}{|@{\hspace{.35em}}c|c|c@{\hspace{.5em}}|}
				\hline
				\multicolumn{3}{|c|}{1-lepton channel, boosted, FCC-hh} \tabularnewline
				\hline
				\multirow{2}{*}{$p_{T}^{h}$ bin [GeV]} & \multicolumn{2}{c|}{Number of expected events}\tabularnewline
				\cline{2-3} &  \rule{0pt}{1.15em}Signal & Background \tabularnewline
				\hline
				$[0-200]$ & $\rule[-.85em]{0pt}{2.35em} 3.63\times10^4 \, + 2.11\times10^5 \,\chqt + 3.09\times10^5 \,\left(\chqt\right)^{2}$ & $(2.9\pm0.2)\times10^6$\tabularnewline
				\hline 
				$[200-400]$ & $\rule[-.85em]{0pt}{2.35em} 2.04\times10^5 \, + 1.97\times 10^6 \,\chqt + 5.43\times 10^6 \,\left(\chqt\right)^{2}$ & $(1.8\pm0.2)\times 10^6 $\tabularnewline
				\hline 
                $[400-600]$ & $\rule[-.85em]{0pt}{2.35em} 1.75\times10^4 \, + 3.96\times10^5 \,\chqt + 2.31\times10^6 \,\left(\chqt\right)^{2}$ & $(3.3\pm0.3)\times10^4 $\tabularnewline
				\hline 
				$[600-800]$ & $\rule[-.85em]{0pt}{2.35em} (210\pm22) \, + (10000\pm1000) \,\chqt + (93000\pm4100) \,\left(\chqt\right)^{2}$ & $350\pm160$ \tabularnewline
				\hline 
				$[800-\infty]$ & $\rule[-.85em]{0pt}{2.35em} (1\pm1) \, + (68\pm175) \,\chqt + (1400\pm600) \,\left(\chqt\right)^{2}$ & $ 33\pm23 $ \tabularnewline
				\hline
		\end{tabular}
        \end{scriptsize}
	\caption{Number of expected signal and background events in the $Wh \rightarrow \nu \ell b\bar{b}$ channel, boosted category, at FCC-hh.
	}
	\label{tab:App_sigma_full_Wh_FCC_boos}
\end{table}

Notice that, in the boosted category, due to the low Monte Carlo statistics, we got a sizeable uncertainty on the signal coefficients and the background in the overflow bin, and on the background in the $[600-800]\,$GeV bin. Varying the results within the $1\sigma$-uncertainty bands, we verified that the change in the bounds is marginal and can be safely ignored.

\subsection{The 2-lepton channel}
\label{app:EvtNumbers_Zh_ll}

In this subsection we report the expected number of signal and background events in the 2-lepton channel. Tables~\ref{tab:App_sigma_full_Zh_lep_HL_LHC_res} and~\ref{tab:App_sigma_full_Zh_lep_HL_LHC_boos} show the results for HL-LHC in the resolved and boosted categories respectively. The FCC-hh results are given in table~\ref{tab:App_sigma_full_Zh_lep_FCC_res} for resolved category and in table~\ref{tab:App_sigma_full_Zh_lep_FCC_boos} for the boosted one.

\begin{table}[t]
	\centering
		\begin{scriptsize}
			\begin{tabular}{|@{\hspace{.35em}}c|c|c@{\hspace{.5em}}|}
				\hline
				\multicolumn{3}{|c|}{2-lepton channel, resolved, HL-LHC} \tabularnewline
				\hline
				\multirow{2}{*}{$p_{T,\mathrm{min}}$ bin [GeV]} & \multicolumn{2}{c|}{Number of expected events}\tabularnewline
				\cline{2-3} &  \rule{0pt}{1.15em}Signal & Background \tabularnewline
				\hline
				 $[175-200]$ &  $\begin{aligned} \rule{0pt}{1.25em} 57 \,& + 277 \,\chqt
					- (3\pm7)\,\chq + (73\pm5) \,\chu - (19\pm4) \,\chd\\
					& + 402 \,\left(\chqt\right)^{2}+ 403 \,\left(\chq\right)^{2}\\
					\rule[-.65em]{0pt}{1em}& + 238\,\left(\chu\right)^{2} +  172 \,\left(\chd\right)^{2} - (141 \pm 47)\,\chqt\,\chq
				\end{aligned}$ & $361\pm21$ \tabularnewline
				\hline 
				 $[200-\infty]$ &  $\begin{aligned} \rule{0pt}{1.25em} 48 \,& + 299 \,\chqt
					- (5\pm6)\,\chq + (65\pm5) \,\chu - (25 \pm 4) \,\chd\\
					& + 580 \,\left(\chqt\right)^{2}+ 560 \,\left(\chq\right)^{2}\\
					\rule[-.65em]{0pt}{1em}& + 324\,\left(\chu\right)^{2} +  256\,\left(\chd\right)^{2} - ( 110 \pm 49)\,\chqt\,\chq
				\end{aligned}$ & $296 \pm 19$ \tabularnewline
				\hline
		\end{tabular}
		\end{scriptsize}
	\caption{ Number of expected signal and background events in the $Zh \rightarrow \ell^+ \ell^- b\bar{b}$ channel, resolved category, at HL-LHC.}
	\label{tab:App_sigma_full_Zh_lep_HL_LHC_res}
\end{table}

\begin{table}[t]
	\begin{center}
		\setlength{\extrarowheight}{0mm}%
		\begin{scriptsize}
			\begin{tabular}{|@{\hspace{.35em}}c|c|c@{\hspace{.5em}}|}
				\hline
				\multicolumn{3}{|c|}{2-lepton channel, boosted, HL-LHC} \tabularnewline
				\hline
				\multirow{2}{*}{$p_{T,\mathrm{min}}$ bin [GeV]} & \multicolumn{2}{c|}{Number of expected events}\tabularnewline
				\cline{2-3} &  \rule{0pt}{1.15em}Signal & Background \tabularnewline
				\hline
				 $[250-\infty]$ &  $\begin{aligned} \rule{0pt}{1.25em} 103 \,& + 974 \,\chqt
					-(53\pm11)\,\chq + 231 \,\chu - (79\pm7) \,\chd\\
					& + 2800 \,\left(\chqt\right)^{2}+ 2850 \,\left(\chq\right)^{2}\\
					\rule[-.65em]{0pt}{1em}& + 1660\,\left(\chu\right)^{2} +  1150\,\left(\chd\right)^{2} - (1070 \pm 93)\,\chqt\,\chq
				\end{aligned}$ & $370\pm21$ \tabularnewline
				\hline
		\end{tabular}
		\end{scriptsize}
	\end{center}
	\caption{ Number of expected signaland background events in the $Zh \rightarrow \ell^+ \ell^- b\bar{b}$ channel, boosted category, at HL-LHC.}
	\label{tab:App_sigma_full_Zh_lep_HL_LHC_boos}
\end{table}

	\begin{table}[t]
		\centering
		\begin{scriptsize}
			\begin{tabular}{|@{\hspace{.35em}}c|c|c@{\hspace{.25em}}|c@{\hspace{.5em}}|}
				\hline
				\multicolumn{4}{|c|}{2-lepton channel, resolved, FCC-hh} \tabularnewline
				\hline
				\multirow{2}{*}{
				\hspace{-2.5em}
				\begin{tabular}{c}
				$p_{T,\mathrm{min}}$ bin\\
				$[$GeV$]$
				\end{tabular}
				\hspace{-2.5em}}  &\multirow{2}{*}{$|y_{Zh}|$ bin} & \multicolumn{2}{c|}{Number of expected events}\tabularnewline
				\cline{3-4} & & \rule{0pt}{1.15em}Signal & Background \tabularnewline
				\hline
				\multirow{2}{*}[-22pt]{$[0-200]$} & $[0-2]$ & 
				$\begin{aligned} \rule{0pt}{1.25em} 8.08\times10^4 \,& + 2.49\times10^5 \,\chqt + (3.9\pm0.5)\times10^4 \,\chq\\
					& + (4.8\pm0.5)\times10^4 \,\chu - (2.3\pm0.4)\times10^4 \,\chd\\
					& + 2.53\times10^5 \,\left(\chqt\right)^{2}+ 2.62\times10^5 \,\left(\chq\right)^{2}\\
					\rule[-.5em]{0pt}{1em}& + (1.31\pm0.09)\times10^5 \,\left(\chu\right)^{2} +  1.27\times10^5 \,\left(\chd\right)^{2}\\
					\rule[-.65em]{0pt}{1em} & + (9.2 \pm 29)\times10^3 \,\chqt\,\chq
				\end{aligned}$ & $2.95\times10^6$\tabularnewline
				\cline{2-4}
				& $[2-6]$ & $\begin{aligned} \rule{0pt}{1.25em} 6.3\times10^4 \,& + 1.88\times10^5 \,\chqt - (500\pm4800) \,\chq \\& 
				    + (4.4\pm0.4)\times10^4 \,\chu - (2.2\pm0.3)\times10^4 \,\chd\\
					& + 1.88\times10^5 \,\left(\chqt\right)^{2}+ 1.95\times10^5 \,\left(\chq\right)^{2}\\
					\rule[-.5em]{0pt}{1em}& + (1.14\pm0.08)\times10^5 \,\left(\chu\right)^{2} + (8.6\pm0.5)\times10^4 \,\left(\chd\right)^{2}\\
					\rule[-.65em]{0pt}{1em}& - ( 5 \pm 3)\times10^4 \,\chqt\,\chq
				\end{aligned}$ & $ 1.41\times10^6 $\tabularnewline
				\hline 
				 \multirow{2}{*}[-22pt]{$[200-400]$} & $[0-2]$ &  $\begin{aligned} \rule{0pt}{1.25em} 5930 \,& + 3.91\times10^4 \,\chqt
					+ (6000\pm1000)  \,\chq\\& + (6900\pm700) \,\chu - (3900\pm700) \,\chd\\
					& + 8.4\times10^4 \,\left(\chqt\right)^{2} + 8.0\times10^4 \,\left(\chq\right)^{2}\\
					\rule[-.5em]{0pt}{1em}& + 4.0\times10^4 \,\left(\chu\right)^{2} +  4.4\times10^4\,\left(\chd\right)^{2}\\
					\rule[-.65em]{0pt}{1em}& + (5.3 \pm 8)\times10^3\,\chqt\,\chq
				\end{aligned}$ & $ 7.94\times10^4 $ \tabularnewline
				\cline{2-4}
				& $[2-6]$ & $\begin{aligned} \rule{0pt}{1.25em} 4230 \,& + 2.75\times10^4 \,\chqt
					- (1090\pm890)\,\chq\\& + (7000\pm700) \,\chu - (1700\pm600) \,\chd\\
					& + 5.9\times10^4 \,\left(\chqt\right)^{2}+ 5.6\times10^4 \,\left(\chq\right)^{2}\\
					\rule[-.5em]{0pt}{1em}& + 3.6\times10^4 \,\left(\chu\right)^{2} +  2.5\times10^4 \,\left(\chd\right)^{2}\\
					\rule[-.65em]{0pt}{1em}& - (1.5 \pm 0.7)\times10^4 \,\chqt\,\chq
				\end{aligned}$ & $3.5\times10^4$\tabularnewline
				\hline 
				\multirow{2}{*}[-22pt]{$[400-600]$} & $[0-2]$ &  $\begin{aligned}  \rule{0pt}{1.25em} (54\pm4) \,& + (730\pm80) \,\chqt
					+ (120\pm100) \,\chq + (130\pm70) \,\chu \\
					& - (90\pm70) \,\chd + 6000 \left(\chqt\right)^{2}+ (8200\pm500) \left(\chq\right)^{2}\\
					\rule[-.5em]{0pt}{1em}& + (4100\pm300) \,\left(\chu\right)^{2} +  (4200\pm300) \,\left(\chd\right)^{2}\\
					\rule[-.65em]{0pt}{1em}& + (1600 \pm 1500) \,\chqt\,\chq
				\end{aligned}$ & $ 260\pm33 $ \tabularnewline
				\cline{2-4}
				& $[2-6]$ & $\begin{aligned} \rule{0pt}{1.25em} (37\pm3) \,& + (410\pm60) \,\chqt
					- (75\pm76) \,\chq + (180 \pm 60) \,\chu \\
					& - (130\pm50) \,\chd + (3300\pm200) \left(\chqt\right)^{2}+ (4800\pm400) \left(\chq\right)^{2}\\
					\rule[-.5em]{0pt}{1em}& + (3000\pm300) \,\left(\chu\right)^{2} +  (1800\pm180) \,\left(\chd\right)^{2}\\
					\rule[-.65em]{0pt}{1em}& - (2200 \pm 1500)\,\chqt\,\chq
				\end{aligned}$ & $ 100\pm20 $\tabularnewline
				\hline 
				\multirow{2}{*}[-22pt]{$[600-\infty]$} & $[0-2]$ &  $\begin{aligned}  \rule{0pt}{1.25em} (4\pm1) \,& + (170\pm60) \,\chqt
					+ (52\pm40) \,\chq + (52\pm20) \,\chu \\
					& + (10\pm24) \,\chd + (2900\pm200) \left(\chqt\right)^{2}+ (3500\pm300) \left(\chq\right)^{2}\\
					\rule[-.5em]{0pt}{1em}& + (1600\pm170) \,\left(\chu\right)^{2} +  (1600\pm160) \,\left(\chd\right)^{2}\\
					\rule[-.65em]{0pt}{1em}& + (1100 \pm 800)\,\chqt\,\chq
				\end{aligned}$ & $ 14\pm3 $ \tabularnewline
				\cline{2-4}
				& $[2-6]$ & $\begin{aligned} \rule{0pt}{1.25em} (4\pm1) \,& + (11\pm80) \,\chqt
					- (26\pm29) \,\chq + (32\pm17) \,\chu \\
					& - (18\pm14) \,\chd + (1700\pm150) \left(\chqt\right)^{2}+ (1300\pm200) \left(\chq\right)^{2}\\
					\rule[-.5em]{0pt}{1em}& + (800\pm120) \,\left(\chu\right)^{2} + (380\pm80) \,\left(\chd\right)^{2}\\
					\rule[-.65em]{0pt}{1em}& - (520 \pm 590)\,\chqt\,\chq
				\end{aligned}$ & $9\pm3$\tabularnewline
				\hline 

			\end{tabular}
			\end{scriptsize}
				\caption{ Number of expected signal and background events in the $Zh \rightarrow \ell^+ \ell^- b\bar{b}$ channel, resolved category, at FCC-hh.
	}
	\label{tab:App_sigma_full_Zh_lep_FCC_res}
	\end{table}

	\begin{table}[t]
		\centering
		\begin{scriptsize}
			\begin{tabular}{|@{\hspace{.35em}}c|c|c|c@{\hspace{.5em}}|}
				\hline
				\multicolumn{4}{|c|}{2-lepton channel, boosted, FCC-hh} \tabularnewline
				\hline
				\multirow{2}{*}{
				\hspace{-2.5em}
				\begin{tabular}{c}
				$p_{T,\mathrm{min}}$ bin\\
				$[$GeV$]$
				\end{tabular}
				\hspace{-2.5em}}  &\multirow{2}{*}{$|y_{Zh}|$ bin} & \multicolumn{2}{c|}{Number of expected events}\tabularnewline
				\cline{3-4} & & \rule{0pt}{1.15em}Signal & Background \tabularnewline
				\hline
				\multirow{2}{*}[-22pt]{$[0-200]$} & $[0-2]$ & 
				$\begin{aligned} \rule{0pt}{1.25em} 5100 \,& + 2.21\times10^4 \,\chqt
					+ (1700\pm1250) \,\chq\\& + (2300\pm1000) \,\chu - (2200\pm900) \,\chd\\
					& + 2.91\times10^4 \,\left(\chqt\right)^{2}+ (2.9\pm0.3)\times10^4 \,\left(\chq\right)^{2}\\
					\rule[-.5em]{0pt}{1em}& + (1.1\pm0.2)\times10^4 \,\left(\chu\right)^{2} +  (1.6\pm0.15)\times10^4\,\left(\chd\right)^{2}\\
					\rule[-.65em]{0pt}{1em}& + (1.2 \pm 0.7)\times10^4\,\chqt\,\chq
				\end{aligned}$ & $ (6.3\pm 0.8)\times10^4$\tabularnewline
				\cline{2-4}
				& $[2-6]$ & $\begin{aligned} \rule{0pt}{1.25em} 3480 \,& + 1.41\times10^4 \,\chqt
					+ (500\pm1000) \,\chq \\ & + (1600\pm800) \,\chu - (1700\pm700) \,\chd\\
					& +  1.81 \times10^4 \,\left(\chqt\right)^{2}+ (1.7\pm0.2)\times10^4 \,\left(\chq\right)^{2}\\
					\rule[-.5em]{0pt}{1em}& + (9400\pm1600) \,\left(\chu\right)^{2} +  (7800\pm1100)\,\left(\chd\right)^{2} \\
					\rule[-.65em]{0pt}{1em}& - (3600 \pm 6200)\,\chqt\,\chq
				\end{aligned}$ & $ (3.3\pm0.6)\times10^4 $\tabularnewline
				\hline 
				 \multirow{2}{*}[-22pt]{$[200-400]$} & $[0-2]$ &  $\begin{aligned} \rule{0pt}{1.25em} 2.53\times10^4 \,& + 2.14\times10^5 \,\chqt
					+ (3.3\pm0.2)\times10^4\,\chq \\ & + 4.1\times10^4 \,\chu - (2.3\pm0.2)\times10^4 \,\chd\\
					& + 5.83\times10^5 \,\left(\chqt\right)^{2}+ 6.13\times10^5 \,\left(\chq\right)^{2}\\
					\rule[-.5em]{0pt}{1em}& + 2.88\times10^5 \,\left(\chu\right)^{2} + 3.2\times10^5 \,\left(\chd\right)^{2}\\
					\rule[-.65em]{0pt}{1em}& + (6.5 \pm 2.0)\times10^4\,\chqt\,\chq
				\end{aligned}$ & $1.63\times10^5$ \tabularnewline
				\cline{2-4}
				& $[2-6]$ & $\begin{aligned} \rule{0pt}{1.25em} 1.69\times10^4 \,& + 1.42\times10^5 \,\chqt
					- (9200\pm1900) \,\chq \\ 
					& + (3.5\pm0.1)\times10^4 \,\chu - (1.3\pm0.1)\times10^4 \,\chd\\
					& + 3.95\times10^5 \,\left(\chqt\right)^{2}+ 4.1\times10^5 \,\left(\chq\right)^{2}\\
					\rule[-.5em]{0pt}{1em}& + 2.42\times10^5\,\left(\chu\right)^{2} + 1.68\times10^5 \,\left(\chd\right)^{2}\\
					\rule[-.65em]{0pt}{1em}& - (1.55 \pm 0.16)\times10^5\,\chqt\,\chq
				\end{aligned}$ & $6.5\times10^4$\tabularnewline
				\hline 
				\multirow{2}{*}[-22pt]{$[400-600]$} & $[0-2]$ &  $\begin{aligned} \rule{0pt}{1.25em} 2850 \,& + 3.77\times10^4 \,\chqt
					+ (8200\pm700)\,\chq + 1.15\times10^4 \,\chu \\
					& - (5700\pm500) \,\chd + 2.30\times10^5 \left(\chqt\right)^{2}+ 3.65\times10^5 \left(\chq\right)^{2}\\
					\rule[-.5em]{0pt}{1em}& + 1.69\times10^5 \,\left(\chu\right)^{2} +  1.79\times10^5 \,\left(\chd\right)^{2} \\
					\rule[-.65em]{0pt}{1em}& + (1.2\pm 1.1)\times10^4\,\chqt\,\chq
				\end{aligned}$ & $ 3800 $ \tabularnewline
				\cline{2-4}
				& $[2-6]$ & $\begin{aligned} \rule{0pt}{1.25em} 1600 \,& + 2.01\times10^4 \,\chqt
					- (4700\pm500) \,\chq
					+ 8800 \,\chu \\
					& - (2100\pm300) \,\chd + 1.29\times10^5 \left(\chqt\right)^{2}+ 2.17\times10^5 \left(\chq\right)^{2}\\
					\rule[-.65em]{0pt}{1em}& + 1.24\times10^5\,\left(\chu\right)^{2} +  7.7\times10^4\,\left(\chd\right)^{2} \\
					\rule[-.5em]{0pt}{1em}& - ( 7.2 \pm 0.8)\times10^4\,\chqt\,\chq
				\end{aligned}$ & $1600$\tabularnewline
				\hline 
				\multirow{2}{*}[-22pt]{$[600-\infty]$} & $[0-2]$ &  $\begin{aligned}  \rule{0pt}{1.25em} (24\pm2) \,& + (720\pm94) \,\chqt
					+ (190\pm63)\,\chq + (190\pm42) \,\chu \\
					& - (30\pm42) \,\chd + 7500 \left(\chqt\right)^{2}+ (8200\pm500) \left(\chq\right)^{2}\\
					\rule[-.5em]{0pt}{1em}& + (4200\pm260)\,\left(\chu\right)^{2} +  (3400\pm200)\,\left(\chd\right)^{2} \\
					\rule[-.65em]{0pt}{1em}& - ( 210 \pm 1500)\,\chqt\,\chq
				\end{aligned}$ & $ 15 \pm 3 $ \tabularnewline
				\cline{2-4}
				& $[2-6]$ & $\begin{aligned} \rule{0pt}{1.25em} (9\pm1) \,& + (310\pm60) \,\chqt
					- (23\pm42)\,\chq \\ & + (120\pm30) \,\chu + (17\pm25) \,\chd  \\
					& + (3500\pm200) \left(\chqt\right)^{2}+ (3800\pm300) \left(\chq\right)^{2}\\
					\rule[-.5em]{0pt}{1em}& + (2500\pm200) \,\left(\chu\right)^{2} +  (1300\pm130)\,\left(\chd\right)^{2} \\
					\rule[-.65em]{0pt}{1em}& - (2400 \pm 950)\,\chqt\,\chq
				\end{aligned}$ & $ 3 \pm 1 $\tabularnewline
				\hline 

			\end{tabular}
			\end{scriptsize}
				\caption{Number of expected signal and background events in the $Zh \rightarrow \ell^+ \ell^- b\bar{b}$ channel, boosted category, at FCC-hh.
	}
	\label{tab:App_sigma_full_Zh_lep_FCC_boos}
	\end{table}

\clearpage

\bibliographystyle{JHEP.bst}
\bibliography{references}
\end{document}